%% file: main.tex
\RequirePackage[bookmarksnumbered,unicode]{hyperref}
\documentclass[acmtog,nonacm]{acmart}

\AtBeginDocument{%
  \providecommand\BibTeX{{%
    \normalfont B\kern-0.5em{\scshape i\kern-0.25em b}\kern-0.8em\TeX}}}

\setcopyright{cc}
\setcctype{by}
\acmJournal{TOG}
\acmYear{2026} \acmVolume{45} \acmNumber{6} \acmArticle{205}
\acmMonth{12} \acmDOI{10.1145/3842503}

\newcommand{\customcitet}[2]{\citet{#2}}

\acmSubmissionID{papers\_1100}

\input{tex/00_aux_preamble}

\begin{document}

\input{tex/01_aux_document}

\title{Cascaded Non-Line-of-Sight Imaging}

\author{Diego Royo}
\authornote{Both authors contributed equally to this research.}
\correspondingauthor 
\email{droyo@unizar.es}
\orcid{0000-0001-6880-322X}
\affiliation{%
   \institution{Universidad de Zaragoza--I3A}
   \country{Spain}
}

\author{María Peña}
\authornotemark[1]
\email{mpena@unizar.es}
\orcid{0009-0008-0296-0917}
\affiliation{%
   \institution{Universidad de Zaragoza--I3A}
   \country{Spain}
}

\author{Forrest B. Peterson}
\email{fpeterson2@wisc.edu}
\orcid{https://orcid.org/0009-0002-1179-2587}
\affiliation{%
   \institution{University of Wisconsin--Madison}
   \country{United States of America}
}

\author{Andreas Velten}
\email{velten@wisc.edu}
\orcid{0000-0001-5591-828X}
\affiliation{%
   \institution{University of Wisconsin--Madison}
   \country{United States of America}
}

\author{Julio Marco}
\email{juliom@unizar.es}
\orcid{0000-0001-9960-8945}
\affiliation{%
   \institution{Universidad de Zaragoza--I3A}
   \country{Spain}
}

\author{Diego Gutiérrez}
\email{diegog@unizar.es}
\orcid{0000-0002-7503-7022}
\affiliation{%
   \institution{Universidad de Zaragoza--I3A}
   \country{Spain}
}

\renewcommand{\shortauthors}{Royo et al.}

\begin{abstract}

\input{tex/02_abstract.tex}

\end{abstract}



\maketitle

\input{tex/10_introduction}

\input{tex/20_related_work}
\input{journal_modified_tex/30_background}

\input{journal_modified_tex/40_cascaded_camera}
\input{journal_modified_tex/51_methods}
\input{journal_modified_tex/52_virtual_reflectance_droyo}
\input{journal_modified_tex/60_results}

\input{journal_modified_tex/70_real_experiments}
\input{journal_modified_tex/80_conclusions_discussion}

\begin{acks}
    \input{tex/90_acks}
\end{acks}

\bibliographystyle{ACM-Reference-Format}
\bibliography{bibliography}

\end{document}


\setcounter{figure}{0}
\renewcommand{\thefigure}{S.\arabic{figure}}
\setcounter{equation}{0}
\renewcommand{\theequation}{S.\arabic{equation}}

\input{tex/01_aux_document}

\title{SUPPLEMENTAL MATERIAL\newline Cascaded Non-Line-of-Sight Imaging}

\author{Diego Royo}
\authornote{Both authors contributed equally to this research.}
\correspondingauthor 
\email{droyo@unizar.es}
\orcid{0000-0001-6880-322X}
\affiliation{%
   \institution{Universidad de Zaragoza--I3A}
   \country{Spain}
}

\author{María Peña}
\authornotemark[1]
\email{mpena@unizar.es}
\orcid{0009-0008-0296-0917}
\affiliation{%
   \institution{Universidad de Zaragoza--I3A}
   \country{Spain}
}

\author{Forrest B. Peterson}
\email{fpeterson2@wisc.edu}
\orcid{https://orcid.org/0009-0002-1179-2587}
\affiliation{%
   \institution{University of Wisconsin--Madison}
   \country{United States of America}
}

\author{Andreas Velten}
\email{velten@wisc.edu}
\orcid{0000-0001-5591-828X}
\affiliation{%
   \institution{University of Wisconsin--Madison}
   \country{United States of America}
}

\author{Julio Marco}
\email{juliom@unizar.es}
\orcid{0000-0001-9960-8945}
\affiliation{%
   \institution{Universidad de Zaragoza--I3A}
   \country{Spain}
}

\author{Diego Gutiérrez}
\email{diegog@unizar.es}
\orcid{0000-0002-7503-7022}
\affiliation{%
   \institution{Universidad de Zaragoza--I3A}
   \country{Spain}
}

\maketitle

\appendix
\input{tex/A6_optimizations}

\input{tex/A1_resolution_limits}
\input{tex/A5_scene_knowledge}

\input{tex/A2_virtual_reflectance}

\input{tex/A2_noise}

\input{tex/A3_setup_pictures}

\bibliographystyle{ACM-Reference-Format}
\bibliography{bibliography}

%% file: tex/00_aux_preamble.tex
\usepackage{soul}
\usepackage{subcaption}
\usepackage{xspace}
\usepackage{cancel}
\usepackage{mathtools}
\usepackage{graphicx}
\graphicspath{{./fig/}}
\usepackage{pifont}
\usepackage{amsmath}
\usepackage[inline]{enumitem}
\usepackage{stmaryrd}
\usepackage{siunitx}
\usepackage[capitalise,noabbrev]{cleveref}
\creflabelformat{equation}{#2\textup{#1}#3}

\AtBeginDocument{\colorlet{defaultcolor}{.}}

%% file: tex/01_aux_document.tex
\newcommand{\fref}[1]{Figure~\ref{#1}}
\newcommand{\ffref}[2]{Figures~\ref{#1}~and~\ref{#2}}
\newcommand{\tref}[1]{Table~\ref{#1}}
\newcommand{\eref}[1]{Equation~\ref{#1}}
\newcommand{\eeref}[2]{Equations~\ref{#1} and \ref{#2}}
\newcommand{\sref}[1]{Section~\ref{#1}}
\newcommand{\ssref}[2]{Sections~\ref{#1}~and~\ref{#2}}
\newcommand{\sssref}[3]{Sections~\ref{#1},~\ref{#2}~and~\ref{#3}}
\newcommand{\aref}[1]{Appendix~\ref{#1}}
\newcommand{\aaref}[2]{Appendices~\ref{#1}~and~\ref{#2}}

\newcommand{\crhide}[1]{#1}
\newcommand{\phide}[1]{#1}

\newcommand{\draft}[1]{#1}

\definecolor{red}{rgb}{0.8,0,0}
\definecolor{purered}{rgb}{1,0,0}
\definecolor{pink}{rgb}{0.9,0,0.9}
\definecolor{darkred}{rgb}{0.6,0,0}
\definecolor{green}{rgb}{0.0,0.5,0}
\definecolor{blue}{rgb}{0,0,0.75}
\definecolor{darkblue}{rgb}{0,0,0.55}
\definecolor{lightcyan}{rgb}{0.5,0.7,0.7}
\definecolor{orange}{rgb}{0.9,0.3,0.1}
\definecolor{purple}{rgb}{0.6,0.0,0.6}
\definecolor{cyan}{rgb}{0.0,0.7,0.7}
\definecolor{darkgray}{rgb}{0.4,0.4,0.4}
\definecolor{bronze}{rgb}{0.7, 0.4, 0.18}
\definecolor{dorange}{rgb}{0.75, 0.4, 0.0}
\definecolor{darkgray}{rgb}{0.25,0.25,0.25}
\definecolor{black}{rgb}{0.0,0.0,0.0}

\newcommand{\new}[1]{\textcolor{cyan}{#1}}
\newcommand{\final}[1]{#1}

\newcommand{\unsure}[1]{\textcolor{bronze}{(#1?)}}

\newcommand{\checkthis}[1]{\textcolor{bronze}{#1}}

\newcommand{\note}[3][magenta]{\textcolor{#1}{\emph{(\textbf{#2}: #3})}}
\newcommand{\diegoc}[1]{\note[purple]{Diego}{#1}}
\newcommand{\diego}[1]{\note[purple]{Diego}{#1}}
\newcommand{\D}[1]{\note[purple]{Diego}{#1}}
\newcommand{\droyo}[1]{\note[blue]{Droyo}{#1}}
\newcommand{\adolfo}[1]{\note[purple]{Adolfo}{#1}}
\newcommand{\question}[1]{\note[green]{Q}{#1}}
\newcommand{\maria}[1]{\note[dorange]{Maria}{#1}}
\newcommand{\julioc}[1]{\note[pink]{Julio}{#1}}

\newcommand{\todo}[1]{\textcolor{cyan}{#1}}

\newcommand{\juliotxt}[1]{\textcolor{pink}{\emph{\textbf{Julio:} #1}}}
\newcommand{\julio}[1]{{\leavevmode\color{black}{#1}}}
\newcommand{\revise}[1]{{\leavevmode\color{purple}{#1}}}
\newcommand{\OLD}[1]{{\leavevmode\color{lightcyan}{\textit{OLD TEXT}: #1}}}

\newcommand{\andreasc}[1]{\textcolor{gray}{\emph{(\textbf{Andreas:} #1)}}}
\newcommand{\andreastxt}[1]{\textcolor{gray}{\emph{\textbf{Andreas:} #1}}}

\newcommand{\talha}[1]{\note[orange]{Talha}{#1}}

\newcommand{\rarr}{\rightarrow}
\newcommand{\larr}{\leftarrow}
\newcommand{\transpose}{{\scriptstyle\top}}
\renewcommand{\transpose}{{\intercal}}
\newcommand{\minv}[1]{{#1}^{-1}}
\newcommand{\diff}{\mathrm{d}}

\newcommand{\Fspace}[1]{\widehat{#1}}
\newcommand{\Fourier}[1]{\mathcal{F}\left\{#1\right\}}
\newcommand{\invFourier}[1]{\mathcal{F}^{-1}\left\{#1\right\}}
\newcommand{\fq}{\Omega}
\newcommand{\dfq}{\diff \fq}
\newcommand{\Fq}{\mathcal{N}}
\newcommand{\wl}{\lambda}
\newcommand{\wlc}{\wl_c}
\newcommand{\conv}{\ast}
\newcommand{\convt}{\conv_t}
\newcommand{\convs}{\conv_s}
\newcommand\given[1][]{\:#1\vert\:}

\let\norm\undefined
\DeclarePairedDelimiter{\norm}{\lvert}{\rvert}
\DeclarePairedDelimiter{\Norm}{\lVert}{\rVert}
\newcommand{\xa}{\mathbf{a}} %
\newcommand{\xb}{\mathbf{b}} %
\newcommand{\xl}{\mathbf{l}_{\scriptstyle 1}} %
\newcommand{\xs}{\mathbf{s}_{\scriptstyle 1}} %
\newcommand{\xv}{\mathbf{v}} %

\newcommand{\dxl}{\diff \xl} %
\newcommand{\dxs}{\diff \xs} %

\newcommand{\xlp}{\mathbf{l}_{\scriptstyle a}} %
\newcommand{\xsp}{\mathbf{s}_{\scriptstyle b}} %

\newcommand{\xlpp}{\xlp} %
\newcommand{\xspp}{\xsp} %

\newcommand{\Iop}{\Phi} %
\newcommand{\Ifop}{\Fspace{\Iop}} %
\newcommand{\Ibp}{\Iop_\text{bp}}
\newcommand{\Icc}{\Iop_\text{cc}}
\newcommand{\Ifcc}{\Ifop_\text{cc}}
\newcommand{\Ipc}{\Iop_\text{pc}}
\newcommand{\Itpc}{\Iop_\text{tpc}}
\newcommand{\Ifpc}{\Ifop_\text{tpc}}
\newcommand{\Ifproj}{\Ifop_\text{tp}}
\newcommand{\Itproj}{\Iop_\text{tp}}
\newcommand{\Ifcam}{\Ifop_\text{tc}}
\newcommand{\Itcam}{\Iop_\text{tc}}

\newcommand{\Hfiltert}{H}
\newcommand{\Hfilterf}{\Fspace{\Hfiltert}}
\newcommand{\Hf}{\Fspace{H}}
\newcommand{\Hp}{H^\prime}
\newcommand{\Hpf}{\Fspace{\Hp}}
\newcommand{\Hpp}{H^{\prime\prime}}
\newcommand{\Hleft}{H_{2,1}}
\newcommand{\Hright}{H_{3,3}}

\newcommand{\fleft}{f_{2,1}}
\newcommand{\ffront}{f_{1,1}}
\newcommand{\fright}{f_{3,3}}
\newcommand{\ftop}{f_{2,1}}
\newcommand{\fnorm}{\dagger}
\newcommand{\fresult}{f_\text{all}}

\newcommand{\Rt}{R}
\newcommand{\Rf}{\Fspace{\Rt}}

\newcommand{\lM}{M}
\newcommand{\lR}{\mathfrak{R}}
\newcommand{\lRp}{\lR^{\prime}}
\newcommand{\lGp}{\lG^{\prime}}
\newcommand{\lL}{{\mathcal{L}_1}}
\newcommand{\lLp}{{\mathcal{L}_2}}
\newcommand{\lS}{{\mathcal{S}_1}}
\newcommand{\lSp}{{\mathcal{S}_2}}
\newcommand{\lV}{\mathcal{V}}
\newcommand{\lW}{\mathcal{W}}

\newcommand{\lRone}{\mathfrak{R}_1}
\newcommand{\lRoneDG}{{\mathcal{R}_1}}
\newcommand{\lRtwo}{\mathfrak{R}_2}
\newcommand{\lRthree}{\mathfrak{R}_3}
\newcommand{\lRleft}{\mathfrak{R}_2}
\newcommand{\lRright}{\mathfrak{R}_3}
\newcommand{\lRtop}{\mathfrak{R}_2}
\newcommand{\lG}{\textcolor{green}{FIXME}}

\newcommand{\pc}{projector-camera }
\newcommand{\tpc}{transient \pc}

%% file: tex/02_abstract.tex
Time-of-flight non-line-of-sight (NLOS) imaging recovers information from hidden objects by analyzing the time of flight of indirect photons scattered on a visible (relay) wall.
Most methods make the simplifying assumption that photons travel exclusively three-bounce paths, thus ignoring other useful information encoded in higher-order photons (with, e.g., four- or five-bounce paths).
We present a novel \emph{cascaded} NLOS imaging approach that leverages higher-order information and allows imaging a broader range of single- and multi-corner scenarios. We combine ultra-fast laser scanning with recent time-gated 2D sensor arrays to capture the scene's impulse response on a visible relay wall.
From the captured impulse response, our method computes an analogous \emph{virtual} impulse response at any other hidden wall.
This effectively allows us to concatenate a second, virtual NLOS imaging system that leverages higher-order illumination.
We validate our cascaded imaging method both in simulation and with a real prototype, demonstrating NLOS imaging with fourth- and fifth-bounce illumination of objects in challenging orientations and hidden around two corners.
\final{We also analyze how wave-based NLOS imaging interacts with rough hidden walls, which explains and helps overcome existing visibility limitations.}
We further illustrate how to image hidden objects from different perspectives, thus observing previously unseen features, by relying on multiple hidden walls.

%% file: tex/10_introduction.tex
  \section{Introduction}\label{sec:introduction}

Non-line-of-sight (NLOS) imaging methods are able to image objects hidden around a corner by capturing and analyzing indirect light on a visible relay wall.
Such NLOS imaging methods take advantage of ultra-fast sensors, able to measure the time of flight of single photons at picosecond resolution (e.g., \citet{Liu2019phasor, OToole2018confocal}). This has led to a wide array of unprecedented imaging capabilities, %
with many potential applications in fields such as autonomous driving, remote sensing or medical imaging \cite{maeda2019recent, Jarabo2017transient}.

\begin{figure}
    \centering
    \captionsetup{skip=-5pt}
    \def\svgwidth{\columnwidth} 
    \begin{small}
    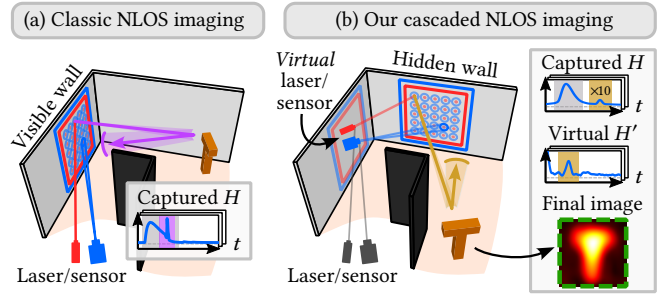
    \end{small}
    \caption{
    (a) Classic time-of-flight NLOS imaging has shown how to image objects hidden around one corner, using time-resolved measurements of third-bounce indirect light (purple) from an ultra-fast laser (red) and sensor (blue).
    (b) Our \textit{cascaded NLOS} imaging method allows imaging objects hidden around two corners by concatenating a second, \textit{virtual} NLOS imaging system and without the need to capture any additional data, effectively using higher-order illumination (e.g., fifth-bounce light in yellow). Our work enables NLOS imaging over a wider, more general set of configurations, helping overcome visibility issues suffered by previous methods.}
    \label{fig:teaser}
\end{figure}

However, most existing methods assume third-bounce-only illumination, where photons interact just once with the hidden scene before reaching the relay wall again (\fref{fig:teaser}a).
Ignoring the broad range of higher-order illumination paths (i.e., fourth- and subsequent-bounce photons) that traverse the hidden scene limits NLOS imaging capabilities to single-corner setups with favorable surface orientations \cite{Liu2019analysis}.
Recent work that exploited higher-order illumination is also restricted to objects in particular locations and orientations, and to paths that follow precise specular directions \cite{royo2023virtual}.    
Thus, in this work, we investigate the following question: can we fully exploit higher-order illumination to extract richer, useful information from hidden scenes?

We introduce a novel \emph{cascaded} NLOS imaging method, showing how higher-order photons can lift some of the main limitations of existing methods. Recent wave-based NLOS systems \cite{Liu2019phasor, Marco2021NLOSvLTM} introduced an imaging paradigm that interprets time-resolved photons captured on the relay wall as light arriving at a virtual aperture. The relay wall could potentially be transformed into a \textit{virtual} line-of-sight (LOS) imaging system; this is the main intuition behind our cascaded approach.

The key idea of cascaded NLOS imaging is to operate on the captured impulse response at the visible relay wall to compute an analogous, \emph{virtual impulse response} at a hidden wall (\fref{fig:teaser}b, left). For this, we introduce two novel virtual LOS imaging tools: our \emph{virtual projector} and \emph{virtual camera} operators, analogous to the ultra-fast laser and sensor from classic NLOS captures.
As a result, we concatenate a second virtual NLOS system \textit{from a single capture} at the visible relay wall, which allows us, for instance, to image objects hidden around \textit{two} corners (\fref{fig:teaser}b, right).

We demonstrate our cascaded imaging method both in simulation and with a real prototype. 
Our acquisition setup pairs an ultra-fast pulsed laser that scans a single, visible relay wall, with a time-gated, $16\times16$ 2D sensor array \cite{riccardoFastGated16162022}.
Ultimately, our cascaded NLOS imaging methods unlock a much broader range of applications. We first employ classic third-bounce imaging to identify potential secondary relay walls within the hidden scene, which we then use to create secondary imaging systems. By leveraging fourth- and fifth-bounce illumination, we showcase NLOS imaging of objects hidden around two corners and objects with limited visibility under existing state-of-the-art methods due to challenging surface orientations. \final{On top of this, we explore how wave-based NLOS imaging methods interact with rough walls, providing novel insights that explain and overcome object visibility limitations.}
Finally, we show how combining cascaded imaging systems at multiple secondary relay walls increases the visibility of hidden objects, enabling us to image them from different points of view.
\final{Code and data for this paper are available at \url{https://graphics.unizar.es/projects/CascadedNLOS/}.}

%% file: fig/teaser2-v3.pdf_tex
\begingroup%
  \makeatletter%
  \providecommand\color[2][]{%
    \errmessage{(Inkscape) Color is used for the text in Inkscape, but the package 'color.sty' is not loaded}%
    \renewcommand\color[2][]{}%
  }%
  \providecommand\transparent[1]{%
    \errmessage{(Inkscape) Transparency is used (non-zero) for the text in Inkscape, but the package 'transparent.sty' is not loaded}%
    \renewcommand\transparent[1]{}%
  }%
  \providecommand\rotatebox[2]{#2}%
  \newcommand*\fsize{\dimexpr\f@size pt\relax}%
  \newcommand*\lineheight[1]{\fontsize{\fsize}{#1\fsize}\selectfont}%
  \ifx\svgwidth\undefined%
    \setlength{\unitlength}{1597.38806056bp}%
    \ifx\svgscale\undefined%
      \relax%
    \else%
      \setlength{\unitlength}{\unitlength * \real{\svgscale}}%
    \fi%
  \else%
    \setlength{\unitlength}{\svgwidth}%
  \fi%
  \global\let\svgwidth\undefined%
  \global\let\svgscale\undefined%
  \makeatother%
  \begin{picture}(1,0.45125927)%
    \lineheight{1}%
    \setlength\tabcolsep{0pt}%
    \put(0,0){\includegraphics[width=\unitlength,page=1]{teaser2-v3.pdf}}%
    \put(0.17940868,0.01683158){\makebox(0,0)[rt]{\smash{\begin{tabular}[t]{r}Laser/sensor\end{tabular}}}}%
    \put(0.59880811,0.01683158){\makebox(0,0)[rt]{\smash{\begin{tabular}[t]{r}Laser/sensor\end{tabular}}}}%
    \put(0.71751046,0.41410758){\makebox(0,0)[t]{\smash{\begin{tabular}[t]{c}(b) Our cascaded NLOS imaging\end{tabular}}}}%
    \put(0.19090236,0.41410758){\makebox(0,0)[t]{\smash{\begin{tabular}[t]{c}(a) Classic NLOS imaging\end{tabular}}}}%
    \put(0,0){\includegraphics[width=\unitlength,page=2]{teaser2-v3.pdf}}%
    \put(0.45494329,0.35146845){\makebox(0,0)[t]{\smash{\begin{tabular}[t]{c}\emph{Virtual}\\laser/\\sensor\end{tabular}}}}%
    \put(0.59026362,0.35492918){\rotatebox{-7.40497917}{\makebox(0,0)[lt]{\smash{\begin{tabular}[t]{l}Hidden wall\end{tabular}}}}}%
    \put(0.01704505,0.21885788){\rotatebox{52.72935911}{\makebox(0,0)[lt]{\smash{\begin{tabular}[t]{l}Visible wall\end{tabular}}}}}%
    \put(0,0){\includegraphics[width=\unitlength,page=3]{teaser2-v3.pdf}}%
    \put(0.90076059,0.12878311){\makebox(0,0)[t]{\smash{\begin{tabular}[t]{c}Final image\end{tabular}}}}%
    \put(0,0){\includegraphics[width=\unitlength,page=4]{teaser2-v3.pdf}}%
    \put(0.27607421,0.13806392){\makebox(0,0)[t]{\smash{\begin{tabular}[t]{c}Captured $H$\end{tabular}}}}%
    \put(0.90142681,0.34755863){\makebox(0,0)[t]{\smash{\begin{tabular}[t]{c}Captured $H$\end{tabular}}}}%
    \put(0.90236437,0.23729707){\makebox(0,0)[t]{\smash{\begin{tabular}[t]{c}Virtual $H'$\end{tabular}}}}%
    \put(0,0){\includegraphics[width=\unitlength,page=5]{teaser2-v3.pdf}}%
    \put(0.34385428,0.06716394){\color[rgb]{0,0,0}\makebox(0,0)[lt]{\lineheight{1.25}\smash{\begin{tabular}[t]{l}$t$\end{tabular}}}}%
    \put(0,0){\includegraphics[width=\unitlength,page=6]{teaser2-v3.pdf}}%
    \put(0.96597819,0.28004994){\color[rgb]{0,0,0}\makebox(0,0)[lt]{\lineheight{1.25}\smash{\begin{tabular}[t]{l}$t$\end{tabular}}}}%
    \put(0,0){\includegraphics[width=\unitlength,page=7]{teaser2-v3.pdf}}%
    \put(0.96597819,0.17001269){\color[rgb]{0,0,0}\makebox(0,0)[lt]{\lineheight{1.25}\smash{\begin{tabular}[t]{l}$t$\end{tabular}}}}%
    \put(0,0){\includegraphics[width=\unitlength,page=8]{teaser2-v3.pdf}}%
    \put(0.9123416,0.30618968){\makebox(0,0)[t]{\smash{\begin{tabular}[t]{c}{\tiny $\times\!10$}\end{tabular}}}}%
  \end{picture}%
\endgroup%

%% file: tex/20_related_work.tex
\section{Related work} 
\paragraph{Time-of-flight (ToF) NLOS imaging:} \customcitet{Kirmani}{kirmani2009looking} originally proposed an NLOS imaging method that emitted ultra-short light pulses towards a relay wall and analyzed the resulting indirect illumination using ToF sensors. 
This idea was later demonstrated by \customcitet{Velten}{Velten2012nc} using ellipsoidal backprojection of signals captured by a streak camera at picosecond resolution. 
Single-photon avalanche diodes (SPADs) emerged as a cheaper alternative \cite{buttafava2015non}, leading to a wide variety of ToF NLOS imaging methods inverting the time of flight of indirect photons through physical \cite{Liu2019phasor, tsai2017geometry, Xin2019theory, pediredla2019snlos,pueyociutad2024polNLOS} or deep learning models \cite{chopite2020CVPR,shen2021NeTF,li2023CVPR,ye2024plug-and-play}.
Some works reduce computational costs using efficient algorithms constrained to confocal acquisition \cite{OToole2018confocal, Lindell2019wave}, performing non-uniform sampling \cite{gu2023fast, sultan2025optimized}, or providing efficient implementations \cite{luesia2023zone}.

The phasor-field formulation \cite{Liu2019phasor} poses NLOS imaging as a wave propagation problem by interpreting the real ToF measurements on the relay wall as \emph{virtual} waves arriving at a wall-sized aperture of a virtual imaging device. This allows exploiting optics-based forward imaging operators to \textit{virtually} illuminate and image the hidden scene \cite{garcia2025forward, dove2019paraxial, dove2020nonparaxial, reza2019wave, teichmanPhasorFieldWaves2019,choi2023selfcalibrating}. 
This formulation has fostered many efficient general NLOS imaging methods \cite{Liu2020phasor,nam2020real,Marco2021NLOSvLTM,luesia2025zerophasepf}.
Our work builds upon the phasor-field formulation, which we extend to reconstruct time-resolved illumination at hidden surfaces different from the visible relay wall, thus leveraging higher-order illumination instead.

\paragraph{NLOS imaging with higher-order illumination:}
Most NLOS imaging methods assume that light follows a three-bounce path: from the relay wall to the hidden scene and back.
Some exceptions include the work by \customcitet{Yi}{Yi2021SIGA}, which used differentiable rendering to optimize an object's position around two corners using fifth-bounce illumination; \customcitet{Marco}{Marco2021NLOSvLTM} separated fourth-bounce illumination by computing virtual light transport matrices (LTMs); and
\customcitet{Sultan}{sultan2024iterating} extended the LTM computation to the time domain, showing time-dependent interreflections in the hidden scene. In our work, we go beyond visualization and parameter optimization tasks, enabling cascaded NLOS imaging in challenging scenes. %

\customcitet{Wang}{wang2024vectorial} rely on polarization to separate third- and fifth-bounce photons. Our work does not require tracking polarization, as it computationally focuses the light to model different bounce paths as needed.
Finally, \customcitet{Royo}{royo2023virtual} leverage fourth- and fifth-bounce illumination to image objects with limited visibility and hidden around two corners, although their method requires planar hidden walls and is limited to hidden objects placed along a subset of specular paths. 
\final{Our cascaded method mitigates these limitations by creating secondary virtual imaging systems. 
This allows us to leverage a broader collection of specular paths, thus imaging a wider range of scene configurations. Moreover, our method does not require these secondary hidden walls to be planar; on the contrary, rough walls allow us to exploit both specular \textit{and non-specular} paths. Lastly, our cascaded method can combine third-, fourth-, and fifth-bounce information simultaneously.}

%% file: journal_modified_tex/30_background.tex
\section{Background on NLOS imaging} 
\label{sec:background}

\begin{figure}
    \centering
    \captionsetup{skip=-1pt}
    \def\svgwidth{\columnwidth} 
    \begin{small}
    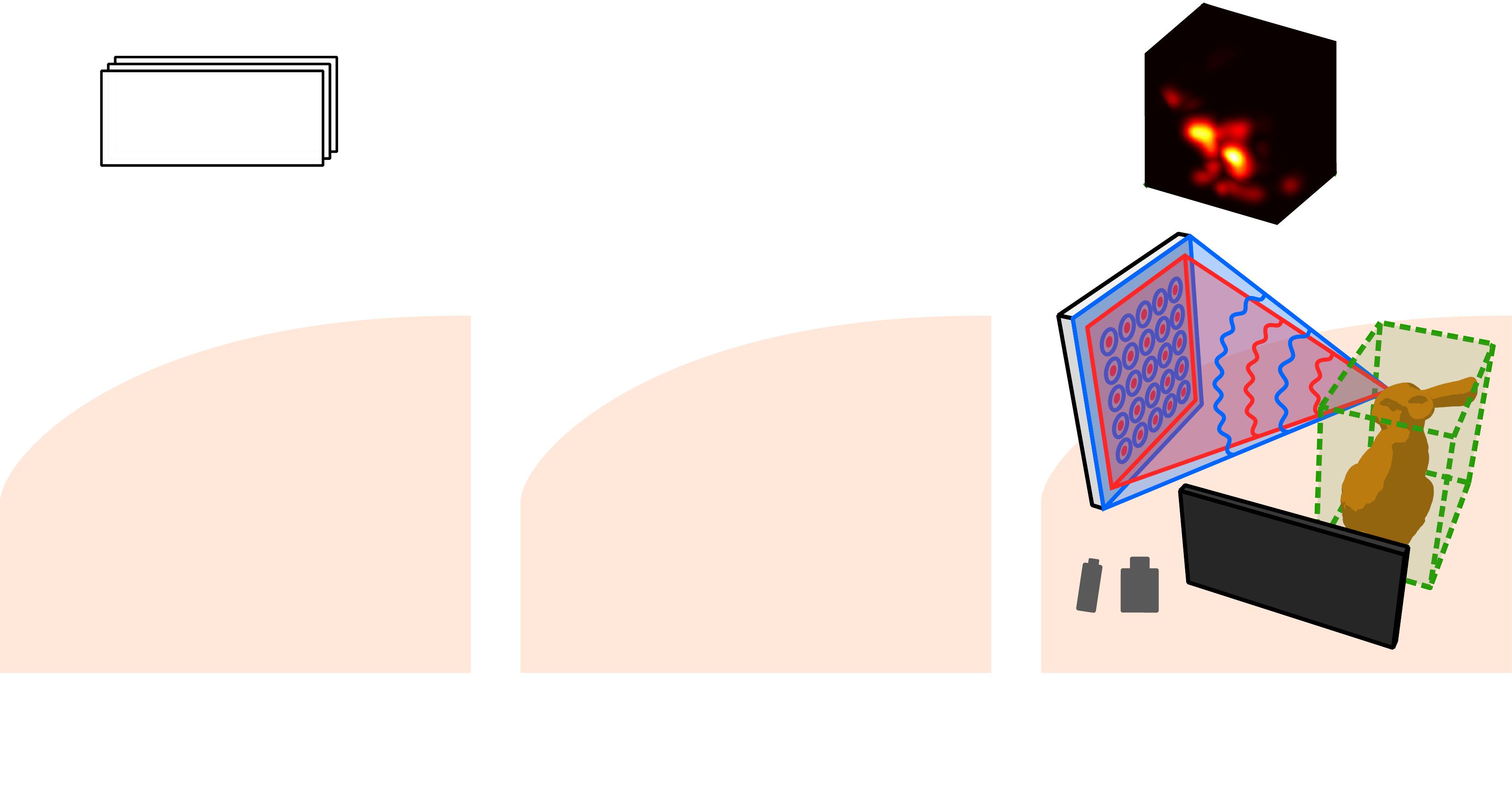
    \end{small}
    \caption{
    (a) A laser device emits delta light pulses $\delta(\xl, t)$, illuminating each point $\xl \in \lL$ on the relay wall $\lRone$.
    (b) An ultra-fast sensing device captures the time-resolved impulse response $H(\xl,\xs,t)$ on multiple visible points $\xs \in \lS$ of the imaging aperture $\lS$ in $\lRone$.
    (c) Imaging operators such as the confocal camera $\Icc(\xv; \Hfiltert)$ transform $H(\xl, \xs, t)$ to obtain an image $f(\xv)$ of the hidden bunny at points $\xv$ on a bounding volume $\lV$.
    }
    \label{fig:RSD_image_formation}
\end{figure}

\fref{fig:RSD_image_formation} illustrates a conventional NLOS imaging setup. A laser individually illuminates points $\xl \in \lL$ on a visible diffuse wall $\lRone$ (the \textit{relay wall}) using ultra-fast light pulses (\fref{fig:RSD_image_formation}a). Light travels to the hidden scene and some is reflected back. An ultra-fast camera captures such indirect illumination from the hidden scene at points $\xs \in \lS$ also on the relay wall (\fref{fig:RSD_image_formation}b). This yields a time-resolved impulse response $H(\xl, \xs, t)$, where $t$ denotes the total time of flight from $\xl$ to $\xs$. \final{We refer the reader to \tref{tab:symbols} for a table of all symbols used in this paper.}

\input{tex/31_symbols_table}

Imaging algorithms then define imaging operators $\Iop(\xv; H)$ that transform $H(\xl, \xs, t)$ to obtain an image $f(\xv)$ of the
hidden geometry at points $\xv$ in a bounding volume $\lV$ (\fref{fig:RSD_image_formation}c) as:
\begin{equation}
f(\xv) = \Phi(\xv; H).
\label{eq:imaging-operator}
\end{equation}
The commonly used backprojection method $\Ibp$ \cite{Velten2012nc, Arellano2017fast} operates by temporally shifting and adding the impulse response:
\begin{equation}
\Ibp(\xv; H) = \int_\lS \int_\lL H(\xl, \xs, t_{\xl \rarr \xv} + t_{\xv \rarr \xs}) \; \dxl \dxs,
\label{eq:backprojection}
\end{equation}
where $t_\mathbf{a \rarr b} = \norm{\xb - \xa}/c$ for any two points $\xa$ and $\xb$, with $c$ the speed of light.
The temporal shift term $t_{\xl \rarr \xv} + t_{\xv \rarr \xs}$ corresponds to the time of flight of the third-bounce path $\xl \rarr \xv \rarr \xs$. Thus, $f(\xv) = \Ibp(\xv; H)$ provides a good estimation of the third-bounce light reflected by the hidden geometry, and thus of the hidden geometry itself.
The majority of time-of-flight NLOS imaging algorithms are based on \eref{eq:backprojection}, proposing different formulations and implementations to improve imaging quality and performance. A notable issue arises from multi-path interference (MPI) from higher-order illumination 
captured in $H$, e.g., fourth or fifth bounce. Methods that assume third-bounce-only illumination thus suffer from imaging artifacts in $f(\xv)$ due to MPI.
A common strategy to improve imaging quality is to filter $f(\xv)$ \cite{Velten2012nc} or $H(\xl, \xs, t)$ \cite{OToole2018confocal, Liu2019phasor} so as to enhance geometric features and mitigate imaging artifacts caused by noise. %

\paragraph{Phasor fields.} In particular, the phasor-field framework \cite{Liu2019phasor} proposes using a band-pass filter with central wavelength $\wlc$ and a standard deviation $\sigma$, denoted by $K(t; \wlc, \sigma)$:
\begin{align}
    K(t; \wlc, \sigma) = e^{i 2 \pi \frac{ct}{\wlc} - \frac{1}{2} \left( \frac{ct}{\sigma} \right)^2  },
    \label{eq:gaussian-pulse}
\end{align}
which is applied to the impulse response function via temporal convolution $K(t; \wlc, \sigma) \convt H(\xl, \xs, t)$. In our work, we apply this filtering strategy to all impulse response functions $H(\xl,\xs,t)$. For notation clarity, we do not explicitly formulate the filtering step throughout the equations, and use $H$ to denote already filtered impulse response functions. We will specify the parameters $\wlc$, $\sigma$ used in each experiment to denote when and how filtering is done.

The phasor-field framework implements \eref{eq:backprojection} in the frequency domain using wave propagation operators, which allows bringing well-known wave-based LOS tools to the NLOS domain.
Most relevant to this paper is their \emph{confocal} camera operator $\Icc$;
while $\Icc$ is analogous to $\Ibp$, its significance lies in its frequency-domain version $\Ifcc$, defined for each imaging frequency $\fq$ as:
\begin{equation}
    \Ifcc(\xv, \fq; \Hfilterf)
    \!=\!\!\int_\lS \!\!\!  \Rf(t_{\xv \rarr \xs}\!, \fq) 
    \!\!\int_\lL \!\!\!\! \Rf(t_{\xl \rarr \xv}, \fq) \Hfilterf(\xl, \xs, \fq) \, \dxl \dxs,
    \label{eq:confocal-camera-freq}
\end{equation}
where $\Hfilterf(\xl, \xs, \fq) = \Fourier{\Hfiltert(\xl, \xs, t)}$ is obtained by applying a Fourier transform over the time domain. Note that this confocal \emph{operator} $\Icc$ is unrelated to confocal \emph{acquisitions} (where $\xl \equiv \xs$). %

The temporal shift in \eref{eq:confocal-camera-freq} can be interpreted using Rayleigh-Sommerfeld Diffraction (RSD) operators $\Rf(t, \fq)$, which express the phase shift of a wave as a function of time $t$ and frequency $\fq$:
\begin{equation}
\Rf(t, \fq) = e^{i 2 \pi \fq t}.
\label{eq:rsd}
\end{equation}
Finally, to obtain the resulting image, one can convert back from the Fourier domain by integrating over all frequencies $\fq$:%
\begin{equation}
    \Icc(\xv; \Hfiltert) = \int_{-\infty}^{+\infty} \Ifcc (\xv, \fq; \Hfilterf ) \, \dfq.
    \label{eq:confocal-camera}
\end{equation}
Using wave-based operators such as the RSD integral allows us to look at the same problem from a different perspective: we can interpret the relay wall $\lRone$ as a computational lens that \emph{focuses} light from points in the illumination aperture $\lL$ (through $\Rf(t_{\xl\rarr \xv}, \fq)$), and in the camera aperture $\lS$ (through $\Rf(t_{\xv\rarr \xs}, \fq)$), both illustrated in \fref{fig:RSD_image_formation}c. Notably, the confocal camera $\Icc(\xv; \Hfiltert)$ focuses light at both apertures $\lL$ and $\lS$ to each point $\xv$ simultaneously, which works for third-bounce illumination.
Throughout the rest of our work we will show how to leverage higher-order illumination in novel imaging algorithms, instead of filtering out this information.

%% file: fig/background-v12.pdf_tex
\begingroup%
  \makeatletter%
  \providecommand\color[2][]{%
    \errmessage{(Inkscape) Color is used for the text in Inkscape, but the package 'color.sty' is not loaded}%
    \renewcommand\color[2][]{}%
  }%
  \providecommand\transparent[1]{%
    \errmessage{(Inkscape) Transparency is used (non-zero) for the text in Inkscape, but the package 'transparent.sty' is not loaded}%
    \renewcommand\transparent[1]{}%
  }%
  \providecommand\rotatebox[2]{#2}%
  \newcommand*\fsize{\dimexpr\f@size pt\relax}%
  \newcommand*\lineheight[1]{\fontsize{\fsize}{#1\fsize}\selectfont}%
  \ifx\svgwidth\undefined%
    \setlength{\unitlength}{1596.75102654bp}%
    \ifx\svgscale\undefined%
      \relax%
    \else%
      \setlength{\unitlength}{\unitlength * \real{\svgscale}}%
    \fi%
  \else%
    \setlength{\unitlength}{\svgwidth}%
  \fi%
  \global\let\svgwidth\undefined%
  \global\let\svgscale\undefined%
  \makeatother%
  \begin{picture}(1,0.5192477)%
    \lineheight{1}%
    \setlength\tabcolsep{0pt}%
    \put(0,0){\includegraphics[width=\unitlength,page=1]{background-v12.pdf}}%
    \put(0.92153778,0.27431134){\color[rgb]{0.12941176,0.61568627,0}\makebox(0,0)[lt]{\lineheight{1.25}\smash{\begin{tabular}[t]{l}\textbf{$\xv$}\end{tabular}}}}%
    \put(0.94048225,0.31477789){\color[rgb]{0.12941176,0.61568627,0}\makebox(0,0)[lt]{\lineheight{1.25}\smash{\begin{tabular}[t]{l}\textbf{$\lV$}\end{tabular}}}}%
    \put(0.89655425,0.46064932){\color[rgb]{0.12941176,0.61568627,0}\makebox(0,0)[lt]{\lineheight{1.25}\smash{\begin{tabular}[t]{l}\textbf{$\lV$}\end{tabular}}}}%
    \put(0.84860331,0.327767){\color[rgb]{0,0,0}\makebox(0,0)[lt]{\lineheight{1.25}\smash{\begin{tabular}[t]{l}\textbf{$\Icc$}\end{tabular}}}}%
    \put(0.89655425,0.4228027){\color[rgb]{0,0,0}\makebox(0,0)[lt]{\lineheight{1.25}\smash{\begin{tabular}[t]{l}\textbf{$f(\xv)$}\end{tabular}}}}%
    \put(0,0){\includegraphics[width=\unitlength,page=2]{background-v12.pdf}}%
    \put(0.49208843,0.34302182){\color[rgb]{0,0.4,1}\makebox(0,0)[lt]{\lineheight{1.25}\smash{\begin{tabular}[t]{l}\textbf{$\lS$}\end{tabular}}}}%
    \put(0.48922704,0.3093488){\color[rgb]{0,0.4,1}\makebox(0,0)[lt]{\lineheight{1.25}\smash{\begin{tabular}[t]{l}\textbf{$\xs$}\end{tabular}}}}%
    \put(0,0){\includegraphics[width=\unitlength,page=3]{background-v12.pdf}}%
    \put(0.14233564,0.34302182){\color[rgb]{1,0.14901961,0.14901961}\makebox(0,0)[lt]{\lineheight{1.25}\smash{\begin{tabular}[t]{l}\textbf{$\lL$}\end{tabular}}}}%
    \put(0.14135494,0.30933966){\color[rgb]{1,0.14901961,0.14901961}\makebox(0,0)[lt]{\lineheight{1.25}\smash{\begin{tabular}[t]{l}\textbf{$\xl$}\end{tabular}}}}%
    \put(0,0){\includegraphics[width=\unitlength,page=4]{background-v12.pdf}}%
    \put(0.15570686,0.02842981){\makebox(0,0)[t]{\lineheight{1.25}\smash{\begin{tabular}[t]{c}(a) Illuminate $\xl$\\with a delta pulse\end{tabular}}}}%
    \put(0.00473183,0.0835533){\makebox(0,0)[lt]{\lineheight{1.25}\smash{\begin{tabular}[t]{l}Laser/sensor\end{tabular}}}}%
    \put(0.00980433,0.34957513){\makebox(0,0)[lt]{\lineheight{1.25}\smash{\begin{tabular}[t]{l}$\lRone$\end{tabular}}}}%
    \put(0.35529905,0.34848336){\makebox(0,0)[lt]{\lineheight{1.25}\smash{\begin{tabular}[t]{l}$\lRone$\end{tabular}}}}%
    \put(0.69959202,0.34848336){\makebox(0,0)[lt]{\lineheight{1.25}\smash{\begin{tabular}[t]{l}$\lRone$\end{tabular}}}}%
    \put(0.16584503,0.13496948){\color[rgb]{1,1,1}\makebox(0,0)[t]{\lineheight{1.25}\smash{\begin{tabular}[t]{c}Occluder\end{tabular}}}}%
    \put(0.34902497,0.0835533){\makebox(0,0)[lt]{\lineheight{1.25}\smash{\begin{tabular}[t]{l}Laser/sensor\end{tabular}}}}%
    \put(0.4999995,0.02799867){\makebox(0,0)[t]{\lineheight{1.25}\smash{\begin{tabular}[t]{c}(b) Capture $H(\xl, \xs, t)$\\at points $\xs$\end{tabular}}}}%
    \put(0.84429214,0.02842981){\makebox(0,0)[t]{\lineheight{1.25}\smash{\begin{tabular}[t]{c}(c) Image the\\hidden scene\end{tabular}}}}%
    \put(0,0){\includegraphics[width=\unitlength,page=5]{background-v12.pdf}}%
    \put(0.5101379,0.13496948){\color[rgb]{1,1,1}\makebox(0,0)[t]{\lineheight{1.25}\smash{\begin{tabular}[t]{c}Occluder\end{tabular}}}}%
    \put(0.85563814,0.13496948){\color[rgb]{1,1,1}\makebox(0,0)[t]{\lineheight{1.25}\smash{\begin{tabular}[t]{c}Occluder\end{tabular}}}}%
    \put(0,0){\includegraphics[width=\unitlength,page=6]{background-v12.pdf}}%
    \put(0.14037354,0.49588388){\color[rgb]{0,0,0}\makebox(0,0)[t]{\lineheight{1.25}\smash{\begin{tabular}[t]{c}$\delta(\xl, t)$\end{tabular}}}}%
    \put(0,0){\includegraphics[width=\unitlength,page=7]{background-v12.pdf}}%
    \put(0.22936438,0.40294931){\color[rgb]{0,0,0}\makebox(0,0)[lt]{\lineheight{1.25}\smash{\begin{tabular}[t]{l}$t$\end{tabular}}}}%
    \put(0,0){\includegraphics[width=\unitlength,page=8]{background-v12.pdf}}%
    \put(0.48331777,0.49640644){\color[rgb]{0,0,0}\makebox(0,0)[t]{\lineheight{1.25}\smash{\begin{tabular}[t]{c}$H(\xl, \xs, t)$\end{tabular}}}}%
    \put(0,0){\includegraphics[width=\unitlength,page=9]{background-v12.pdf}}%
    \put(0.57510197,0.40294931){\color[rgb]{0,0,0}\makebox(0,0)[lt]{\lineheight{1.25}\smash{\begin{tabular}[t]{l}$t$\end{tabular}}}}%
    \put(0,0){\includegraphics[width=\unitlength,page=10]{background-v12.pdf}}%
  \end{picture}%
\endgroup%

%% file: tex/31_symbols_table.tex
\newcommand{\stack}[1]{%
  \begin{tabular}[c]{@{}l@{}}\renewcommand{\arraystretch}{1}#1\end{tabular}}

\begin{table}[t]
  \centering
  \caption{\final{Symbols used throughout this paper. A hat ($\hat{\cdot}$) denotes
  the temporal Fourier transform, with the time $t$ replaced
  by the frequency $\fq$.}}
  \label{tab:symbols}
  \footnotesize
  \setlength{\tabcolsep}{5pt}
  \renewcommand{\arraystretch}{1.3}
  \final{%
  \begin{tabular}{|l|m{0.72\columnwidth}|}
    \hline
    \rule{0pt}{2.7ex}\textbf{Symbol} & \textbf{Description}\rule[-1.1ex]{0pt}{0pt} \\
    \hline
    \noalign{\vskip 3pt}
    \hline
    $\lR_n,\ n=1$                   & The visible relay wall \\ \hline
    $\lR_n,\ n>1$                   & A hidden, secondary relay wall \\ \hline
    $\xlp,\xsp$                     & Illuminated and measured points on $\lR_a$ and $\lR_b$ \\ \hline
    $\mathcal{L}_a,\mathcal{S}_b$   & Illumination and camera apertures on $\lR_a$ and $\lR_b$ \\ \hline
    $H(\xl,\xs,t)$                  & \emph{Captured} impulse response \\ \hline
    $H'_{a,b}(\xlp,\xsp,t)$         & Our \emph{virtual} impulse response \\ \hline
    $K(t;\lambda_c,\sigma)$         & Filter function with wavelength $\lambda_c$ and pulse width $\sigma$ \\ \hline
    $\hat{R}(t,\fq)$                & RSD operator \\ \hline
    \stack{$\Phi_\text{bp}(\xv; H),$\\ $\Phi_\text{cc}(\xv; H)$}
      & Third-bounce imaging operators: backprojection and confocal camera \\ \hline
    \stack{$\Phi_\text{tp}(\xlp, t; H),$\\ $\Phi_\text{tc}(\xsp, t; H)$}
      & Our transient projector and transient camera operators \\ \hline
    $\xv,\ \lV$                     & Imaged point and volume in the hidden scene \\ \hline
    \stack{$f_{a,b}(\xv),$\\ $f_\text{all}(\xv)$}
      & Imaging result using $H_{a,b}$ or $H'_{a,b}$, or a combination of multiple imaging results \\ \hline
  \end{tabular}}
\end{table}

%% file: journal_modified_tex/40_cascaded_camera.tex
\section{Cascaded NLOS imaging}
\label{sec:cascaded}

The key idea of our novel \emph{cascaded NLOS imaging} method is that we can use the virtual imaging system at the visible relay wall to create and leverage \emph{secondary} virtual imaging systems at different hidden walls in the scene (\fref{fig:teaser}b). These secondary systems allow us to take into account higher-order illumination, breaking the third-bounce-only assumption of most conventional methods.%

Here we introduce the building blocks of our method.
First, we introduce the transient camera $\Itcam$ and transient projector $\Itproj$ operators in \sref{sec:cascaded:projector-camera}; in \sref{sec:cascaded:virtual-impulse-response}, we show how to use them to compute the \emph{virtual impulse response} $\Hp$ of a hidden wall, which we then use in \sref{sec:cascaded:concatenation} to turn it into a secondary imaging system.

\subsection{Transient camera and transient projector operators}
\label{sec:cascaded:projector-camera}

\begin{figure}
    \centering
    \captionsetup{skip=0pt}
    \def\svgwidth{\columnwidth} 
    \begin{small}
    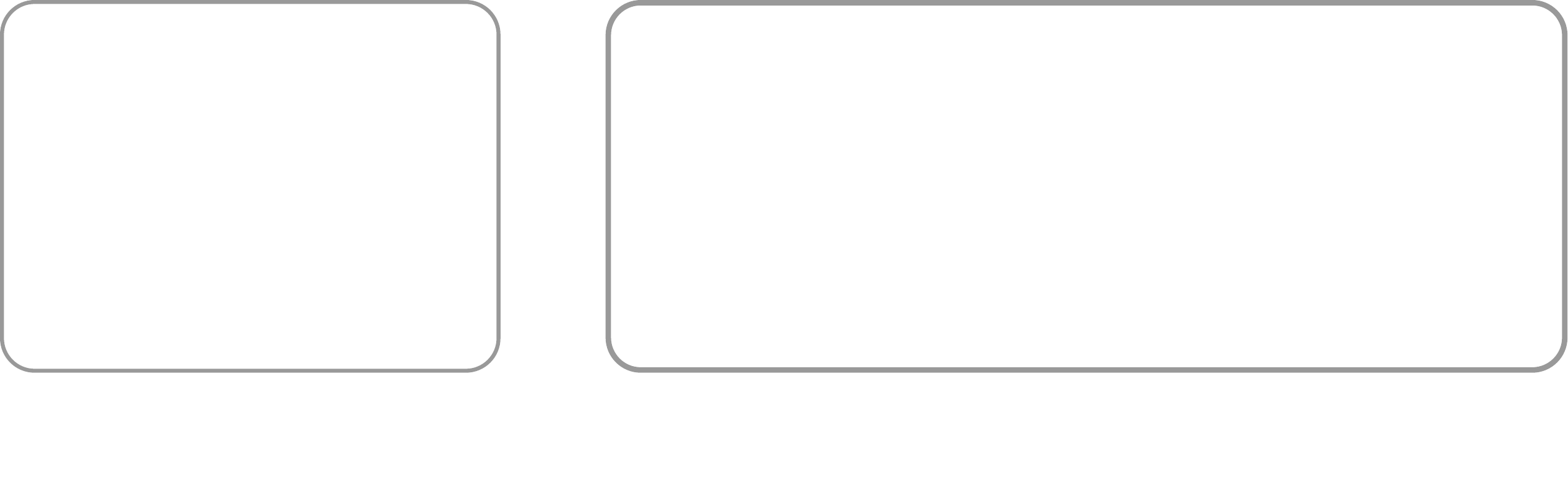
    \end{small}
    \caption{
    (a) The confocal camera operator $\Icc$ computationally focuses light at the illumination aperture $\lL$ and at the camera aperture $\lS$ of $\lRone$ on the same point $\xv$. We separate this procedure into (b) the transient camera $\Itcam$ and (c) our transient projector $\Itproj$ operators. Each focuses the light at $\lS$ or $\lL$ to individual points $\xsp$ and $\xlp$ respectively.}
    \label{fig:cascaded-operators}
\end{figure}

We introduce the transient camera operator $\Itcam$ from work by \customcitet{Liu}{Liu2019phasor}, and our transient projector operator $\Itproj$ as a generalized  virtual projector. We formalize these operators as the building blocks for phasor-based NLOS imaging methods.

As illustrated in \fref{fig:cascaded-operators}a, the confocal camera operator $\Icc$ (\sref{sec:background}) focuses all points $\xs$ of the camera aperture $\lS$ and all points $\xl$ of the illumination aperture $\lL$ simultaneously on the same point $\xv$ of the hidden scene (\eref{eq:confocal-camera-freq}).
Our general paradigm independently formulates the focusing operations of $\lS$ and $\lL$ using the transient camera operator $\Itcam$ (\fref{fig:cascaded-operators}b) and transient projector operator $\Itproj$ (\fref{fig:cascaded-operators}c) respectively; each focuses light to individual points $\xsp$ and $\xlp$ in the hidden scene.
Note that $\xsp$ and $\xlp$ are analogous to $\xv$ in the general case; we use a different notation to highlight that these operations have a different purpose.
This formulation allows us to model light paths from the relay wall to the hidden scene and vice versa, and to reconstruct the time-resolved light transport through the hidden scene. 

The \emph{transient camera operator} $\Itcam$ (\fref{fig:cascaded-operators}b) focuses light from all $\xs \in \lS$ to each $\xsp$. We first define its frequency-domain counterpart $\Ifcam$ for each imaging frequency $\fq$:
\begin{equation}
    \Ifcam(\xsp, \fq; \Hfilterf) = \int_\lS \Rf(t_{\xsp\rarr \xs}, \fq) \, \Hfilterf( \xl, \xs, \fq ) \, \dxs,
    \label{eq:transient-camera-freq}
\end{equation}
with $\Hfilterf( \xl, \xs, \fq ) = \mathcal{F}\left\{ H( \xl, \xs, t ) \right\}$ the Fourier transform in time, and only using one of the two RSD operators $\Rf$ (\eref{eq:rsd}) from the confocal camera operator $\Ifcc$ (\eref{eq:confocal-camera-freq}), since the illuminated points $\xl \in \lL$ do not vary.
We obtain the time-resolved signal via an inverse Fourier transform over all imaging frequencies $\fq$:
\begin{equation}
    \Itcam(\xsp, t; \Hfiltert) = \invFourier{ \; \Ifcam \left(\xsp, \fq; \Hfilterf \right) \; }.
    \label{eq:transient-camera}
\end{equation}
Our \emph{transient projector operator} $\Itproj$ (\fref{fig:cascaded-operators}c) is similar to $\Itcam$, except that $\Itproj$ focuses  light from $\lL$ to  $\xlp$, leaving $\lS$ unchanged. We define its frequency-domain counterpart $\Ifproj$ as:
\begin{equation}
    \Ifproj(\xlp, \fq; \Hfilterf) = \int_\lL \Rf(t_{\xl\rarr \xlp}, \fq) \, \Hfilterf( \xl, \xs, \fq ) \, \dxl,
    \label{eq:transient-projector-freq}
\end{equation}
using the other RSD operator $\Rf$ from \eref{eq:confocal-camera-freq}. Its time-domain version becomes:
\begin{equation}
    \Itproj(\xlp, t; \Hfiltert) = \invFourier{ \; \Ifproj \left(\xlp, \fq; \Hfilterf \right) \; }.
    \label{eq:transient-projector}
\end{equation}

\paragraph{Capture procedure.}
Our transient projector $\Itproj$ and transient camera $\Itcam$ operators enable focusing light onto the hidden scene, analogous to beamforming operators. Effectively focusing light onto a single point requires illuminating and capturing multiple points $\xl$ and $\xs$ in the impulse response $H(\xl, \xs, t)$ over finite apertures $\lL$ and $\lS$, respectively.
To enable single-point focusing using both operators, we illuminate a 2D grid of points $\xl$ on the relay wall with an ultra-fast laser. For each illuminated point $\xl$ we exhaustively capture the resulting indirect photons at a 2D grid of points $\xs$ with a time-gated sensor array. This results in a five-dimensional $H(\xl, \xs, t)$, which can be reduced to 3D in applications that only require single-point focusing using either $\Itcam$ or $\Itproj$. 

\subsection{Virtual impulse response function}
\label{sec:cascaded:virtual-impulse-response}

\begin{figure}
    \centering
    \captionsetup{skip=0pt}
    \def\svgwidth{\columnwidth} 
    \begin{small}
    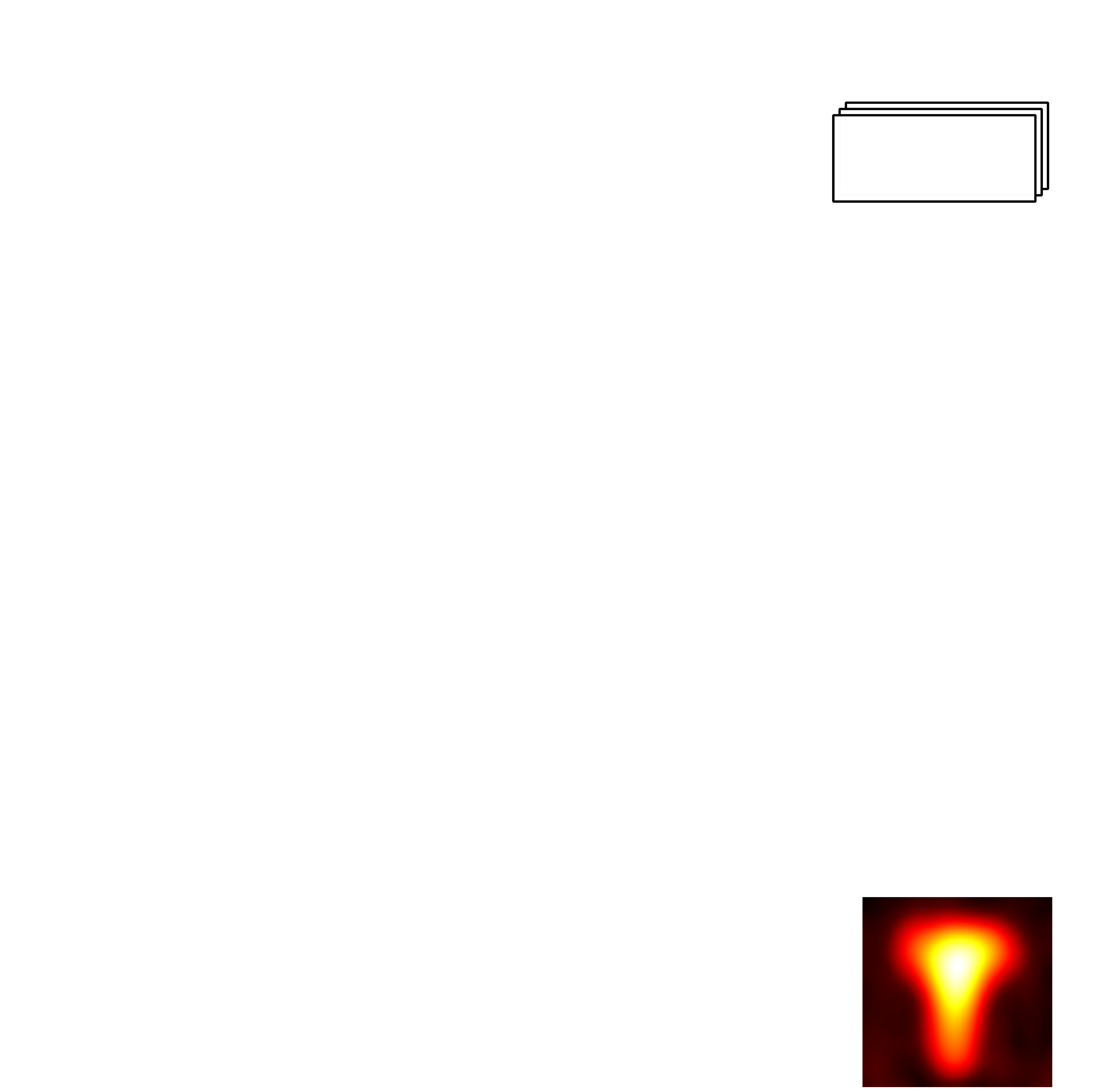 %
    \end{small}
    \caption{
    Cascading NLOS imaging systems. (a) A laser emits pulses to each point $\xl \in \lL$ on $\lRone$ and an ultra-fast camera captures $H(\xl, \xs, t)$ at points $\xs \in \lS$ also on $\lRone$. Here $H(\xl,\xs,t)$ contains third- and fifth-bounce illumination from $\lRtwo$ and T, respectively.
    (b) We compute our virtual impulse response $\Hp_{a, b}(\xlp, \xsp, t)$ on $\lRtwo$ from $H$, i.e., $\Hp_{2,2}(\mathbf{l}_2, \mathbf{s}_2, t)$, using the transient projector $\Itproj$ and camera $\Itcam$ operators to focus light at $\lL$ or $\lS$ to each point $\mathbf{l}_2 \in \lLp$ or $\mathbf{s}_2 \in \lSp$ in $\lRtwo$ respectively. Here $\Hp_{2,2}$ shows light reflected by the T as third-bounce illumination.
    (c) Finally we concatenate a confocal camera operator $\Icc$ with $\Hp_{2,2}$, resulting in an image $f_{2,2}(\xv)$ for $\xv \in \lV$ of the T effectively computed using fifth-bounce illumination.}
    \label{fig:cascaded-camera-setup}
\end{figure}

We next introduce our \emph{virtual impulse response} $\Hp$ of the hidden scene, computed from the  impulse response $H$ captured at the relay wall. For reference, $H(\xl, \xs, t)$ contains the temporal responses at $\xs \in \lS$ of light pulses emitted towards each $\xl \in \lL$ of the relay wall $\lRone$.
Throughout our work, we will use $\Hp_{a,b}(\xlp, \xsp, t)$ to denote our virtual impulse response, which analogously to $H$ contains the \emph{reconstructed} temporal response at $\xsp$ on  $\lR_b$ of light focused at $\xlp$ on $\lR_a$ (note that $\lR_a$ and $\lR_b$ can be different). %
This will allow us in \sref{sec:cascaded:concatenation} to turn walls $\lR_a$ and $\lR_b$ into secondary virtual imaging systems. We use the apostrophe in $\Hp$ to highlight that it is \textit{computed} from $H$ and not captured, with the special case $a = b = 1$ corresponding to $H_{1,1}\equiv H$ in $\lRone$.

We formulate our virtual impulse response $\Hp_{a,b}$ as: 
\begin{equation}
\begin{aligned}
    \Hp_{a, b}(\xlp, \xsp, t) = \Itcam \left( \xsp, t; \Itproj \left(\xlp, t; \Hfiltert \right) \right) ,
    \label{eq:virtual-impulse-response}
\end{aligned}
\end{equation}
using $\Itcam$ (\eref{eq:transient-camera}) to compute at $\xsp$ the light that $\Itproj$ focuses at $\xlp$ (\eref{eq:transient-projector}).
In Appendix A of our Supplemental Material we detail an alternative convolution-based formulation, optimized for computational efficiency.

Note that $\Hp_{a, b} ( \xlp, \xsp, t )$ is effectively computed by temporally shifting and adding the filtered impulse response $\Hfiltert(\xl, \xs, t)$ using the distances of paths $\xl \rarr \xlp$ and $\xsp \rarr \xs$. Given that light follows higher-order paths of the form $\xl \rarr \xlp \rarr \cdots \rarr \xsp \rarr \xs$, the time $t$ of $\Hp_{a, b}(\xlp, \xsp, t)$ represents the time of flight of light from $\xlp$ to $\xsp$, i.e., light subpaths of the form $\xlp \rarr \cdots \rarr \xsp$.

In the special case of $b=1$, we only focus the illumination at $\xlp$ while keeping the camera aperture $\lS$ untouched, which produces a time of flight $t$ associated with subpaths $\xlp \rarr \cdots \rarr \xs$ as:
\begin{equation}
    \Hp_{a, 1}(\xlp, \xs, t) = \Itproj \left(\xlp, t; \Hfiltert \right).
    \label{eq:virtual-impulse-response-onlyL}
\end{equation}
Similarly, when $a=1$, we keep the illumination aperture $\lL$ untouched and measure other points $\xsp$ of the hidden scene, with the time of flight $t$ corresponding to subpaths $\xl \rarr \cdots \rarr \xsp$, as:
\begin{equation}
    \Hp_{1, b}(\xl, \xsp, t) = \Itcam \left(\xsp, t; \Hfiltert \right).
    \label{eq:virtual-impulse-response-onlyS}
\end{equation}

\begin{figure*}[t]
    \centering
    \captionsetup{skip=-2pt}
    \def\svgwidth{\textwidth} 
    \begin{small}
    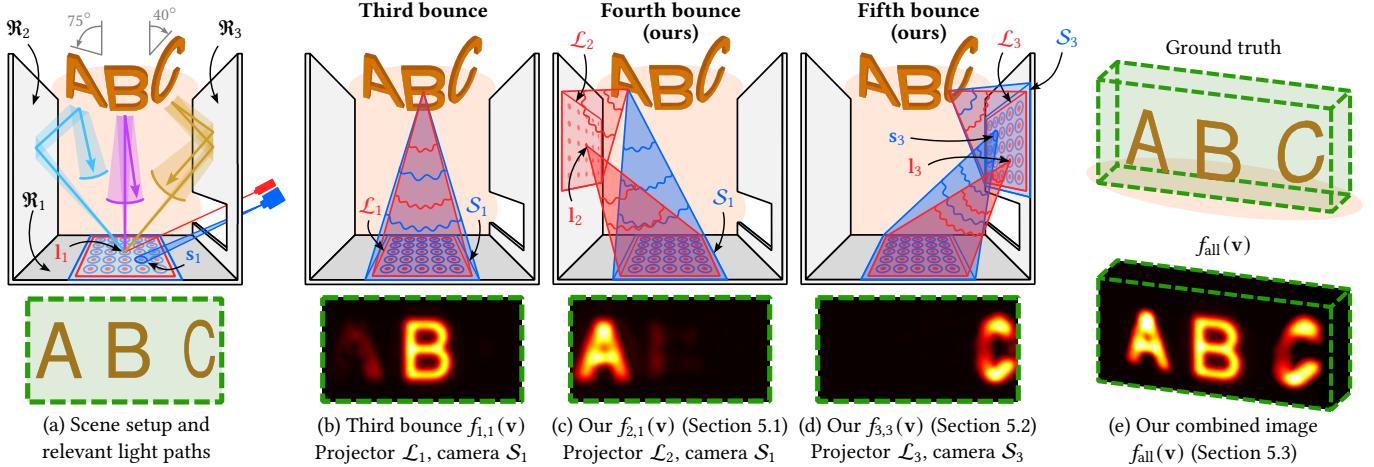
    \end{small}
    \caption{
    Cascaded NLOS imaging methods.
    (a) The simulated impulse response $H(\xl, \xs, t)$ at $\lRone$ contains light that has scattered on the relay wall $\lRone$, the three letters A, B, and C (oriented differently), and on the left $\lRtwo$ and right $\lRthree$ walls. Hence, $H$ contains third- (purple), fourth- (cyan) and fifth-bounce (yellow) illumination. (b) Third-bounce imaging methods use apertures $\lL$ and $\lS$ and thus can only use part of $H$. (c) In this example, our fourth-bounce imaging method concatenates our transient projector operator $\Itproj$ focused at $\mathbf{l}_2\in\mathcal{L}_2$ on the left wall $\lRleft$ along with the confocal camera operator $\Icc$. (d) Our fifth-bounce method concatenates our transient projector $\Itproj$ and camera $\Itcam$ operators focused at $\mathbf{l}_3\in\mathcal{L}_3$ and $\mathbf{s}_3\in\mathcal{S}_3$ respectively on the right wall $\lRright$ along with the confocal camera operator $\Icc$. (e) Combined result.}
    \label{fig:multiview}
\end{figure*}
\fref{fig:cascaded-camera-setup} shows an illustrative example with a letter T hidden around two corners, thus invisible to third-bounce methods. To image it, we leverage $\lRtwo$ as a secondary relay wall.
We illustrate the impulse response $H(\xl, \xs, t)$ at a grid of points for both $\xl$ and $\xs$ on $\lRone$ (\fref{fig:cascaded-camera-setup}a).
As shown in the rightmost graphic, $H$ contains third-bounce illumination from $\lRtwo$ and, importantly, fifth-bounce illumination from the hidden T. From this, we compute the virtual impulse response $\Hp_{2,2}(\mathbf{l}_2, \mathbf{s}_2, t)$ (\eref{eq:virtual-impulse-response}), effectively shifting the illumination and capture positions to $\mathbf{l}_2$ and $\mathbf{s}_2$ in $\lRtwo$ (\fref{fig:cascaded-camera-setup}b). We show how the fifth-bounce illumination in the captured $H$ now corresponds to third-bounce illumination in the reconstructed $\Hp_{2,2}$.

\subsection{Cascading virtual imaging systems}
\label{sec:cascaded:concatenation}
We can now use our virtual impulse response $\Hp_{a,b}(\xlp, \xsp, t)$ with the same imaging operators as the conventional captured impulse response $H(\xl, \xs, t)$. In particular, we can use conventional imaging operators such as the confocal camera $\Icc$ (\eref{eq:confocal-camera}). As this will be common throughout our work, we use $f_{a,b}$ notation (adapted from \eref{eq:imaging-operator}) to refer to images computed using $\Icc$ with $\Hp_{a,b}$ as input:
\begin{equation}
    f_{a,b}(\xv) \equiv \Icc(\xv; \Hp_{a,b}).
    \label{eq:second-reconstruction}
\end{equation}
In \fref{fig:cascaded-camera-setup}c, we use $\Icc$ with $\Hp_{2,2}(\mathbf{l}_2, \mathbf{s}_2, t)$, effectively using fifth-bounce illumination to compute an image $f_{2,2}(\xv)$ of the hidden T. Alternatively, we could also use our $\Itcam$ and $\Itproj$ operators in Equations~\ref{eq:virtual-impulse-response},~\ref{eq:virtual-impulse-response-onlyL}~and~\ref{eq:virtual-impulse-response-onlyS}, which would yield an additional cascaded virtual impulse response $H''$ from $\Hp$, further discussed in \sref{sec:discussion:reconstruction-limits}.  

\paragraph{Cascaded imaging resolution.} Compared to third-bounce imaging, higher-order imaging is a harder problem with more challenging conditions.
Intuitively, the effective imaging resolution obtained from concatenated NLOS setups primarily depends on two factors: it increases with the area and resolution of \emph{all} apertures $\mathcal{L}$ and $\mathcal{S}$, and decreases with the average distance along the paths $\mathcal{L} \rightarrow \mathcal{V} \rightarrow \mathcal{S}$ (i.e., between the apertures and the imaged volume).
We refer the reader to Appendix B in our Supplemental Material for more details, where we reason about the theoretical imaging resolution and provide experimental validation using our hardware prototype.

\paragraph{Secondary relay wall.}
Our method uses a hidden wall as a secondary relay wall. However, note that we do \emph{not} require prior knowledge of the room layout. Instead, we can leverage existing third-bounce methods to reveal such secondary walls with just 1--3 cm of error both in simulated and real captures, which does not affect our cascaded imaging results. We provide several examples in \sssref{sec:results:missing_cone}{sec:results:tracking}{sec:real-experiments:rough} of our results. In Appendix C of our Supplemental Material, we show how using third-bounce methods to find unknown secondary relay walls leads to cascaded imaging results that are practically identical to using exact prior knowledge about the location of such secondary relay walls. This highlights the feasibility of our cascaded approach, without relying on any prior scene information. Finally, to sample points $\xlp$ and $\xsp$ in secondary walls, we use the same sampling spacing as in the primary wall unless stated otherwise. 

%% file: fig/operators-v4.pdf_tex
\begingroup%
  \makeatletter%
  \providecommand\color[2][]{%
    \errmessage{(Inkscape) Color is used for the text in Inkscape, but the package 'color.sty' is not loaded}%
    \renewcommand\color[2][]{}%
  }%
  \providecommand\transparent[1]{%
    \errmessage{(Inkscape) Transparency is used (non-zero) for the text in Inkscape, but the package 'transparent.sty' is not loaded}%
    \renewcommand\transparent[1]{}%
  }%
  \providecommand\rotatebox[2]{#2}%
  \newcommand*\fsize{\dimexpr\f@size pt\relax}%
  \newcommand*\lineheight[1]{\fontsize{\fsize}{#1\fsize}\selectfont}%
  \ifx\svgwidth\undefined%
    \setlength{\unitlength}{1215.92562805bp}%
    \ifx\svgscale\undefined%
      \relax%
    \else%
      \setlength{\unitlength}{\unitlength * \real{\svgscale}}%
    \fi%
  \else%
    \setlength{\unitlength}{\svgwidth}%
  \fi%
  \global\let\svgwidth\undefined%
  \global\let\svgscale\undefined%
  \makeatother%
  \begin{picture}(1,0.30494819)%
    \lineheight{1}%
    \setlength\tabcolsep{0pt}%
    \put(0,0){\includegraphics[width=\unitlength,page=1]{operators-v4.pdf}}%
    \put(0.51240418,0.02908995){\color[rgb]{0,0,0}\makebox(0,0)[t]{\lineheight{1.25}\smash{\begin{tabular}[t]{c}(b) Transient camera\\($\Ifcam$, \eref{eq:transient-camera-freq}) \end{tabular}}}}%
    \put(0.84194956,0.0290891){\color[rgb]{0,0,0}\makebox(0,0)[t]{\lineheight{1.25}\smash{\begin{tabular}[t]{c}(c) Transient projector\\($\Ifproj$, \eref{eq:transient-projector-freq}) \end{tabular}}}}%
    \put(0.67081205,0.1659123){\color[rgb]{0,0.4,1}\makebox(0,0)[t]{\lineheight{1.25}\smash{\begin{tabular}[t]{c}\textbf{$\xsp$}\end{tabular}}}}%
    \put(0,0){\includegraphics[width=\unitlength,page=2]{operators-v4.pdf}}%
    \put(0.57798032,0.25409088){\color[rgb]{0,0,0}\makebox(0,0)[lt]{\lineheight{1.25}\smash{\begin{tabular}[t]{l}\textbf{$\Itcam$}\end{tabular}}}}%
    \put(0,0){\includegraphics[width=\unitlength,page=3]{operators-v4.pdf}}%
    \put(0.8788923,0.22910576){\color[rgb]{0,0,0}\makebox(0,0)[lt]{\lineheight{1.25}\smash{\begin{tabular}[t]{l}\textbf{$\Itproj$}\end{tabular}}}}%
    \put(0.97557861,0.12198692){\color[rgb]{1,0.14901961,0.14901961}\makebox(0,0)[t]{\lineheight{1.25}\smash{\begin{tabular}[t]{c}\textbf{$\xlp$}\end{tabular}}}}%
    \put(0,0){\includegraphics[width=\unitlength,page=4]{operators-v4.pdf}}%
    \put(0.15929205,0.02908936){\color[rgb]{0,0,0}\makebox(0,0)[t]{\lineheight{1.25}\smash{\begin{tabular}[t]{c}(a) Confocal camera\\($\Ifcc$, \eref{eq:confocal-camera-freq}) \end{tabular}}}}%
    \put(0.29080351,0.14320608){\color[rgb]{0.12941176,0.61568627,0}\makebox(0,0)[t]{\lineheight{1.25}\smash{\begin{tabular}[t]{c}\textbf{$\xv$}\end{tabular}}}}%
    \put(0,0){\includegraphics[width=\unitlength,page=5]{operators-v4.pdf}}%
    \put(0.20398919,0.25038966){\color[rgb]{0,0,0}\makebox(0,0)[lt]{\lineheight{1.25}\smash{\begin{tabular}[t]{l}\textbf{$\Icc$}\end{tabular}}}}%
    \put(0.03773127,0.25925001){\color[rgb]{0,0,0}\makebox(0,0)[lt]{\lineheight{1.25}\smash{\begin{tabular}[t]{l}\textbf{$\lRone$}\end{tabular}}}}%
    \put(0.41172197,0.25925001){\color[rgb]{0,0,0}\makebox(0,0)[lt]{\lineheight{1.25}\smash{\begin{tabular}[t]{l}\textbf{$\lRone$}\end{tabular}}}}%
    \put(0.7132828,0.25773605){\color[rgb]{0,0,0}\makebox(0,0)[lt]{\lineheight{1.25}\smash{\begin{tabular}[t]{l}\textbf{$\lRone$}\end{tabular}}}}%
    \put(0,0){\includegraphics[width=\unitlength,page=6]{operators-v4.pdf}}%
    \put(0.01732032,0.15166915){\color[rgb]{1,0.14901961,0.14901961}\makebox(0,0)[lt]{\lineheight{1.25}\smash{\begin{tabular}[t]{l}\textbf{$\lL$}\end{tabular}}}}%
    \put(0.69687259,0.11761708){\color[rgb]{1,0.14901961,0.14901961}\makebox(0,0)[lt]{\lineheight{1.25}\smash{\begin{tabular}[t]{l}\textbf{$\lL$}\end{tabular}}}}%
    \put(0.0278379,0.11309228){\color[rgb]{0,0.4,1}\makebox(0,0)[lt]{\lineheight{1.25}\smash{\begin{tabular}[t]{l}\textbf{$\lS$}\end{tabular}}}}%
    \put(0.40102809,0.11386662){\color[rgb]{0,0.4,1}\makebox(0,0)[lt]{\lineheight{1.25}\smash{\begin{tabular}[t]{l}\textbf{$\lS$}\end{tabular}}}}%
  \end{picture}%
\endgroup%

%% file: fig/cascaded-method-v15.pdf_tex
\begingroup%
  \makeatletter%
  \providecommand\color[2][]{%
    \errmessage{(Inkscape) Color is used for the text in Inkscape, but the package 'color.sty' is not loaded}%
    \renewcommand\color[2][]{}%
  }%
  \providecommand\transparent[1]{%
    \errmessage{(Inkscape) Transparency is used (non-zero) for the text in Inkscape, but the package 'transparent.sty' is not loaded}%
    \renewcommand\transparent[1]{}%
  }%
  \providecommand\rotatebox[2]{#2}%
  \newcommand*\fsize{\dimexpr\f@size pt\relax}%
  \newcommand*\lineheight[1]{\fontsize{\fsize}{#1\fsize}\selectfont}%
  \ifx\svgwidth\undefined%
    \setlength{\unitlength}{1727.18157911bp}%
    \ifx\svgscale\undefined%
      \relax%
    \else%
      \setlength{\unitlength}{\unitlength * \real{\svgscale}}%
    \fi%
  \else%
    \setlength{\unitlength}{\svgwidth}%
  \fi%
  \global\let\svgwidth\undefined%
  \global\let\svgscale\undefined%
  \makeatother%
  \begin{picture}(1,0.99173902)%
    \lineheight{1}%
    \setlength\tabcolsep{0pt}%
    \put(0,0){\includegraphics[width=\unitlength,page=1]{cascaded-method-v15.pdf}}%
    \put(0.84883913,0.91783359){\color[rgb]{0,0,0}\makebox(0,0)[t]{\lineheight{1.25}\smash{\begin{tabular}[t]{c}$H(\xl, \xs, t)$\end{tabular}}}}%
    \put(0.96378437,0.80079326){\color[rgb]{0,0,0}\makebox(0,0)[lt]{\lineheight{1.25}\smash{\begin{tabular}[t]{l}$t$\end{tabular}}}}%
    \put(0,0){\includegraphics[width=\unitlength,page=2]{cascaded-method-v15.pdf}}%
    \put(0.10554768,0.89957529){\makebox(0,0)[rt]{\lineheight{1.25}\smash{\begin{tabular}[t]{r}$\lRone$\end{tabular}}}}%
    \put(0.01507599,0.79918942){\rotatebox{90}{\makebox(0,0)[t]{\lineheight{1.25}\smash{\begin{tabular}[t]{c}(a) Capture\end{tabular}}}}}%
    \put(0.01543804,0.47798986){\rotatebox{90}{\makebox(0,0)[t]{\lineheight{1.25}\smash{\begin{tabular}[t]{c}(b) $\Itproj$ then $\Itcam$\end{tabular}}}}}%
    \put(0.01626156,0.15661826){\rotatebox{90}{\makebox(0,0)[t]{\lineheight{1.25}\smash{\begin{tabular}[t]{c}(c) $\Icc$\end{tabular}}}}}%
    \put(0.20988378,0.97700541){\makebox(0,0)[t]{\lineheight{1.25}\smash{\begin{tabular}[t]{c}Laser/projector\end{tabular}}}}%
    \put(0.55995894,0.97694792){\makebox(0,0)[t]{\lineheight{1.25}\smash{\begin{tabular}[t]{c}Sensor/camera\end{tabular}}}}%
    \put(0.86951328,0.97700541){\makebox(0,0)[t]{\lineheight{1.25}\smash{\begin{tabular}[t]{c}Output\end{tabular}}}}%
    \put(0.78999659,0.73667721){\makebox(0,0)[t]{\smash{\begin{tabular}[t]{c}$\lRtwo$\\(third\\bounce)\end{tabular}}}}%
    \put(0.91975071,0.73656806){\makebox(0,0)[t]{\smash{\begin{tabular}[t]{c}T\\(fifth\\bounce)\end{tabular}}}}%
    \put(0.869697,0.27853619){\makebox(0,0)[t]{\lineheight{1.25}\smash{\begin{tabular}[t]{c}\emph{Fifth-bounce}\\image\\$f_{2,2}(\xv)$\end{tabular}}}}%
    \put(0.32589796,0.92943708){\makebox(0,0)[lt]{\lineheight{1.25}\smash{\begin{tabular}[t]{l}$\lRtwo$\end{tabular}}}}%
    \put(0.07077979,0.7877174){\color[rgb]{1,0.14901961,0.14901961}\makebox(0,0)[lt]{\lineheight{1.25}\smash{\begin{tabular}[t]{l}$\lL$\end{tabular}}}}%
    \put(0,0){\includegraphics[width=\unitlength,page=3]{cascaded-method-v15.pdf}}%
    \put(0.45562315,0.89957529){\makebox(0,0)[rt]{\lineheight{1.25}\smash{\begin{tabular}[t]{r}$\lRone$\end{tabular}}}}%
    \put(0.67597344,0.92943708){\makebox(0,0)[lt]{\lineheight{1.25}\smash{\begin{tabular}[t]{l}$\lRtwo$\end{tabular}}}}%
    \put(0.41641478,0.8032143){\color[rgb]{0,0.4,1}\makebox(0,0)[lt]{\lineheight{1.25}\smash{\begin{tabular}[t]{l}$\lS$\end{tabular}}}}%
    \put(0,0){\includegraphics[width=\unitlength,page=4]{cascaded-method-v15.pdf}}%
    \put(0.10554799,0.57820372){\makebox(0,0)[rt]{\lineheight{1.25}\smash{\begin{tabular}[t]{r}$\lRone$\end{tabular}}}}%
    \put(0.32589826,0.60806552){\makebox(0,0)[lt]{\lineheight{1.25}\smash{\begin{tabular}[t]{l}$\lRtwo$\end{tabular}}}}%
    \put(0.07078009,0.46634581){\color[rgb]{1,0.14901961,0.14901961}\makebox(0,0)[lt]{\lineheight{1.25}\smash{\begin{tabular}[t]{l}$\lL$\end{tabular}}}}%
    \put(0.29782058,0.54884638){\color[rgb]{0.12941176,0.61568627,0}\makebox(0,0)[lt]{\lineheight{1.25}\smash{\begin{tabular}[t]{l}$\lLp$\end{tabular}}}}%
    \put(0,0){\includegraphics[width=\unitlength,page=5]{cascaded-method-v15.pdf}}%
    \put(0.45562315,0.57820372){\makebox(0,0)[rt]{\lineheight{1.25}\smash{\begin{tabular}[t]{r}$\lRone$\end{tabular}}}}%
    \put(0.67597344,0.60806552){\makebox(0,0)[lt]{\lineheight{1.25}\smash{\begin{tabular}[t]{l}$\lRtwo$\end{tabular}}}}%
    \put(0.65486087,0.55428138){\color[rgb]{0.12941176,0.61568627,0}\makebox(0,0)[lt]{\lineheight{1.25}\smash{\begin{tabular}[t]{l}$\lSp$\end{tabular}}}}%
    \put(0.41641478,0.48184273){\color[rgb]{0,0.4,1}\makebox(0,0)[lt]{\lineheight{1.25}\smash{\begin{tabular}[t]{l}$\lS$\end{tabular}}}}%
    \put(0,0){\includegraphics[width=\unitlength,page=6]{cascaded-method-v15.pdf}}%
    \put(0.10554799,0.2568322){\makebox(0,0)[rt]{\lineheight{1.25}\smash{\begin{tabular}[t]{r}$\lRone$\end{tabular}}}}%
    \put(0.32589826,0.28669399){\makebox(0,0)[lt]{\lineheight{1.25}\smash{\begin{tabular}[t]{l}$\lRtwo$\end{tabular}}}}%
    \put(0.29868905,0.2274746){\color[rgb]{1,0.14901961,0.14901961}\makebox(0,0)[lt]{\lineheight{1.25}\smash{\begin{tabular}[t]{l}$\lLp$\end{tabular}}}}%
    \put(0.33370842,0.01517805){\color[rgb]{0.12941176,0.61568627,0}\makebox(0,0)[lt]{\lineheight{1.25}\smash{\begin{tabular}[t]{l}$\lV$\end{tabular}}}}%
    \put(0.29375838,0.12302478){\color[rgb]{0,0,0}\makebox(0,0)[lt]{\lineheight{1.25}\smash{\begin{tabular}[t]{l}$\Icc$\end{tabular}}}}%
    \put(0.4814832,0.43468843){\color[rgb]{0,0,0}\makebox(0,0)[lt]{\lineheight{1.25}\smash{\begin{tabular}[t]{l}$\Itcam$\end{tabular}}}}%
    \put(0.12992978,0.44186747){\color[rgb]{0,0,0}\makebox(0,0)[lt]{\lineheight{1.25}\smash{\begin{tabular}[t]{l}$\Itproj$\end{tabular}}}}%
    \put(0,0){\includegraphics[width=\unitlength,page=7]{cascaded-method-v15.pdf}}%
    \put(0.45562315,0.2568322){\makebox(0,0)[rt]{\lineheight{1.25}\smash{\begin{tabular}[t]{r}$\lRone$\end{tabular}}}}%
    \put(0.67597344,0.28669399){\makebox(0,0)[lt]{\lineheight{1.25}\smash{\begin{tabular}[t]{l}$\lRtwo$\end{tabular}}}}%
    \put(0.6557293,0.2329096){\color[rgb]{0,0.4,1}\makebox(0,0)[lt]{\lineheight{1.25}\smash{\begin{tabular}[t]{l}$\lSp$\end{tabular}}}}%
    \put(0.6837836,0.01517779){\color[rgb]{0.12941176,0.61568627,0}\makebox(0,0)[lt]{\lineheight{1.25}\smash{\begin{tabular}[t]{l}$\lV$\end{tabular}}}}%
    \put(0,0){\includegraphics[width=\unitlength,page=8]{cascaded-method-v15.pdf}}%
    \put(0.83980067,0.39372865){\makebox(0,0)[t]{\smash{\begin{tabular}[t]{c}T (now\\third bounce)\end{tabular}}}}%
    \put(0,0){\includegraphics[width=\unitlength,page=9]{cascaded-method-v15.pdf}}%
    \put(0.64557055,0.12476121){\color[rgb]{0,0,0}\makebox(0,0)[lt]{\lineheight{1.25}\smash{\begin{tabular}[t]{l}$\Icc$\end{tabular}}}}%
    \put(0,0){\includegraphics[width=\unitlength,page=10]{cascaded-method-v15.pdf}}%
    \put(0.84883912,0.57949019){\color[rgb]{0,0,0}\makebox(0,0)[t]{\lineheight{1.25}\smash{\begin{tabular}[t]{c}$\Hp_{2,2}(\mathbf{l}_2, \mathbf{s}_2, t)$\end{tabular}}}}%
    \put(0,0){\includegraphics[width=\unitlength,page=11]{cascaded-method-v15.pdf}}%
    \put(0.96378447,0.46245035){\color[rgb]{0,0,0}\makebox(0,0)[lt]{\lineheight{1.25}\smash{\begin{tabular}[t]{l}$t$\end{tabular}}}}%
    \put(0,0){\includegraphics[width=\unitlength,page=12]{cascaded-method-v15.pdf}}%
    \put(0.52423275,0.90186796){\color[rgb]{0,0.4,1}\makebox(0,0)[lt]{\lineheight{1.25}\smash{\begin{tabular}[t]{l}\textbf{$\xs$}\end{tabular}}}}%
    \put(0.65448976,0.50534306){\color[rgb]{0,0.4,1}\makebox(0,0)[lt]{\lineheight{1.25}\smash{\begin{tabular}[t]{l}\textbf{$\mathbf{s}_2$}\end{tabular}}}}%
    \put(0.17184228,0.90185954){\color[rgb]{1,0.14901961,0.14901961}\makebox(0,0)[lt]{\lineheight{1.25}\smash{\begin{tabular}[t]{l}\textbf{$\xl$}\end{tabular}}}}%
    \put(0.30361613,0.5049009){\color[rgb]{1,0.14901961,0.14901961}\makebox(0,0)[lt]{\lineheight{1.25}\smash{\begin{tabular}[t]{l}\textbf{$\mathbf{l}_2$}\end{tabular}}}}%
    \put(0,0){\includegraphics[width=\unitlength,page=13]{cascaded-method-v15.pdf}}%
    \put(0.88916846,0.83821002){\makebox(0,0)[t]{\smash{\begin{tabular}[t]{c}{\scriptsize $\times 10$}\end{tabular}}}}%
  \end{picture}%
\endgroup%

%% file: fig/multiview-v20.pdf_tex
\begingroup%
  \makeatletter%
  \providecommand\color[2][]{%
    \errmessage{(Inkscape) Color is used for the text in Inkscape, but the package 'color.sty' is not loaded}%
    \renewcommand\color[2][]{}%
  }%
  \providecommand\transparent[1]{%
    \errmessage{(Inkscape) Transparency is used (non-zero) for the text in Inkscape, but the package 'transparent.sty' is not loaded}%
    \renewcommand\transparent[1]{}%
  }%
  \providecommand\rotatebox[2]{#2}%
  \newcommand*\fsize{\dimexpr\f@size pt\relax}%
  \newcommand*\lineheight[1]{\fontsize{\fsize}{#1\fsize}\selectfont}%
  \ifx\svgwidth\undefined%
    \setlength{\unitlength}{2689.27578663bp}%
    \ifx\svgscale\undefined%
      \relax%
    \else%
      \setlength{\unitlength}{\unitlength * \real{\svgscale}}%
    \fi%
  \else%
    \setlength{\unitlength}{\svgwidth}%
  \fi%
  \global\let\svgwidth\undefined%
  \global\let\svgscale\undefined%
  \makeatother%
  \begin{picture}(1,0.33295374)%
    \lineheight{1}%
    \setlength\tabcolsep{0pt}%
    \put(0,0){\includegraphics[width=\unitlength,page=1]{multiview-v20.pdf}}%
    \put(0.89736475,0.29775354){\makebox(0,0)[t]{\lineheight{1.25}\smash{\begin{tabular}[t]{c}Ground truth\end{tabular}}}}%
    \put(0,0){\includegraphics[width=\unitlength,page=2]{multiview-v20.pdf}}%
    \put(0.0880435,0.01902733){\makebox(0,0)[t]{\lineheight{1.25}\smash{\begin{tabular}[t]{c}(a) Scene setup and\\relevant light paths\end{tabular}}}}%
    \put(0.89786706,0.15204967){\makebox(0,0)[t]{\lineheight{1.25}\smash{\begin{tabular}[t]{c}$\fresult(\xv)$\end{tabular}}}}%
    \put(0.88982613,0.01830677){\makebox(0,0)[t]{\lineheight{1.25}\smash{\begin{tabular}[t]{c}(e) Our combined image\\$\fresult(\xv)$ (\sref{sec:methods:merge})\end{tabular}}}}%
    \put(0.67181936,0.01877688){\makebox(0,0)[t]{\lineheight{1.25}\smash{\begin{tabular}[t]{c}(d) Our $\fright(\xv)$ (Section~\ref{sec:methods:fifth})\\Projector $\mathcal{L}_3$, camera $\mathcal{S}_3$ \end{tabular}}}}%
    \put(0.48898646,0.01827388){\makebox(0,0)[t]{\lineheight{1.25}\smash{\begin{tabular}[t]{c}(c) Our $\fleft(\xv)$ (Section~\ref{sec:methods:fourth})\\Projector $\mathcal{L}_2$, camera $\lS$ \end{tabular}}}}%
    \put(0.30615126,0.01827388){\makebox(0,0)[t]{\lineheight{1.25}\smash{\begin{tabular}[t]{c}(b) Third bounce $\ffront(\xv)$\\Projector $\lL$, camera $\lS$\end{tabular}}}}%
    \put(0.04075521,0.14656181){\color[rgb]{1,0.14901961,0.14901961}\makebox(0,0)[t]{\lineheight{1.25}\smash{\begin{tabular}[t]{c}$\xl$\end{tabular}}}}%
    \put(0.26802021,0.18099326){\color[rgb]{1,0.14901961,0.14901961}\makebox(0,0)[t]{\lineheight{1.25}\smash{\begin{tabular}[t]{c}$\lL$\end{tabular}}}}%
    \put(0.3468063,0.18056205){\color[rgb]{0,0.4,1}\makebox(0,0)[t]{\lineheight{1.25}\smash{\begin{tabular}[t]{c}$\lS$\end{tabular}}}}%
    \put(0.13807362,0.14537104){\color[rgb]{0,0.4,1}\makebox(0,0)[t]{\lineheight{1.25}\smash{\begin{tabular}[t]{c}$\xs$\end{tabular}}}}%
    \put(0.0204849,0.18481284){\makebox(0,0)[t]{\lineheight{1.25}\smash{\begin{tabular}[t]{c}$\lRone$\end{tabular}}}}%
    \put(0,0){\includegraphics[width=\unitlength,page=3]{multiview-v20.pdf}}%
    \put(-0.00034623,0.31166167){\makebox(0,0)[lt]{\lineheight{1.25}\smash{\begin{tabular}[t]{l}$\lRleft$\end{tabular}}}}%
    \put(0.1741734,0.3114128){\makebox(0,0)[rt]{\lineheight{1.25}\smash{\begin{tabular}[t]{r}$\lRright$\end{tabular}}}}%
    \put(0.41961164,0.17370679){\color[rgb]{1,0.14901961,0.14901961}\makebox(0,0)[t]{\lineheight{1.25}\smash{\begin{tabular}[t]{c}$\mathbf{l}_2$\end{tabular}}}}%
    \put(0.42383079,0.30371166){\color[rgb]{1,0.14901961,0.14901961}\makebox(0,0)[t]{\lineheight{1.25}\smash{\begin{tabular}[t]{c}$\mathcal{L}_2$\end{tabular}}}}%
    \put(0,0){\includegraphics[width=\unitlength,page=4]{multiview-v20.pdf}}%
    \put(0.73912907,0.30423869){\color[rgb]{1,0.14901961,0.14901961}\makebox(0,0)[t]{\lineheight{1.25}\smash{\begin{tabular}[t]{c}$\mathcal{L}_3$\end{tabular}}}}%
    \put(0,0){\includegraphics[width=\unitlength,page=5]{multiview-v20.pdf}}%
    \put(0.5286541,0.18672733){\color[rgb]{0,0.4,1}\makebox(0,0)[t]{\lineheight{1.25}\smash{\begin{tabular}[t]{c}$\lS$\end{tabular}}}}%
    \put(0.78188219,0.3048005){\color[rgb]{0,0.4,1}\makebox(0,0)[t]{\lineheight{1.25}\smash{\begin{tabular}[t]{c}$\mathcal{S}_3$\end{tabular}}}}%
    \put(0,0){\includegraphics[width=\unitlength,page=6]{multiview-v20.pdf}}%
    \put(0.67017251,0.20965472){\color[rgb]{1,0.14901961,0.14901961}\makebox(0,0)[t]{\lineheight{1.25}\smash{\begin{tabular}[t]{c}$\mathbf{l}_3$\end{tabular}}}}%
    \put(0.65674521,0.23232123){\color[rgb]{0,0.4,1}\makebox(0,0)[t]{\lineheight{1.25}\smash{\begin{tabular}[t]{c}$\mathbf{s}_3$\end{tabular}}}}%
    \put(0,0){\includegraphics[width=\unitlength,page=7]{multiview-v20.pdf}}%
    \put(0.11677297,0.32228735){\color[rgb]{0.50196078,0.50196078,0.50196078}\makebox(0,0)[t]{\lineheight{1.25}\smash{\begin{tabular}[t]{c}{\scriptsize $40^\circ$}\end{tabular}}}}%
    \put(0.05369206,0.31838293){\color[rgb]{0.50196078,0.50196078,0.50196078}\makebox(0,0)[t]{\lineheight{1.25}\smash{\begin{tabular}[t]{c}{\scriptsize $75^\circ$}\end{tabular}}}}%
    \put(0,0){\includegraphics[width=\unitlength,page=8]{multiview-v20.pdf}}%
    \put(0.48966284,0.32447826){\makebox(0,0)[t]{\smash{\begin{tabular}[t]{c}\textbf{Fourth bounce}\\\textbf{(ours)}\end{tabular}}}}%
    \put(0.67253933,0.32441834){\makebox(0,0)[t]{\smash{\begin{tabular}[t]{c}\textbf{Fifth bounce}\\\textbf{(ours)}\end{tabular}}}}%
    \put(0.30721636,0.32447826){\makebox(0,0)[t]{\smash{\begin{tabular}[t]{c}\textbf{Third bounce}\end{tabular}}}}%
    \put(0,0){\includegraphics[width=\unitlength,page=9]{multiview-v20.pdf}}%
  \end{picture}%
\endgroup%

%% file: journal_modified_tex/51_methods.tex
\section{Imaging methods beyond the third bounce}
\label{sec:methods}

We use our building blocks from \sref{sec:cascaded} to design new NLOS imaging methods that leverage illumination beyond the third bounce. Here, we present two methods that use fourth- and fifth-bounce information, respectively, and then discuss how to obtain a combined image that merges third-, fourth-, and fifth-bounce imaging results.

To illustrate our methods, we use the scene in \fref{fig:multiview}a, which contains three letters hidden from direct view: A, B, and C, each one with a different orientation. 
The captured impulse response $H$ contains indirect light from all three letters.
Conventional NLOS imaging methods (\fref{fig:multiview}b) use only third-bounce illumination (purple arrows in \fref{fig:multiview}a), and can only image the letter B. Due to their unfavorable orientation, the letters A and C have limited visibility. 
\final{This is related to the finite size of the relay wall, which prevents certain surfaces from being fully imaged depending on their position and orientation \cite{Liu2019analysis,pena2025flatland}}. \final{We discuss this problem and our approach to it in detail in \sref{sec:virtual-reflectance-main-paper}.}

To mitigate this, we leverage light coming from the left $\lRleft$ or right $\lRright$ walls inside the hidden scene (which will act as secondary relay walls). %
This light then interacts with letters A and C, and finally is captured in the impulse response $H$ as fourth- and fifth-bounce illumination, respectively (cyan and yellow arrows in \fref{fig:multiview}a).

In the following, we formulate our methods using this scene as an example, illustrated in \fref{fig:multiview}c and \ref{fig:multiview}d. 
Note that all methods use the same input data: the impulse response $H$, only changing the computations. %
In \ssref{sec:results}{sec:real-experiments} we show and discuss the results of our methods in simulated and real scenarios.

\subsection{Fourth-bounce imaging method}
\label{sec:methods:fourth}
We are interested in leveraging fourth-bounce illumination (cyan arrow in \fref{fig:multiview}a). Generally, we can model either $\xl \rarr \xlp \rarr \xv \rarr \xs$ or $\xl \rarr \xv \rarr \xsp \rarr \xs$, with $\xlp$ or $\xsp$ on a secondary wall $\lR_a$ or $\lR_b$, and $\xv$ on the target geometry. For the first case, we use our transient projector operator $\Itproj$ to compute $\Hp_{a,1}(\xlp, \xs, t)$ as per \eref{eq:virtual-impulse-response-onlyL}. Similarly, for the second case, we use our transient camera operator $\Itcam$ to compute $\Hp_{1,b}(\xl, \xsp, t)$ as per \eref{eq:virtual-impulse-response-onlyS}. Note that both cases would give the same results in our ABC scene, as the time of flight is the same for both paths.

Our fourth-bounce method consists of computing one of these two virtual impulse responses, then using it as input for the confocal camera operator $\Icc$ as per \eref{eq:second-reconstruction}. \fref{fig:multiview}c illustrates an example using \eeref{eq:virtual-impulse-response-onlyL}{eq:second-reconstruction} with $a=2$ and $b=1$.

\subsection{Fifth-bounce imaging method}
\label{sec:methods:fifth}
To leverage fifth-bounce illumination (yellow arrow in \fref{fig:multiview}a), we model paths of the form $\xl \rarr \xlp \rarr \xv \rarr \xsp \rarr \xs$, with $\xlp$ and $\xsp$ lying on secondary walls $\lR_a$ and $\lR_b$, respectively, and then use both $\Itproj$ and $\Itcam$ to compute $\Hp_{a, b}$ as per \eref{eq:virtual-impulse-response}.
Our fifth-bounce method uses our virtual impulse response $\Hp_{a, b}$ as input for the confocal camera operator $\Icc$ as per \eref{eq:second-reconstruction}, which in this case will yield a fifth-bounce image. \fref{fig:multiview}d illustrates an example using \eeref{eq:virtual-impulse-response}{eq:second-reconstruction} with $a=b=3$.

\subsection{Combining multiple images}
\label{sec:methods:merge}
Last, we can add up the third-, fourth- and fifth-bounce images to obtain the combined image $\fresult(\xv)$, which in our example would be equal to $f_{1,1}(\xv) + f_{2,1}(\xv) + f_{3,3}(\xv)$. Note that our method is phasor-based: specifically, we add the amplitudes of each image.

%% file: journal_modified_tex/52_virtual_reflectance_droyo.tex
\section{\final{Virtual surface reflectance and surface visibility}}
\label{sec:virtual-reflectance-main-paper}

\final{Here, we provide a theoretical and practical analysis of visibility limits of previous NLOS methods, and how our cascaded imaging overcomes such limits. We first introduce the concept of \emph{virtual reflectance} of a surface; although all surfaces in our setups are diffuse under visible light, they may exhibit different reflectance properties when illuminated by virtual phasor waves. We explore here how virtual reflectance affects both the hidden objects' visibility and the imaging results when using a secondary virtual imaging system, and provide an extended analysis in Appendix D of our Supplemental Material. The insights developed here underpin the NLOS imaging results presented later in \ssref{sec:results}{sec:real-experiments}.}

\subsection{\final{Virtual specular and diffuse behavior}} 
\begin{figure}
    \centering
    \captionsetup{skip=-3pt}
    \def\svgwidth{\columnwidth} 
    \begin{small}
    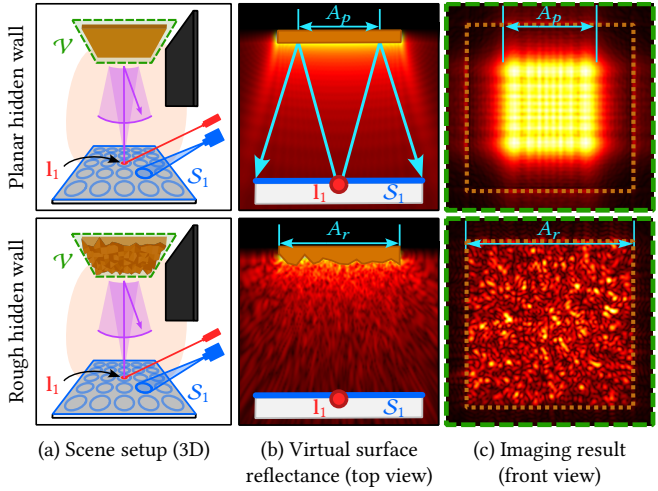
    \end{small}
    \caption{\final{Virtual surface reflectance. Top: planar hidden wall. Bottom: rough hidden wall with \qtyproduct{3 x 3}{cm} facets. (a) Scene setup: we illuminate one point $\xl$, and capture the impulse response in $\lS$. We mark the imaged region as $\lV$. (b) Simulated propagation of a phasor wave of wavelength $\lambda = \SI{3}{cm}$ emitted from $\xl$ and reflected by the hidden wall. The planar wall reflects light specularly (cyan lines);  only the area $A_p$ returns light to the aperture $\lS$. In contrast, the rough wall reflects diffusely, returning light to the aperture from a larger area $A_r$. (c) Resulting image $f_{1,1}(\xv) = \Icc(\xv; \Hfiltert)$ over $\lV$. The orange dotted line marks the expected wall location.}}
    \label{fig:reflectance-analysis}
\end{figure}

\final{\fref{fig:reflectance-analysis}a shows two NLOS setups, each with a single hidden wall: planar (top) and rough with $\SI{3}{cm}$ facets (bottom). In simulation, we illuminate one point $\xl$ and capture the impulse response $H$ at all points in $\lS$. Although this capture uses visible light, the phasor-field formulation convolves $H$ with a virtual illumination function $K(t; \lambda_c, \sigma)$ (\eref{eq:gaussian-pulse}), producing phasor waves whose wavelength is set by $\lambda_c$. Practical $\lambda_c$ values lie in the centimeter range \cite{Liu2019phasor,Liu2020phasor,royo2023virtual} (see Appendix D).}

\final{To better understand how each hidden wall reflects these phasor waves, we simulate how they reflect a spherical wavefront of $\lambda = \SI{3}{cm}$ emitted from $\xl$ (\fref{fig:reflectance-analysis}b). The top wall is planar with respect to $\lambda$ and exhibits virtual \emph{specular} behavior, as observed by \customcitet{Royo}{royo2023virtual}. The wall reflects light from $\xl$ into the specular direction (cyan lines). Only an area $A_p$ redirects light towards the camera aperture $\lS$, which corresponds to the area recovered in the NLOS imaging result (\fref{fig:reflectance-analysis}c, computed with the confocal camera operator $\Icc$ and filtering $\wlc = \sigma = \SI{3}{cm}$). The rough wall instead exhibits virtual \emph{diffuse} behavior: reflected light reaches $\lS$ from all of its points, so the resulting image recovers a larger area $A_r$, but also introduces speckle-like artifacts from the wall roughness.}%

\final{This relates to the missing-cone problem, and illustrates why rough walls have a greater coverage compared to planar walls, since an NLOS imaging system can only image parts of the surface that reflect phasor waves towards its sensor aperture \cite{Liu2019analysis,royo2023virtual,pena2025flatland}. }

\begin{figure}
    \centering
    \captionsetup{skip=-3pt}
    \def\svgwidth{\columnwidth} 
    \begin{small}
    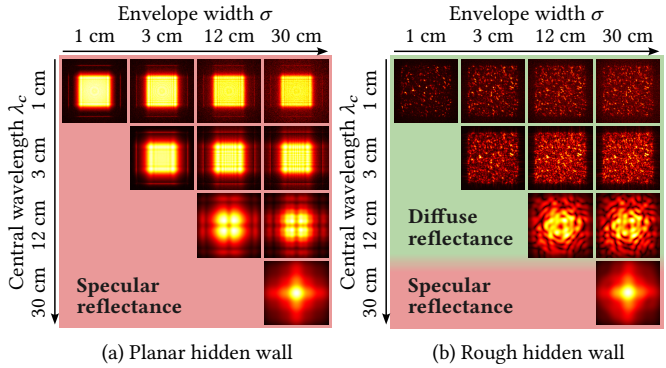
    \end{small}
    \caption{\final{Effect of the filter parameters $\lambda_c$ and $\sigma$ of $K(t; \lambda_c, \sigma)$ (\eref{eq:gaussian-pulse}) for the setups in \fref{fig:reflectance-analysis}. (a) The hidden wall is planar with respect to every $\lambda_c$ and is thus always virtually specular; its edges have limited visibility. (b) The rough wall is virtually diffuse for $\lambda_c$ around $\SI{3}{cm}$ (the size of its facets), and its full surface is imaged. For larger $\lambda_c$, it becomes planar with respect to the wavelength and converges to the same virtually specular behavior. Larger $\sigma$, corresponding to a narrower bandwidth of $K$, makes the NLOS imaging result more uniform, suggesting increased diffuseness.}}
    \label{fig:planar-facets-reconstruction}
\end{figure}
\final{\paragraph{Effect of the phasor-field  parameters.} The virtual reflectance of a surface depends on the  parameters $\lambda_c$ and $\sigma$ of $K(t; \lambda_c, \sigma)$. In \fref{fig:planar-facets-reconstruction}, we explore how $\lambda_c$ and $\sigma$ affect the NLOS imaging result. When $\lambda_c$ is close to the rough wall's facet size, it behaves as virtually diffuse, enhancing its visibility. As $\lambda_c$ grows, the phasor waves become planar with respect to both walls, which converge to virtual specular behavior. In this regime, parts of the wall disappear at the edges, where phasor waves no longer reach $\lS$. Also, increasing $\sigma$ narrows the filter bandwidth, which further enhances diffuseness when $\lambda_c$ is near the facet size.}

\subsection{\final{Surface visibility under higher-order illumination}}

\final{In the third-bounce case, a hidden surface is visible only if it virtually reflects the illumination from $\lL$ towards the aperture $\lS$. Our insights on virtual reflectance and surface visibility extend to higher-order illumination: along every step of the path, each surface must virtually reflect light towards the next one, and the last one back towards $\lS$.
Visibility therefore depends on the virtual reflectance of every surface along the path, not just on the surface being imaged.}

\final{Leveraging the virtual diffuse behavior of rough walls allows us to turn hidden surfaces into secondary virtual imaging systems, whose angular visibility is much wider than the specular directions available to previous works \cite{royo2023virtual}. This improvement entails a trade-off, since the speckle-like artifacts from virtual diffuse reflection limit the resolution of secondary virtual imaging systems.} 

\final{Note that (virtually diffuse) rough walls are \textit{not} a requirement of our method. On virtually specular surfaces our transient projector $\Itproj$ and camera $\Itcam$ operators form new apertures on secondary walls. This allows us to leverage more specular paths than previous works, which place apertures $\lL$ and $\lS$ only on the primary relay wall. 
We provide further analysis and illustrations in Appendix D.1.}

%% file: fig/facets-v4.pdf_tex
\begingroup%
  \makeatletter%
  \providecommand\color[2][]{%
    \errmessage{(Inkscape) Color is used for the text in Inkscape, but the package 'color.sty' is not loaded}%
    \renewcommand\color[2][]{}%
  }%
  \providecommand\transparent[1]{%
    \errmessage{(Inkscape) Transparency is used (non-zero) for the text in Inkscape, but the package 'transparent.sty' is not loaded}%
    \renewcommand\transparent[1]{}%
  }%
  \providecommand\rotatebox[2]{#2}%
  \newcommand*\fsize{\dimexpr\f@size pt\relax}%
  \newcommand*\lineheight[1]{\fontsize{\fsize}{#1\fsize}\selectfont}%
  \ifx\svgwidth\undefined%
    \setlength{\unitlength}{1538.20860375bp}%
    \ifx\svgscale\undefined%
      \relax%
    \else%
      \setlength{\unitlength}{\unitlength * \real{\svgscale}}%
    \fi%
  \else%
    \setlength{\unitlength}{\svgwidth}%
  \fi%
  \global\let\svgwidth\undefined%
  \global\let\svgscale\undefined%
  \makeatother%
  \begin{picture}(1,0.73906912)%
    \lineheight{1}%
    \setlength\tabcolsep{0pt}%
    \put(0,0){\includegraphics[width=\unitlength,page=1]{facets-v4.pdf}}%
    \put(0.02169178,0.57580828){\rotatebox{90}{\makebox(0,0)[t]{\lineheight{1.25}\smash{\begin{tabular}[t]{c}Planar hidden wall\end{tabular}}}}}%
    \put(0.17803025,0.03658067){\makebox(0,0)[t]{\lineheight{1.25}\smash{\begin{tabular}[t]{c}(a) Scene setup (3D)\end{tabular}}}}%
    \put(0.06873297,0.13731733){\color[rgb]{1,0.14901961,0.14901961}\makebox(0,0)[t]{\lineheight{1.25}\smash{\begin{tabular}[t]{c}$\xl$\end{tabular}}}}%
    \put(0.08326084,0.32655796){\color[rgb]{0.12941176,0.61568627,0}\makebox(0,0)[t]{\lineheight{1.25}\smash{\begin{tabular}[t]{c}$\lV$\end{tabular}}}}%
    \put(0.08326084,0.65924214){\color[rgb]{0.12941176,0.61568627,0}\makebox(0,0)[t]{\lineheight{1.25}\smash{\begin{tabular}[t]{c}$\lV$\end{tabular}}}}%
    \put(0.06873297,0.46937414){\color[rgb]{1,0.14901961,0.14901961}\makebox(0,0)[t]{\lineheight{1.25}\smash{\begin{tabular}[t]{c}$\xl$\end{tabular}}}}%
    \put(0,0){\includegraphics[width=\unitlength,page=2]{facets-v4.pdf}}%
    \put(0.51134052,0.03540961){\makebox(0,0)[t]{\lineheight{1.20000005}\smash{\begin{tabular}[t]{c}(b) Virtual surface\\reflectance (top view)\end{tabular}}}}%
    \put(0.5110995,0.71221331){\color[rgb]{0.14901961,0.91372549,1}\makebox(0,0)[t]{\smash{\begin{tabular}[t]{c}$A_p$\end{tabular}}}}%
    \put(0.83707577,0.03540961){\makebox(0,0)[t]{\lineheight{1.20000005}\smash{\begin{tabular}[t]{c}(c) Imaging result\\(front view)\end{tabular}}}}%
    \put(0.02169178,0.24417799){\rotatebox{90}{\makebox(0,0)[t]{\lineheight{1.25}\smash{\begin{tabular}[t]{c}Rough hidden wall\end{tabular}}}}}%
    \put(0,0){\includegraphics[width=\unitlength,page=3]{facets-v4.pdf}}%
    \put(0.83745731,0.71221331){\color[rgb]{0.14901961,0.91372549,1}\makebox(0,0)[t]{\smash{\begin{tabular}[t]{c}$A_p$\end{tabular}}}}%
    \put(0,0){\includegraphics[width=\unitlength,page=4]{facets-v4.pdf}}%
    \put(0.51122081,0.37751808){\color[rgb]{0.14901961,0.91372549,1}\makebox(0,0)[t]{\smash{\begin{tabular}[t]{c}$A_r$\end{tabular}}}}%
    \put(0,0){\includegraphics[width=\unitlength,page=5]{facets-v4.pdf}}%
    \put(0.48226689,0.10140767){\color[rgb]{1,0.14901961,0.14901961}\makebox(0,0)[t]{\lineheight{1.25}\smash{\begin{tabular}[t]{c}$\xl$\end{tabular}}}}%
    \put(0.59611877,0.10132329){\color[rgb]{0,0.4,1}\makebox(0,0)[t]{\lineheight{1.25}\smash{\begin{tabular}[t]{c}$\lS$\end{tabular}}}}%
    \put(0.29308539,0.12448798){\color[rgb]{0,0.4,1}\makebox(0,0)[t]{\lineheight{1.25}\smash{\begin{tabular}[t]{c}$\lS$\end{tabular}}}}%
    \put(0.29308539,0.45799286){\color[rgb]{0,0.4,1}\makebox(0,0)[t]{\lineheight{1.25}\smash{\begin{tabular}[t]{c}$\lS$\end{tabular}}}}%
    \put(0,0){\includegraphics[width=\unitlength,page=6]{facets-v4.pdf}}%
    \put(0.48226761,0.43344896){\color[rgb]{1,0.14901961,0.14901961}\makebox(0,0)[t]{\lineheight{1.25}\smash{\begin{tabular}[t]{c}$\xl$\end{tabular}}}}%
    \put(0.5961195,0.43323964){\color[rgb]{0,0.4,1}\makebox(0,0)[t]{\lineheight{1.25}\smash{\begin{tabular}[t]{c}$\lS$\end{tabular}}}}%
    \put(0,0){\includegraphics[width=\unitlength,page=7]{facets-v4.pdf}}%
    \put(0.83708668,0.37751808){\color[rgb]{0.14901961,0.91372549,1}\makebox(0,0)[t]{\smash{\begin{tabular}[t]{c}$A_r$\end{tabular}}}}%
  \end{picture}%
\endgroup%

%% file: fig/wavelengths-v3.pdf_tex
\begingroup%
  \makeatletter%
  \providecommand\color[2][]{%
    \errmessage{(Inkscape) Color is used for the text in Inkscape, but the package 'color.sty' is not loaded}%
    \renewcommand\color[2][]{}%
  }%
  \providecommand\transparent[1]{%
    \errmessage{(Inkscape) Transparency is used (non-zero) for the text in Inkscape, but the package 'transparent.sty' is not loaded}%
    \renewcommand\transparent[1]{}%
  }%
  \providecommand\rotatebox[2]{#2}%
  \newcommand*\fsize{\dimexpr\f@size pt\relax}%
  \newcommand*\lineheight[1]{\fontsize{\fsize}{#1\fsize}\selectfont}%
  \ifx\svgwidth\undefined%
    \setlength{\unitlength}{2456.986489bp}%
    \ifx\svgscale\undefined%
      \relax%
    \else%
      \setlength{\unitlength}{\unitlength * \real{\svgscale}}%
    \fi%
  \else%
    \setlength{\unitlength}{\svgwidth}%
  \fi%
  \global\let\svgwidth\undefined%
  \global\let\svgscale\undefined%
  \makeatother%
  \begin{picture}(1,0.53690832)%
    \lineheight{1}%
    \setlength\tabcolsep{0pt}%
    \put(0,0){\includegraphics[width=\unitlength,page=1]{wavelengths-v3.pdf}}%
    \put(0.11959623,0.49307667){\makebox(0,0)[t]{\lineheight{1.25}\smash{\begin{tabular}[t]{c}1 cm\end{tabular}}}}%
    \put(0.27603995,0.52763155){\makebox(0,0)[t]{\lineheight{1.25}\smash{\begin{tabular}[t]{c}Envelope width $\sigma$\end{tabular}}}}%
    \put(0.27892553,0.00293327){\makebox(0,0)[t]{\lineheight{1.25}\smash{\begin{tabular}[t]{c}(a) Planar hidden wall \end{tabular}}}}%
    \put(0.79181,0.00293327){\makebox(0,0)[t]{\lineheight{1.25}\smash{\begin{tabular}[t]{c}(b) Rough hidden wall \end{tabular}}}}%
    \put(0.00927677,0.26156476){\rotatebox{90}{\makebox(0,0)[t]{\lineheight{1.25}\smash{\begin{tabular}[t]{c}Central wavelength $\wlc$\end{tabular}}}}}%
    \put(0.22411961,0.49301407){\makebox(0,0)[t]{\lineheight{1.25}\smash{\begin{tabular}[t]{c}3 cm\end{tabular}}}}%
    \put(0.0922077,0.10673364){\color[rgb]{0.10196078,0.10196078,0.10196078}\makebox(0,0)[lt]{\smash{\begin{tabular}[t]{l}\textbf{Specular}\\\textbf{reflectance}\end{tabular}}}}%
    \put(0.60512366,0.10692479){\color[rgb]{0.10196078,0.10196078,0.10196078}\makebox(0,0)[lt]{\smash{\begin{tabular}[t]{l}\textbf{Specular}\\\textbf{reflectance}\end{tabular}}}}%
    \put(0.60512366,0.21335332){\color[rgb]{0.10196078,0.10196078,0.10196078}\makebox(0,0)[lt]{\smash{\begin{tabular}[t]{l}\textbf{Diffuse}\\\textbf{reflectance}\end{tabular}}}}%
    \put(0.32801693,0.49301407){\makebox(0,0)[t]{\lineheight{1.25}\smash{\begin{tabular}[t]{c}12 cm\end{tabular}}}}%
    \put(0.4325403,0.49300812){\makebox(0,0)[t]{\lineheight{1.25}\smash{\begin{tabular}[t]{c}30 cm\end{tabular}}}}%
    \put(0.04265352,0.41775213){\rotatebox{90}{\makebox(0,0)[t]{\lineheight{1.25}\smash{\begin{tabular}[t]{c}1 cm\end{tabular}}}}}%
    \put(0.04271614,0.31385477){\rotatebox{90}{\makebox(0,0)[t]{\lineheight{1.25}\smash{\begin{tabular}[t]{c}3 cm\end{tabular}}}}}%
    \put(0.04271614,0.20933143){\rotatebox{90}{\makebox(0,0)[t]{\lineheight{1.25}\smash{\begin{tabular}[t]{c}12 cm\end{tabular}}}}}%
    \put(0.04272209,0.10543396){\rotatebox{90}{\makebox(0,0)[t]{\lineheight{1.25}\smash{\begin{tabular}[t]{c}30 cm\end{tabular}}}}}%
    \put(0,0){\includegraphics[width=\unitlength,page=2]{wavelengths-v3.pdf}}%
    \put(0.63248057,0.49307667){\makebox(0,0)[t]{\lineheight{1.25}\smash{\begin{tabular}[t]{c}1 cm\end{tabular}}}}%
    \put(0.78892444,0.52763155){\makebox(0,0)[t]{\lineheight{1.25}\smash{\begin{tabular}[t]{c}Envelope width $\sigma$\end{tabular}}}}%
    \put(0.52216098,0.26156476){\rotatebox{90}{\makebox(0,0)[t]{\lineheight{1.25}\smash{\begin{tabular}[t]{c}Central wavelength $\wlc$\end{tabular}}}}}%
    \put(0.73700398,0.49301407){\makebox(0,0)[t]{\lineheight{1.25}\smash{\begin{tabular}[t]{c}3 cm\end{tabular}}}}%
    \put(0.84090152,0.49301407){\makebox(0,0)[t]{\lineheight{1.25}\smash{\begin{tabular}[t]{c}12 cm\end{tabular}}}}%
    \put(0.94542486,0.49300812){\makebox(0,0)[t]{\lineheight{1.25}\smash{\begin{tabular}[t]{c}30 cm\end{tabular}}}}%
    \put(0.55553786,0.41775213){\rotatebox{90}{\makebox(0,0)[t]{\lineheight{1.25}\smash{\begin{tabular}[t]{c}1 cm\end{tabular}}}}}%
    \put(0.55560046,0.31385477){\rotatebox{90}{\makebox(0,0)[t]{\lineheight{1.25}\smash{\begin{tabular}[t]{c}3 cm\end{tabular}}}}}%
    \put(0.55560046,0.20933143){\rotatebox{90}{\makebox(0,0)[t]{\lineheight{1.25}\smash{\begin{tabular}[t]{c}12 cm\end{tabular}}}}}%
    \put(0.55560641,0.10543396){\rotatebox{90}{\makebox(0,0)[t]{\lineheight{1.25}\smash{\begin{tabular}[t]{c}30 cm\end{tabular}}}}}%
    \put(0,0){\includegraphics[width=\unitlength,page=3]{wavelengths-v3.pdf}}%
  \end{picture}%
\endgroup%

%% file: journal_modified_tex/60_results.tex
\section{Results in simulation}
\label{sec:results}
\label{sec:simulated-experiments}

We show the versatility of our cascaded NLOS imaging in three different scenarios that represent novel imaging possibilities: overcoming visibility limits from third-bounce methods (\sref{sec:results:missing_cone}), enhanced multi-view imaging of complex geometry (\sref{sec:results:complex_geom}), and imaging and tracking around two corners \final{in non-specular directions} (\sref{sec:results:tracking}). The results in this section are obtained from simulations using publicly available software \cite{royo2024mitransient, pena2025flatland}, \final{and include realistic noise as detailed in Appendix E of our Supplemental Material. We match SNR, calibration errors, and temporal uncertainty to values reported in previous works and measured in our own real-world captures.}
Later, \sref{sec:real-experiments} discusses additional results obtained with our real prototype. We reiterate that, both for simulated and real-world results, the impulse response $H$ needs to be captured only once.

\subsection{Overcoming third-bounce visibility limits}
\label{sec:results:missing_cone}
\label{sec:simulated-experiments:abc}

The bottom row of \fref{fig:multiview} shows results of the scene previously introduced in \sref{sec:methods}. In our simulation, we exhaustively capture the impulse response $H(\xl, \xs, t)$ by scanning a $32 \times 32$ grid of illumination points $\xl \in \lL$ over a \qtyproduct{2 x 2}{m} area, while measuring the reflected signal at an equivalent, co-located grid of sensing points $\xs \in \lS$.
The impulse response $H$ captures interreflections from the \textsc{ABC} letters and the left $\lRtwo$ and right $\lRthree$ walls. \final{Third-bounce photons account for $90\%$ of the total, fourth-bounce for $7.5\%$, and fifth-bounce for just $2.5\%$}. We show a classic, third-bounce image $f_{1,1}(\xv)$ in \fref{fig:multiview}b from the confocal camera operator $\Icc$ (\eref{eq:confocal-camera}).
The image only shows the B letter, since A and C exhibit limited visibility due to their orientation. To image these, we leverage higher-order illumination from $\lRtwo$ and  $\lRthree$.

First, we estimate the position of $\lRtwo$ and  $\lRthree$. Inspired by work by \customcitet{Royo}{royo2023virtual}, we image the \textit{reflection} $\xl^\prime$ of illuminated points $\xl$ produced by $\lRtwo$ and $\lRthree$, and estimate each wall as the perpendicular bisector of $\xl^\prime$ and $\xl$. This estimate is accurate to about $\SI{1}{cm}$.

Then, we create secondary imaging systems on $\lRtwo$ and $\lRthree$ to image the A and C letters. \fref{fig:multiview}c shows the results using fourth-bounce illumination (\sref{sec:methods:fourth}) from the left wall $\lRleft$. We apply $\Itproj$ from \eref{eq:virtual-impulse-response-onlyL} to compute $\Hp_{2,1}$ from $H$, then compute $f_{2,1}(\xv)$ as per \eref{eq:second-reconstruction}. \fref{fig:multiview}d shows results using fifth-bounce illumination (\sref{sec:methods:fifth}) using the right wall $\lRright$, applying both $\Itproj$ and $\Itcam$ (\eref{eq:virtual-impulse-response}) to compute $\Hp_{3,3}$ from $H$, and then computing $f_{3,3}(\xv)$ as per \eref{eq:second-reconstruction}. The combined result $\fresult(\xv) = f_{1,1}(\xv) + f_{2,1}(\xv) + f_{3,3}(\xv)$ in \fref{fig:multiview}e shows all three letters. We use $\wlc = \sigma = \SI{10}{cm}$ for imaging with third and fourth bounces, and $\wlc = \sigma = \SI{14}{cm}$ for the fifth bounce, due to increased noise in fifth-bounce photons. %

\subsection{Enhanced multi-view imaging of complex geometry}
\label{sec:results:complex_geom}

\begin{figure}
    \centering
    \captionsetup{skip=-6pt}
    \def\svgwidth{\columnwidth} 
    \begin{small}
    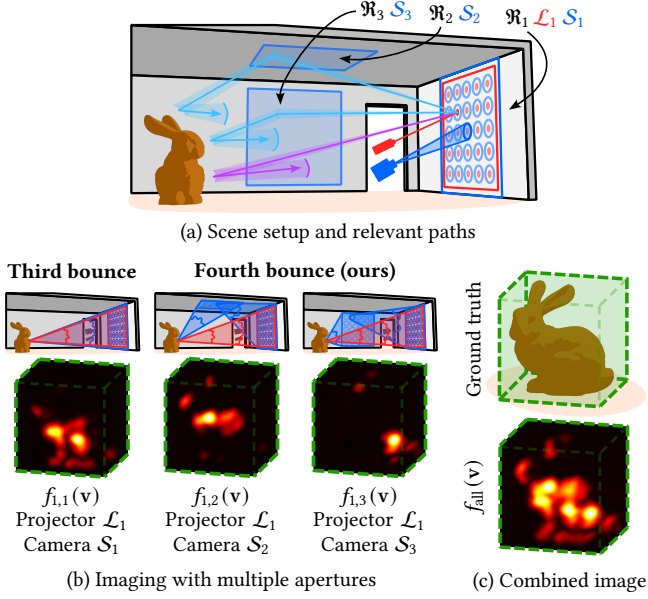
    \end{small}
    \caption{
    Enhanced multi-view imaging of complex geometry. (a)  The simulated capture $H(\xl, \xs, t)$ contains third-bounce illumination from the target object (\textsc{Bunny}) and also higher-order interreflections that arise from interactions with the top $\lRtwo$ and right $\lRright$ walls. (b) We show that fourth-bounce illumination extends the coverage of conventional NLOS imaging methods. Here we focus the camera aperture $\lS$ of the first relay wall $\lRone$ on the top $\lRtwo$ and right $\lRthree$ walls (creating virtual imaging devices with apertures $\lSp$ and $\mathcal{S}_3$, respectively). Each yields a different image $f_{1,1}(\xv),f_{1,2}(\xv),f_{1,3}(\xv)$ of the target object, and (c) we combine all of them in $\fresult(\xv)$ to show the enhanced visibility of our cascaded NLOS imaging paradigm. 
    }
    \label{fig:bunny-cascaded}
\end{figure}

The bunny in \fref{fig:bunny-cascaded}a presents self-occlusions and surface features oriented in a wide range of directions. Again, we simulate the capture of $H$ at the \qtyproduct{2 x 2}{m} relay wall $\lRone$ using a grid of $8\times8$ laser $\lL$ and $128\times128$ sensor $\lS$ points. The relay wall $\lRone$ receives indirect illumination from the object (purple arrow), and from the top $\lRtop$ and right $\lRright$ walls (cyan arrows). %

\fref{fig:bunny-cascaded}b shows our results. The first column shows the image obtained from classic third-bounce methods $f_{1,1}(\xv) = \Icc(\xv; H)$, where many features are missing from the bunny. The second and third columns show images obtained using fourth-bounce illumination (\sref{sec:methods:fourth}), from the top $\lRtop$ and right $\lRright$ walls respectively, applying $\Itcam$ from \eref{eq:virtual-impulse-response-onlyS} to compute $\Hp_{1,2}$ and $\Hp_{1,3}$ from $H$, and computing $f_{1,2}(\xv)$ and $f_{1,3}(\xv)$ as per \eref{eq:second-reconstruction}. 
We can see how, adding the top and right virtual imaging systems, we recover many of the previously missing features, such as the ears and the tail. The combined result in \fref{fig:bunny-cascaded}c shows a greater coverage of the hidden geometry with respect to third-bounce methods, reducing depth RMSE from $\SI{3.5}{cm}$ to $\SI{2.9}{cm}$. We use $\wlc = \sigma = \SI{8}{cm}$ for all cases.

\subsection{Imaging and tracking around two corners}
\label{sec:real-experiments:tracking}
\label{sec:results:tracking}

\begin{figure}[t]
    \centering
    \captionsetup{skip=-6pt}
    \def\svgwidth{\columnwidth}
    \begin{small}
        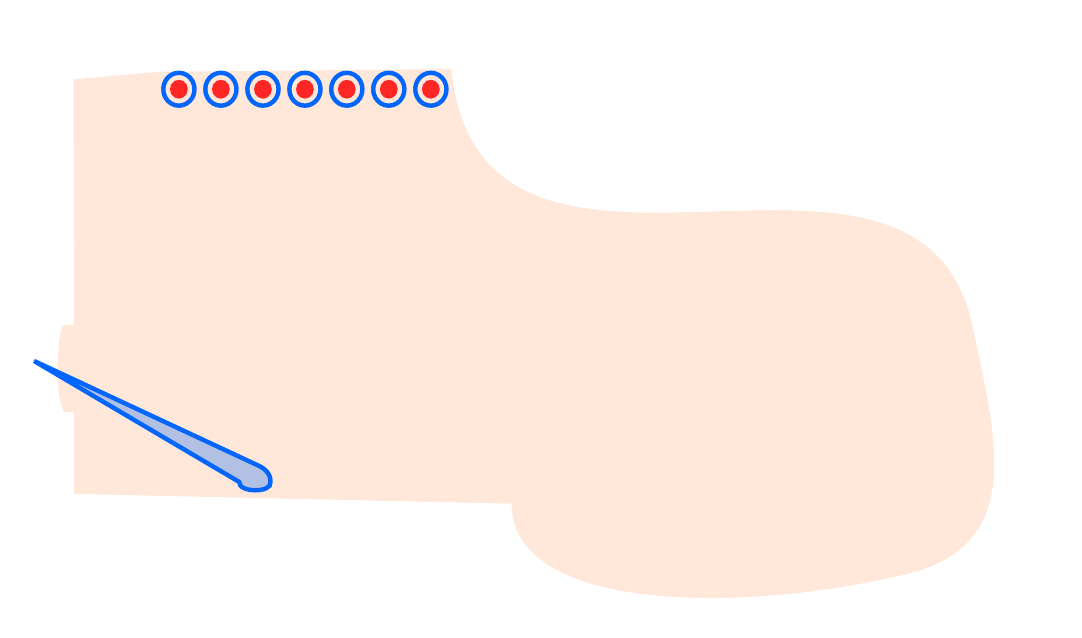
    \end{small}
    \caption{
    We simulate tracking an object in a region $\lV_\text{target}$ hidden around two corners, not visible from the relay wall $\lRone$. (a) Light from the target object is reflected on the rough hidden wall $\lRtwo$ before reaching $\lRone$, resulting in fifth-bounce illumination (yellow arrow). 
    (b) First, we image the secondary wall $\lRtwo$ from $\lRone$ using third-bounce illumination (purple arrow).
    (c) \final{The virtual mirrors (VM) approach \cite{royo2023virtual} fails due to the roughness of $\lRtwo$.} Our cascaded NLOS imaging yields a good approximation of the trajectory of the hidden object by implementing a secondary imaging system at $\lRtwo$.}
    \label{fig:twocorner-tracking}
\end{figure}

\fref{fig:twocorner-tracking}a shows an NLOS setup where we detect and track an object hidden around two corners in the presence of a rough secondary relay wall $\lRtwo$, with the centimeter-scale features one may find in regular brick or stone walls.
Light from the  object needs to bounce on $\lRtwo$ before reaching the relay wall $\lRone$, resulting in fifth-bounce illumination (yellow arrow) captured on $H$. 
We observe that the interaction of light with rough walls produces distinctive transport that affects NLOS imaging, \final{as explained in \sref{sec:virtual-reflectance-main-paper}}. Concretely, rough surfaces decrease spatial imaging resolution but allow us to image parts of the hidden scene that would otherwise remain invisible to other NLOS methods, similar to how diffuse surfaces scatter the reflected light across a broader angular range.

We simulate the capture of $H$ for $30$ points in $\lL$ and $\lS$ on the \SI{2}{m} relay wall $\lRone$. 
Our method works in two steps. First, we image $\lRtwo$ using conventional third-bounce illumination (purple arrow). In particular, we compute $f_{1,1}(\xv) = \Icc(\xv; H)$ (\eref{eq:confocal-camera}) on points $\xv$ in the bounding volume $\lV_2$ using $\Icc$.
We assign apertures $\mathcal{L}_2$ and $\mathcal{S}_2$ that follow our estimation of $\lRtwo$. Note that $\mathcal{L}_2$ and $\mathcal{S}_2$ can be planar, i.e., they do not need to follow the irregularities of $\lRtwo$.

Our second step uses the fifth-bounce method from \sref{sec:methods:fifth} to image the target object. We compute \eeref{eq:virtual-impulse-response}{eq:second-reconstruction} with $a = b = 2$ to create a second virtual imaging system at $\lRtwo$. However, the roughness of $\lRtwo$ also introduces virtual diffraction, which would create strong speckle patterns in the resulting image $f_{2,2}(\xv)$. To mitigate this, we instead take the \textit{absolute value} of the virtual impulse response $\Hp_{2,2}$, which effectively removes the phase of the reconstructed wave, retaining only its magnitude. We use $\lambda_c = \sigma = \SI{10}{cm}$ for $H$ and $\lambda_c = \sigma = \SI{25}{cm}$ for $\Hp_{2,2}$.
We repeat this for different positions of the target object, tracking its location by finding a nearby peak in the image $f_{2,2}(\xv)$ at points $\xv \in \lV_\text{target}$. \fref{fig:twocorner-tracking}c shows that we accurately follow the object's trajectory.
\final{The figure also shows how the virtual mirrors approach from \customcitet{Royo}{royo2023virtual}, which requires a planar secondary relay wall and images the mirrored region behind $\lRtwo$, fails to accurately track the same object.}

%% file: fig/moreresults-v14.pdf_tex
\begingroup%
  \makeatletter%
  \providecommand\color[2][]{%
    \errmessage{(Inkscape) Color is used for the text in Inkscape, but the package 'color.sty' is not loaded}%
    \renewcommand\color[2][]{}%
  }%
  \providecommand\transparent[1]{%
    \errmessage{(Inkscape) Transparency is used (non-zero) for the text in Inkscape, but the package 'transparent.sty' is not loaded}%
    \renewcommand\transparent[1]{}%
  }%
  \providecommand\rotatebox[2]{#2}%
  \newcommand*\fsize{\dimexpr\f@size pt\relax}%
  \newcommand*\lineheight[1]{\fontsize{\fsize}{#1\fsize}\selectfont}%
  \ifx\svgwidth\undefined%
    \setlength{\unitlength}{1278.55011227bp}%
    \ifx\svgscale\undefined%
      \relax%
    \else%
      \setlength{\unitlength}{\unitlength * \real{\svgscale}}%
    \fi%
  \else%
    \setlength{\unitlength}{\svgwidth}%
  \fi%
  \global\let\svgwidth\undefined%
  \global\let\svgscale\undefined%
  \makeatother%
  \begin{picture}(1,0.90593487)%
    \lineheight{1}%
    \setlength\tabcolsep{0pt}%
    \put(0.50005451,0.54712741){\makebox(0,0)[t]{\lineheight{1.25}\smash{\begin{tabular}[t]{c}(a) Scene setup and relevant paths\end{tabular}}}}%
    \put(0,0){\includegraphics[width=\unitlength,page=1]{moreresults-v14.pdf}}%
    \put(0.81450677,0.87839356){\color[rgb]{1,0.14901961,0.14901961}\makebox(0,0)[lt]{\lineheight{1.25}\smash{\begin{tabular}[t]{l}$\lL$\end{tabular}}}}%
    \put(0.70062191,0.88403429){\color[rgb]{0,0.4,1}\makebox(0,0)[lt]{\lineheight{1.25}\smash{\begin{tabular}[t]{l}$\lSp$\end{tabular}}}}%
    \put(0.59451764,0.88967512){\color[rgb]{0,0.4,1}\makebox(0,0)[lt]{\lineheight{1.25}\smash{\begin{tabular}[t]{l}$\mathcal{S}_3$\end{tabular}}}}%
    \put(0.8614199,0.87839356){\color[rgb]{0,0.4,1}\makebox(0,0)[lt]{\lineheight{1.25}\smash{\begin{tabular}[t]{l}$\lS$\end{tabular}}}}%
    \put(0.77500713,0.87839356){\makebox(0,0)[lt]{\lineheight{1.25}\smash{\begin{tabular}[t]{l}$\lRone$\end{tabular}}}}%
    \put(0.65701816,0.88403429){\makebox(0,0)[lt]{\lineheight{1.25}\smash{\begin{tabular}[t]{l}$\lRtop$\end{tabular}}}}%
    \put(0.55086399,0.88967512){\makebox(0,0)[lt]{\lineheight{1.25}\smash{\begin{tabular}[t]{l}$\lRright$\end{tabular}}}}%
    \put(0,0){\includegraphics[width=\unitlength,page=2]{moreresults-v14.pdf}}%
    \put(0.10348587,0.13806495){\makebox(0,0)[t]{\lineheight{1.14999998}\smash{\begin{tabular}[t]{c}$f_{1,1}(\xv)$\\Projector $\lL$\\Camera $\lS$\end{tabular}}}}%
    \put(0.10383209,0.48514168){\makebox(0,0)[t]{\lineheight{1.25}\smash{\begin{tabular}[t]{c}\textbf{Third bounce}\end{tabular}}}}%
    \put(0.44914435,0.48609059){\makebox(0,0)[t]{\lineheight{1.25}\smash{\begin{tabular}[t]{c}\textbf{Fourth bounce (ours)}\end{tabular}}}}%
    \put(0.33499391,0.13806495){\makebox(0,0)[t]{\lineheight{1.14999998}\smash{\begin{tabular}[t]{c}$f_{1,2}(\xv)$\\Projector $\mathcal{L}_1$\\Camera $\mathcal{S}_2$\end{tabular}}}}%
    \put(0.56410509,0.13806495){\makebox(0,0)[t]{\lineheight{1.14999998}\smash{\begin{tabular}[t]{c}$f_{1,3}(\xv)$\\Projector $\mathcal{L}_1$\\Camera $\mathcal{S}_3$\end{tabular}}}}%
    \put(0.85866045,0.00702884){\makebox(0,0)[t]{\lineheight{1.25}\smash{\begin{tabular}[t]{c}(c) Combined image\end{tabular}}}}%
    \put(0.33516285,0.00829358){\makebox(0,0)[t]{\lineheight{1.25}\smash{\begin{tabular}[t]{c}(b) Imaging with multiple apertures\end{tabular}}}}%
    \put(0,0){\includegraphics[width=\unitlength,page=3]{moreresults-v14.pdf}}%
    \put(0.73400639,0.16213932){\rotatebox{90}{\makebox(0,0)[t]{\lineheight{1.25}\smash{\begin{tabular}[t]{c}$\fresult(\xv)$\end{tabular}}}}}%
    \put(0.73609931,0.38986388){\rotatebox{90}{\makebox(0,0)[t]{\lineheight{1.25}\smash{\begin{tabular}[t]{c}Ground truth\end{tabular}}}}}%
  \end{picture}%
\endgroup%

%% file: fig/twocorner-tracking-v11.pdf_tex
\begingroup%
  \makeatletter%
  \providecommand\color[2][]{%
    \errmessage{(Inkscape) Color is used for the text in Inkscape, but the package 'color.sty' is not loaded}%
    \renewcommand\color[2][]{}%
  }%
  \providecommand\transparent[1]{%
    \errmessage{(Inkscape) Transparency is used (non-zero) for the text in Inkscape, but the package 'transparent.sty' is not loaded}%
    \renewcommand\transparent[1]{}%
  }%
  \providecommand\rotatebox[2]{#2}%
  \newcommand*\fsize{\dimexpr\f@size pt\relax}%
  \newcommand*\lineheight[1]{\fontsize{\fsize}{#1\fsize}\selectfont}%
  \ifx\svgwidth\undefined%
    \setlength{\unitlength}{518.57034518bp}%
    \ifx\svgscale\undefined%
      \relax%
    \else%
      \setlength{\unitlength}{\unitlength * \real{\svgscale}}%
    \fi%
  \else%
    \setlength{\unitlength}{\svgwidth}%
  \fi%
  \global\let\svgwidth\undefined%
  \global\let\svgscale\undefined%
  \makeatother%
  \begin{picture}(1,0.58998073)%
    \lineheight{1}%
    \setlength\tabcolsep{0pt}%
    \put(0,0){\includegraphics[width=\unitlength,page=1]{twocorner-tracking-v11.pdf}}%
    \put(0.26692602,0.08664108){\color[rgb]{1,0.14901961,0.14901961}\makebox(0,0)[t]{\lineheight{1.25}\smash{\begin{tabular}[t]{c}$\lL$\end{tabular}}}}%
    \put(0.32049619,0.08664108){\color[rgb]{0,0.4,1}\makebox(0,0)[t]{\lineheight{1.25}\smash{\begin{tabular}[t]{c}$\lS$\end{tabular}}}}%
    \put(0.2133558,0.08664108){\color[rgb]{0,0,0}\makebox(0,0)[t]{\lineheight{1.25}\smash{\begin{tabular}[t]{c}$\lRone$\end{tabular}}}}%
    \put(0.73545587,0.00555374){\color[rgb]{0,0,0}\makebox(0,0)[t]{\lineheight{1.25}\smash{\begin{tabular}[t]{c}(c) Step 2: tracking with $f_{2,2}(\xv_\text{target})$\end{tabular}}}}%
    \put(0.66584634,0.35776912){\color[rgb]{0,0,0}\makebox(0,0)[t]{\lineheight{1.25}\smash{\begin{tabular}[t]{c}Initial position of the\\\emph{moving} target object\end{tabular}}}}%
    \put(0.26681438,0.04026456){\color[rgb]{0,0,0}\makebox(0,0)[t]{\lineheight{1.25}\smash{\begin{tabular}[t]{c}(a) Setup overview\end{tabular}}}}%
    \put(0.76016787,0.41261521){\color[rgb]{0,0,0}\makebox(0,0)[t]{\lineheight{1.25}\smash{\begin{tabular}[t]{c}(b) Step 1: image $\lRtwo$ in $f_{1,1}(\xv_2)$\end{tabular}}}}%
    \put(0,0){\includegraphics[width=\unitlength,page=2]{twocorner-tracking-v11.pdf}}%
    \put(0.9300176,0.49955268){\color[rgb]{1,1,1}\makebox(0,0)[lt]{\lineheight{1.25}\smash{\begin{tabular}[t]{l}$\lV_2$\end{tabular}}}}%
    \put(0,0){\includegraphics[width=\unitlength,page=3]{twocorner-tracking-v11.pdf}}%
    \put(0.77588067,0.24220767){\color[rgb]{0,0,0}\makebox(0,0)[lt]{\lineheight{1.25}\smash{\begin{tabular}[t]{l}True trajectory\end{tabular}}}}%
    \put(0.77588083,0.19935655){\color[rgb]{0,0,0}\makebox(0,0)[lt]{\lineheight{1.25}\smash{\begin{tabular}[t]{l}Estimate (VM)\end{tabular}}}}%
    \put(0,0){\includegraphics[width=\unitlength,page=4]{twocorner-tracking-v11.pdf}}%
    \put(0.31125714,0.57239393){\color[rgb]{1,0.14901961,0.14901961}\makebox(0,0)[t]{\lineheight{1.25}\smash{\begin{tabular}[t]{c}$\lLp$\end{tabular}}}}%
    \put(0.41441137,0.57239393){\color[rgb]{0.12941176,0.61568627,0}\makebox(0,0)[t]{\lineheight{1.25}\smash{\begin{tabular}[t]{c}$\lV_2$\end{tabular}}}}%
    \put(0.70211488,0.07474972){\color[rgb]{0.12941176,0.61568627,0}\makebox(0,0)[lt]{\lineheight{1.25}\smash{\begin{tabular}[t]{l}$\lV_\text{target}$\end{tabular}}}}%
    \put(0.36401735,0.57239393){\color[rgb]{0,0.4,1}\makebox(0,0)[t]{\lineheight{1.25}\smash{\begin{tabular}[t]{c}$\lSp$\end{tabular}}}}%
    \put(0.28028355,0.57239391){\color[rgb]{0,0,0}\makebox(0,0)[rt]{\lineheight{1.25}\smash{\begin{tabular}[t]{r}\emph{Rough} $\lRtwo$\end{tabular}}}}%
    \put(0,0){\includegraphics[width=\unitlength,page=5]{twocorner-tracking-v11.pdf}}%
    \put(0.92148978,0.49892912){\color[rgb]{1,1,1}\makebox(0,0)[lt]{\lineheight{1.25}\smash{\begin{tabular}[t]{l}$\lV_2$\end{tabular}}}}%
    \put(0,0){\includegraphics[width=\unitlength,page=6]{twocorner-tracking-v11.pdf}}%
    \put(0.77588267,0.15398149){\color[rgb]{0,0,0}\makebox(0,0)[lt]{\lineheight{1.25}\smash{\begin{tabular}[t]{l}Estimate (Ours)\end{tabular}}}}%
    \put(0,0){\includegraphics[width=\unitlength,page=7]{twocorner-tracking-v11.pdf}}%
  \end{picture}%
\endgroup%

%% file: journal_modified_tex/70_real_experiments.tex
\section{Results with our real prototype}
\label{sec:real-experiments}

We now present real-world results captured with our hardware prototype, imaging scenes using higher-order illumination scattered by secondary hidden surfaces. We evaluate our fourth-bounce imaging method in \sref{sec:real-experiments:fourth}, and our fifth-bounce imaging method in \sref{sec:real-experiments:smooth}, including imaging through rough walls, and without any prior knowledge of the hidden scene.

\begin{figure}[t]
    \centering
    \captionsetup{skip=-4pt}
    \def\svgwidth{\columnwidth}
    \begin{small}
        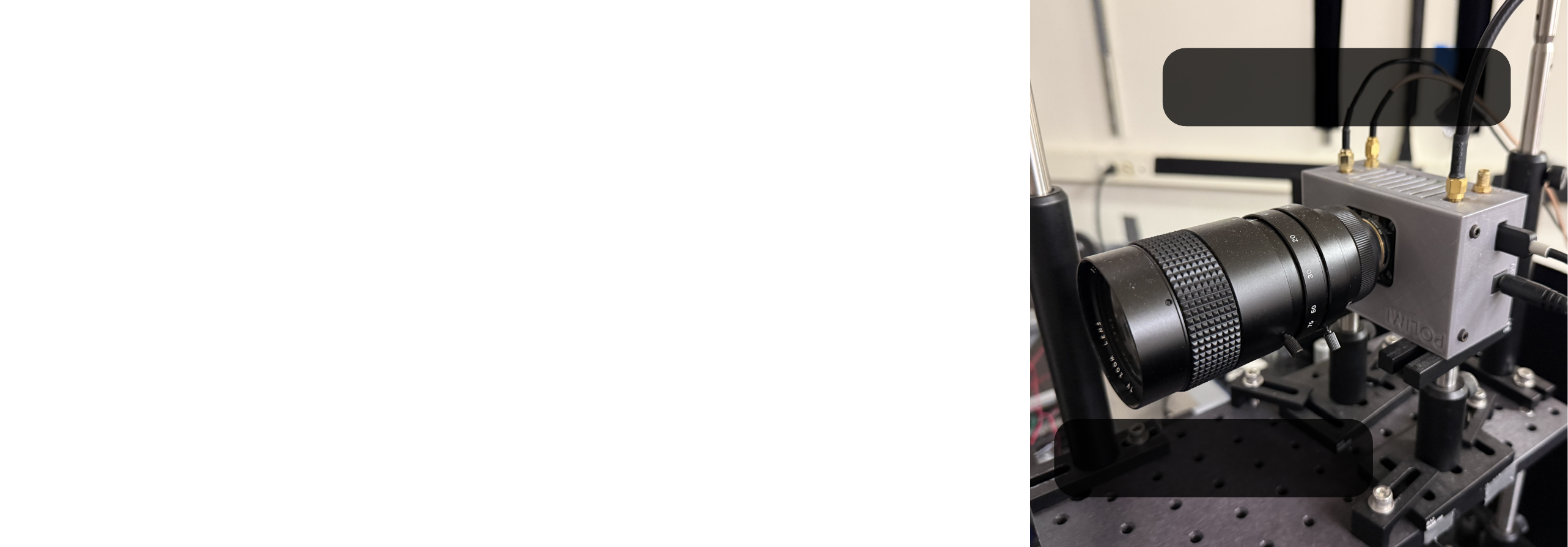
    \end{small}
    \caption{Our hardware prototype. We use a $16\times16$ SPAD array, focused on the relay wall using a $6\times$ Edmund Optics lens, and a two-mirror galvanometer (Thorlabs GVS012) to guide an ultra-fast laser emitter towards a $190\times190$ grid of points on the relay wall.}
    \label{fig:hardware}
\end{figure}
\paragraph{Hardware setup.} Our NLOS imaging system (\fref{fig:hardware}) consists of a $16\times16$ SPAD array sensor \cite{riccardoFastGated16162022}, a laser emitter (Onefive KATANA HP) and a two-mirror galvanometer (Thorlabs GVS012). The galvanometer guides the laser towards a grid of $190 \times 190$ points on a \qtyproduct{1.9 x 1.9}{m} region of the relay wall, resulting in \SI{1}{cm} spacing between adjacent laser points. We use an Edmund Optics 6$\times$ lens (\SI{12.5}{mm}, f/1.2) to focus the sensor array on a \qtyproduct{50 x 32.5}{cm} region on the relay wall. The laser emits \SI{532}{nm} pulses with a maximum pulse width of \SI{40}{ps}, at an average repetition rate of \SI{10}{MHz}. The sensor has a temporal resolution of approximately \SI{60}{ps} Full-Width at Half Maximum (FWHM).
\final{Pictures of all our scenes are included in Appendix F of our Supplemental Material.}

\subsection{Fourth-bounce imaging results}
\label{sec:real-experiments:fourth}
\begin{figure}[t]
    \centering
    \captionsetup{skip=-4pt}
    \def\svgwidth{\columnwidth} 
    \begin{small}
    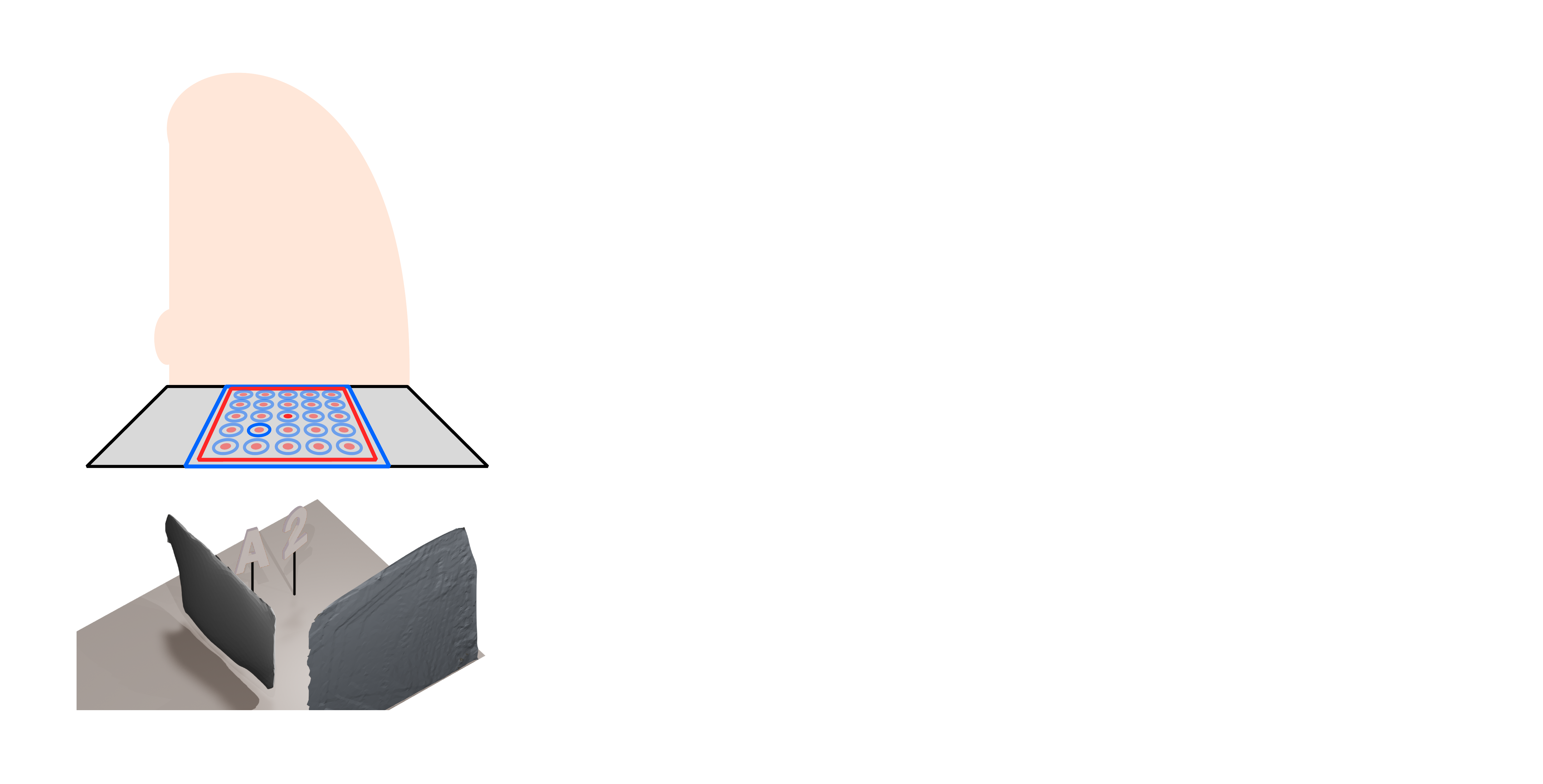
    \end{small}
    \caption{
    Real-world demonstration of NLOS imaging with fourth-bounce light. (a) We use a setup similar to \fref{fig:multiview}. (b) While the 2 can be imaged using third-bounce methods ($f_{1,1}(\xv)$), the orientation of the A requires fourth-bounce illumination ($f_{1,2}(\xv)$), enabled by our cascaded approach (analogous to our simulations in  \fref{fig:multiview}b and \fref{fig:multiview}c). (c) Combined image $\fresult(\xv)= f_{1,1}(\xv)+ f_{1,2}(\xv)$.}
    \label{fig:real-fourth-bounce}
\end{figure}

In \fref{fig:real-fourth-bounce}a, we show a scene similar to \fref{fig:multiview}.
We focus on two target hidden objects: a \emph{2}-shaped object which can be imaged from the relay wall $\lRone$ using third-bounce methods, and an \emph{A}-shaped object which has limited visibility to third-bounce methods due to its orientation. Again, we capture the impulse response $H$ at the relay wall $\lRone$ (see hardware setup). To image the hidden A, we leverage fourth-bounce illumination from $H$ coming from the left wall $\lRleft$.

We use our fourth-bounce imaging method: first, we compute our virtual impulse response $H'_{1,2}$ from $H$ by applying \eref{eq:virtual-impulse-response-onlyS}. We sample a grid of $40 \times 40$ points $\mathbf{s}_2 \in \mathcal{S}_2$ on $\lRleft$. \fref{fig:real-fourth-bounce}b shows our imaging results: while the third-bounce-method image $f_{1,1}(\xv)$ computed from $H$ only shows the 2 at points $\xv$ in the bounding volume $\lV$, our fourth-bounce-method $f_{1,2}(\xv)$ computed from $H'_{1,2}$ successfully images the A. We use $\wlc=\sigma=\SI{6}{cm}$ for computing $f_{1,1}(\xv)$, and $\wlc=\sigma=\SI{8}{cm}$ for computing $H'_{1,2}$ and $f_{1,2}$.

\paragraph{Resolution analysis.} Imaging quality depends heavily on the resolution and size of the captured laser $\lL$ and sensor $\lS$ grids, which we discuss further in Appendix B of our Supplemental Material.
Our simulations in \sref{sec:results} mimic a  $32\times 32$ SPAD array covering a \qtyproduct{2 x 2}{m} area of the relay wall.
While higher-resolution SPAD arrays are progressively becoming available, our prototype uses a lower-resolution, state-of-the-art SPAD array with $16\times16$ points, covering a smaller area on the relay wall of \qtyproduct{50 x 32.5}{cm}.
These limitations affect the fidelity of the real-world results, introducing factors such as higher noise and reduced imaging resolution that account for the observed simulation-to-real gap.

\begin{figure}[t]
    \centering
    \captionsetup{skip=6pt}
    \def\svgwidth{0.85\columnwidth}
    \begin{small}
        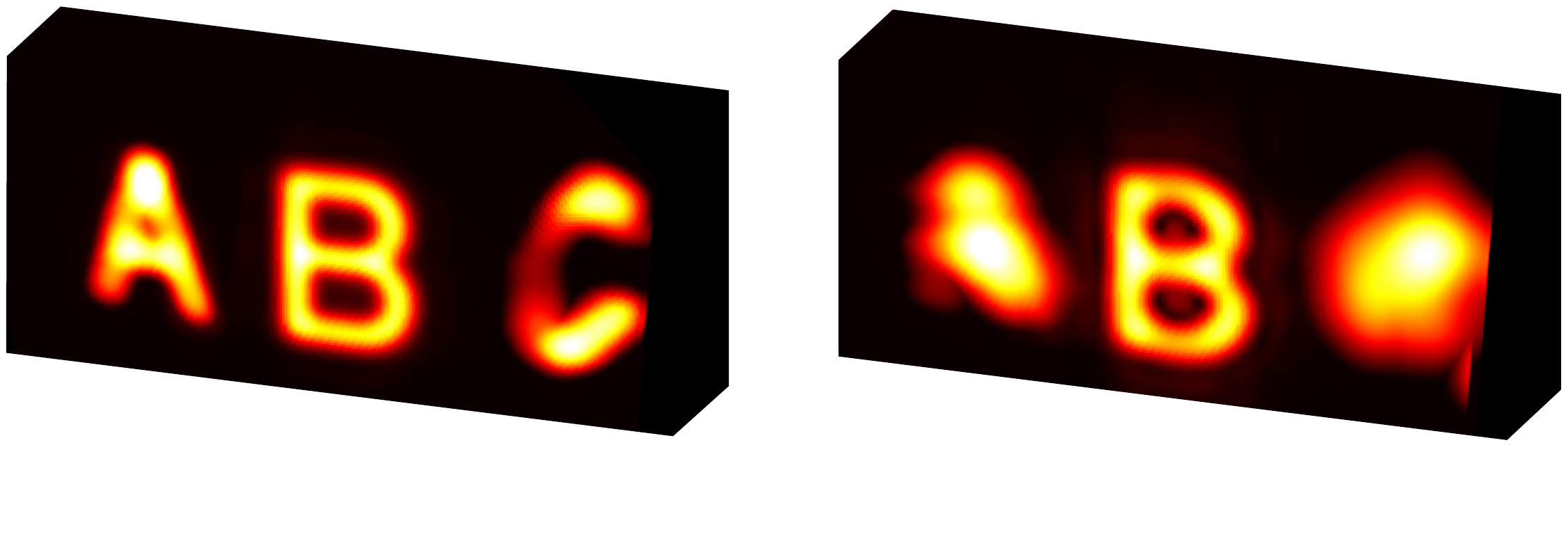
    \end{small}
    \caption{We repeat in simulation the same imaging procedure from \fref{fig:multiview}e, changing the size and resolution of the illumination aperture $\lL$ to match the SPAD array from our prototype. (a) Original result. (b) Reducing both the resolution to $16\times 16$ points and covered area to \qtyproduct{50 x 32.5}{cm}. The fourth-bounce imaging results of the A letter match the resolution of our real capture in \fref{fig:real-fourth-bounce}. Moreover, as explained in Section \ref{sec:real-experiments:detection}, the fifth-bounce results of the C letter also match  our real-world, fifth-bounce experiments.}
    \label{fig:fourthbounce-degradation}
\end{figure}
In \fref{fig:fourthbounce-degradation}, we validate our method by modifying our simulation parameters in the scene of \fref{fig:multiview} to match our real prototype, both in aperture size and SPAD resolution. \fref{fig:fourthbounce-degradation}a shows the combined result of our original simulation with a $32\times 32$ SPAD array capturing a \qtyproduct{2 x 2}{m} area on the relay wall. In \fref{fig:fourthbounce-degradation}b, we change our simulated SPAD array to match the $16\times 16$ resolution and \qtyproduct{50 x 32.5}{cm} area covered by our real prototype. The simulated result degrades consistently. The B and A letters closely match the captured third- and fourth-bounce results in \fref{fig:real-fourth-bounce}. Moreover, as we will see in \sref{sec:real-experiments:detection}, the fifth-bounce image of the C letter also matches the resolution of our real fifth-bounce results.
\vspace{2em}

\subsection{Fifth-bounce imaging results}
\paragraph{Two-corner imaging in unfavorable orientations. }
\label{sec:real-experiments:detection}
\label{sec:real-experiments:smooth}

\begin{figure}[t]
    \centering
    \captionsetup{skip=0pt}
    \def\svgwidth{\columnwidth}
    \begin{small}
        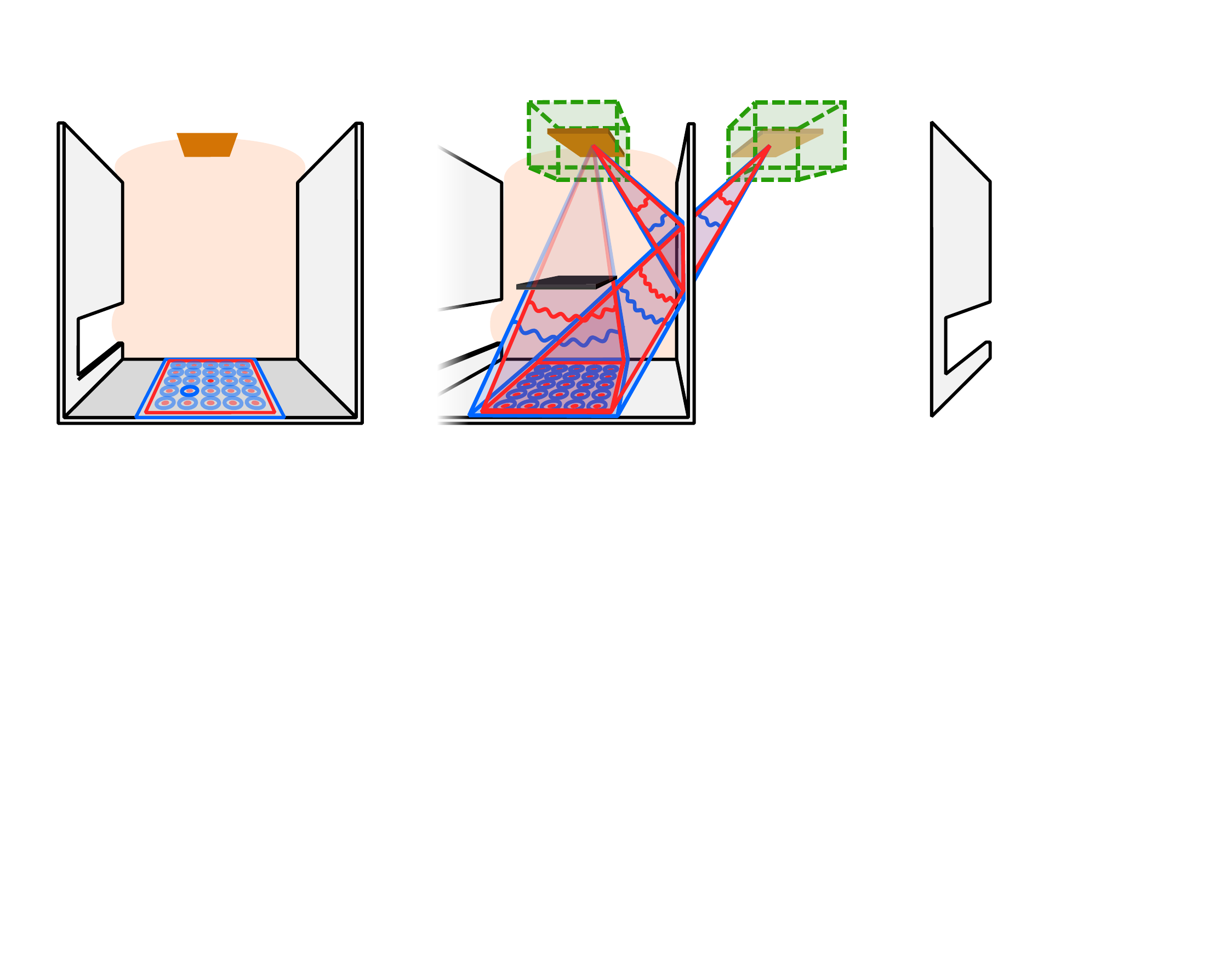
    \end{small}
    \caption{Overcoming two-corner visibility limits using our hardware prototype. (a) The left $\lRtwo$ and right $\lRthree$ walls reflect higher-order illumination to and from the target object. (b) Due to the walls' position and orientation, neither third-bounce methods nor the virtual mirrors approach  \cite{royo2023virtual} can image the target object (marked by $\lV$ and $\lV^m$). (c) Our cascaded NLOS fifth-bounce method images the object at $\lV$ using $\mathcal{L}_2$ and $\mathcal{S}_3$.}
    \label{fig:detection_planar}
\end{figure}
\begin{figure}[t]
    \centering
    \captionsetup{skip=-6pt}
    \def\svgwidth{\columnwidth} 
    \begin{small}
    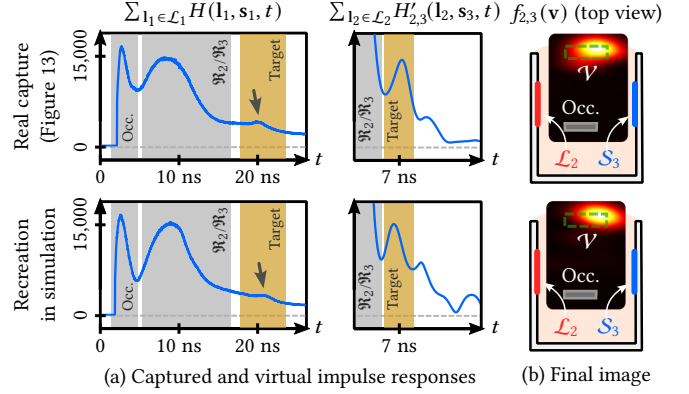
    \end{small}
    \caption{
    We compare our real-world experiment (top) from \fref{fig:detection_planar} with its simulated counterpart (bottom), mimicking our $16\times16$ SPAD. Both the captured $H$ and virtual $\Hp_{2,3}$ impulse responses (a), as well as the resulting image $f_{2,3}(\xv)$ (b) closely match, with small differences due to geometric discrepancies between the real  and the simulated scenes.}
    \label{fig:simulation-32x32spad}
\end{figure}
\fref{fig:detection_planar}a illustrates an object hidden around the corner (highlighted in orange) with an additional occluder (black) in front, preventing third- and even fourth-bounce illumination from reaching $\lRone$ and the sensor. We capture the impulse response $H$ on $\lRone$, which includes fifth-bounce illumination from walls $\lRtwo$ and $\lRthree$.

As Figure~\ref{fig:detection_planar}b shows, conventional third-bounce methods fail to image the object, which should appear in the volume $\lV$. Moreover, the recent \textit{virtual mirrors} approach \cite{royo2023virtual}, which images the mirrored volume $\lV^m$ and relies on fifth-bounce illumination, also fails: their method imposes strict conditions on the position and orientation of both the hidden objects and the secondary walls, which need to align with favorable mirror directions. Note that Royo et al. originally demonstrated their approach using only a single illumination point $\xl$. To provide a fair comparison, we have enhanced their method to use our $16 \times 16$ illumination grid $\lL$.

\final{In contrast, our fifth-bounce method (\sref{sec:methods:fifth}) leverages a broader range of specular paths, and thus is able to image the hidden object at the correct location marked by $\lV$, by creating both a virtual illumination aperture $\mathcal{L}_2$ on $\lRtwo$, and a virtual camera aperture $\mathcal{S}_3$ on a different wall $\lRthree$ (\fref{fig:detection_planar}c). }
Specifically, we compute our virtual impulse response $\Hp_{2,3}(\mathbf{l}_2, \textbf{s}_3, t)$ (\eref{eq:virtual-impulse-response} with $a=2$ and $b=3$), effectively illuminating points $\mathbf{l}_2 \in \mathcal{L}_2$ and computing the time-resolved response at points $\mathbf{s}_3 \in \mathcal{S}_3$. For both \emph{secondary} apertures, we sample grids of $8\times6$ points with \SI{12.5}{cm} spacing. We then use $\Hp_{2,3}$ as input for the confocal camera operator $\Icc$ (\eref{eq:second-reconstruction}), obtaining an image $f_{2,3}(\xv)$, which reveals the target object at the correct location. We use $\wlc = \sigma = \SI{30}{cm}$ for $H$ and $\Hp_{2,3}$.

\fref{fig:simulation-32x32spad} further validates our method,  recreating the same scene in simulation, mimicking the $16\times16$ SPAD from our prototype.
The captured impulse response $H$ and virtual impulse response $\Hp_{2,3}$, as well as the resulting image $f_{2,3}(\xv)$, closely match our physical capture.
\final{Note that we show the sum over all illumination positions for $H$ and $\Hp_{2,3}$, which smooths SPAD shot noise. Early-arriving photons preceding the third-bounce signal are gated out following standard practice.} Other small differences are mainly due to geometric discrepancies between the real and the simulated scenes.

\paragraph{Two-corner imaging through rough walls.}
\label{sec:real-experiments:rough}

One key advantage of our method is that it works in the presence of rough relay walls.
We illustrate this in our next example, imaging a scene around two corners without previous knowledge of the secondary relay wall. 

\begin{figure}[t]
    \centering
    \captionsetup{skip=0pt}
    \def\svgwidth{\columnwidth}
    \begin{small}
        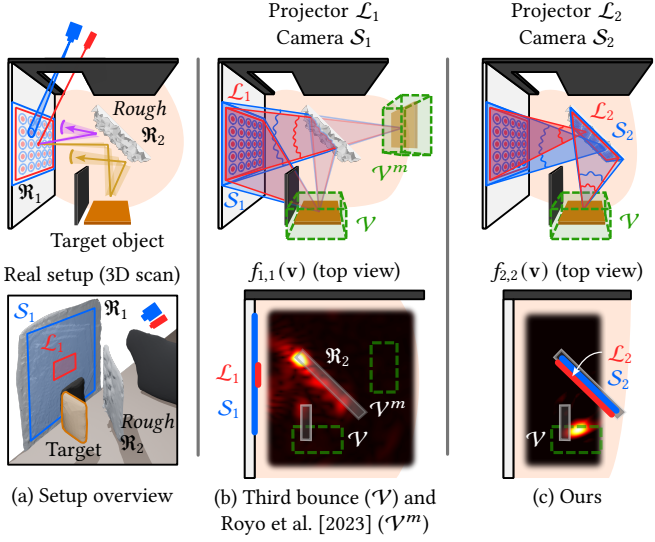
    \end{small}
    \caption{Imaging around two corners using our hardware prototype. (a) The rough hidden wall $\lRtwo$ reflects light to and from the target object. (b) Both conventional methods and virtual mirrors \cite{royo2023virtual} fail to image the target object (marked by $\lV$ and $\lV^m$, respectively). (c) In contrast, our method identifies the target at the correct location $\lV$ by creating a secondary virtual imaging system at $\lRtwo$.}
    \label{fig:detection_nonplanar}
\end{figure}
\fref{fig:detection_nonplanar}a shows our setup.
As in the previous example, both third-bounce methods and our enhanced version of virtual mirrors \cite{royo2023virtual} (which leverages our $16\times16$ points for $\lL$) fail to image the target at $\lV$ and $\lV^m$, respectively (Figure~\ref{fig:detection_nonplanar}b).

We first image the secondary wall $\lRtwo$ by computing a conventional third-bounce image $f_{1,1}(\xv)$ on points $\xv \in \lV$ (as shown in \fref{fig:detection_nonplanar}b).
Leveraging our cascaded approach, we then compute a secondary impulse response $\Hp_{2,2}$ (\eref{eq:virtual-impulse-response}) by creating virtual illumination and camera apertures $\mathcal{L}_2$ and $\mathcal{S}_2$ (\fref{fig:detection_nonplanar}c, top) at the revealed location of $\lRtwo$ in $f_{1,1}(\xv)$ (see Appendix C of our Supplemental Material for further details). We sample grids of $6\times6$ points for both apertures with $\SI{12.5}{cm}$ spacing. Finally, we use $\Hp_{2,2}$ to image the target object by computing $f_{2,2}(\xv)$ (\eref{eq:second-reconstruction}).
\fref{fig:detection_nonplanar}c, bottom, shows that $f_{2,2}(\xv)$ reveals the target object hidden around two corners on its location marked by $\lV$. 
Our algorithm leverages roughness in the secondary relay wall by modulating $\lambda_c$ accordingly; in particular, we use $\lambda_c = \sigma = \SI{15}{cm}$ for $H$, and a coarser $\lambda_c = \sigma = \SI{25}{cm}$ for $\Hp_{2,2}$.

%% file: fig/hardware-v3.pdf_tex
\begingroup%
  \makeatletter%
  \providecommand\color[2][]{%
    \errmessage{(Inkscape) Color is used for the text in Inkscape, but the package 'color.sty' is not loaded}%
    \renewcommand\color[2][]{}%
  }%
  \providecommand\transparent[1]{%
    \errmessage{(Inkscape) Transparency is used (non-zero) for the text in Inkscape, but the package 'transparent.sty' is not loaded}%
    \renewcommand\transparent[1]{}%
  }%
  \providecommand\rotatebox[2]{#2}%
  \newcommand*\fsize{\dimexpr\f@size pt\relax}%
  \newcommand*\lineheight[1]{\fontsize{\fsize}{#1\fsize}\selectfont}%
  \ifx\svgwidth\undefined%
    \setlength{\unitlength}{1230.85491511bp}%
    \ifx\svgscale\undefined%
      \relax%
    \else%
      \setlength{\unitlength}{\unitlength * \real{\svgscale}}%
    \fi%
  \else%
    \setlength{\unitlength}{\svgwidth}%
  \fi%
  \global\let\svgwidth\undefined%
  \global\let\svgscale\undefined%
  \makeatother%
  \begin{picture}(1,0.348335)%
    \lineheight{1}%
    \setlength\tabcolsep{0pt}%
    \put(0,0){\includegraphics[width=\unitlength,page=1]{hardware-v3.pdf}}%
    \put(0.85215017,0.2832655){\color[rgb]{1,1,1}\makebox(0,0)[t]{\lineheight{1.25}\smash{\begin{tabular}[t]{c}SPAD array\end{tabular}}}}%
    \put(0.77388719,0.0453505){\color[rgb]{1,1,1}\makebox(0,0)[t]{\lineheight{1.25}\smash{\begin{tabular}[t]{c}$6\times$ lens\end{tabular}}}}%
    \put(0,0){\includegraphics[width=\unitlength,page=2]{hardware-v3.pdf}}%
    \put(0.44526433,0.0273){\color[rgb]{1,1,1}\makebox(0,0)[t]{\lineheight{1.25}\smash{\begin{tabular}[t]{c}Galvanometer\end{tabular}}}}%
    \put(0,0){\includegraphics[width=\unitlength,page=3]{hardware-v3.pdf}}%
    \put(0.16751039,0.22060738){\color[rgb]{1,1,1}\makebox(0,0)[t]{\lineheight{1.25}\smash{\begin{tabular}[t]{c}Laser emitter\end{tabular}}}}%
  \end{picture}%
\endgroup%

%% file: fig/fourthbounce-real-v8.pdf_tex
\begingroup%
  \makeatletter%
  \providecommand\color[2][]{%
    \errmessage{(Inkscape) Color is used for the text in Inkscape, but the package 'color.sty' is not loaded}%
    \renewcommand\color[2][]{}%
  }%
  \providecommand\transparent[1]{%
    \errmessage{(Inkscape) Transparency is used (non-zero) for the text in Inkscape, but the package 'transparent.sty' is not loaded}%
    \renewcommand\transparent[1]{}%
  }%
  \providecommand\rotatebox[2]{#2}%
  \newcommand*\fsize{\dimexpr\f@size pt\relax}%
  \newcommand*\lineheight[1]{\fontsize{\fsize}{#1\fsize}\selectfont}%
  \ifx\svgwidth\undefined%
    \setlength{\unitlength}{2485.95466674bp}%
    \ifx\svgscale\undefined%
      \relax%
    \else%
      \setlength{\unitlength}{\unitlength * \real{\svgscale}}%
    \fi%
  \else%
    \setlength{\unitlength}{\svgwidth}%
  \fi%
  \global\let\svgwidth\undefined%
  \global\let\svgscale\undefined%
  \makeatother%
  \begin{picture}(1,0.50826646)%
    \lineheight{1}%
    \setlength\tabcolsep{0pt}%
    \put(0,0){\includegraphics[width=\unitlength,page=1]{fourthbounce-real-v8.pdf}}%
    \put(0.18627343,0.00662338){\makebox(0,0)[t]{\lineheight{1.25}\smash{\begin{tabular}[t]{c}(a) Scene setup\end{tabular}}}}%
    \put(0.52000687,0.00618259){\makebox(0,0)[t]{\lineheight{1.25}\smash{\begin{tabular}[t]{c}(b) Imaging results\end{tabular}}}}%
    \put(0.86238023,0.00472443){\makebox(0,0)[t]{\lineheight{1.25}\smash{\begin{tabular}[t]{c}(c) Combined image\end{tabular}}}}%
    \put(0.52013447,0.48397723){\makebox(0,0)[t]{\smash{\begin{tabular}[t]{c}\textbf{Third bounce}\end{tabular}}}}%
    \put(0.26499566,0.22408353){\color[rgb]{1,0.14901961,0.14901961}\makebox(0,0)[t]{\lineheight{1.25}\smash{\begin{tabular}[t]{c}$\xl$\end{tabular}}}}%
    \put(0,0){\includegraphics[width=\unitlength,page=2]{fourthbounce-real-v8.pdf}}%
    \put(0.11111112,0.225836){\color[rgb]{0,0.4,1}\makebox(0,0)[t]{\lineheight{1.25}\smash{\begin{tabular}[t]{c}$\xs$\end{tabular}}}}%
    \put(0.29076277,0.29435187){\color[rgb]{0,0,0}\makebox(0,0)[t]{\lineheight{1.25}\smash{\begin{tabular}[t]{c}$\lRone$\end{tabular}}}}%
    \put(0,0){\includegraphics[width=\unitlength,page=3]{fourthbounce-real-v8.pdf}}%
    \put(0.04896306,0.48973789){\makebox(0,0)[lt]{\lineheight{1.25}\smash{\begin{tabular}[t]{l}$\lRleft$\end{tabular}}}}%
    \put(0,0){\includegraphics[width=\unitlength,page=4]{fourthbounce-real-v8.pdf}}%
    \put(0.13374093,0.50028248){\color[rgb]{0.50196078,0.50196078,0.50196078}\makebox(0,0)[t]{\lineheight{1.25}\smash{\begin{tabular}[t]{c}{\scriptsize $75^\circ$}\end{tabular}}}}%
    \put(0,0){\includegraphics[width=\unitlength,page=5]{fourthbounce-real-v8.pdf}}%
    \put(0.63725334,0.17712466){\color[rgb]{1,1,1}\makebox(0,0)[rt]{\lineheight{1.25}\smash{\begin{tabular}[t]{r}$f_{1,2}(\xv)$\end{tabular}}}}%
    \put(0.40084049,0.42040293){\color[rgb]{1,1,1}\makebox(0,0)[lt]{\lineheight{1.25}\smash{\begin{tabular}[t]{l}$f_{1,1}(\xv)$\end{tabular}}}}%
    \put(0,0){\includegraphics[width=\unitlength,page=6]{fourthbounce-real-v8.pdf}}%
    \put(0.5194951,0.23951335){\makebox(0,0)[t]{\smash{\begin{tabular}[t]{c}\textbf{Fourth bounce (ours)}\end{tabular}}}}%
    \put(0.86183849,0.23951335){\makebox(0,0)[t]{\smash{\begin{tabular}[t]{c}$\fresult(\xv)$\end{tabular}}}}%
    \put(0.86167508,0.48384388){\makebox(0,0)[t]{\smash{\begin{tabular}[t]{c}Reference (3D scan)\end{tabular}}}}%
    \put(0,0){\includegraphics[width=\unitlength,page=7]{fourthbounce-real-v8.pdf}}%
    \put(0.06476639,0.14781626){\makebox(0,0)[lt]{\lineheight{1.25}\smash{\begin{tabular}[t]{l}$\lRleft$\end{tabular}}}}%
    \put(0,0){\includegraphics[width=\unitlength,page=8]{fourthbounce-real-v8.pdf}}%
    \put(0.26009893,0.08749547){\color[rgb]{1,1,1}\makebox(0,0)[t]{\lineheight{1.25}\smash{\begin{tabular}[t]{c}$\lRone$\end{tabular}}}}%
    \put(0,0){\includegraphics[width=\unitlength,page=9]{fourthbounce-real-v8.pdf}}%
    \put(0.26198406,0.15811649){\color[rgb]{0.99215686,0.99215686,0.99215686}\makebox(0,0)[t]{\lineheight{1.25}\smash{\begin{tabular}[t]{c}Targets\end{tabular}}}}%
    \put(0,0){\includegraphics[width=\unitlength,page=10]{fourthbounce-real-v8.pdf}}%
  \end{picture}%
\endgroup%

%% file: fig/abc-sim2real-v2.pdf_tex
\begingroup%
  \makeatletter%
  \providecommand\color[2][]{%
    \errmessage{(Inkscape) Color is used for the text in Inkscape, but the package 'color.sty' is not loaded}%
    \renewcommand\color[2][]{}%
  }%
  \providecommand\transparent[1]{%
    \errmessage{(Inkscape) Transparency is used (non-zero) for the text in Inkscape, but the package 'transparent.sty' is not loaded}%
    \renewcommand\transparent[1]{}%
  }%
  \providecommand\rotatebox[2]{#2}%
  \newcommand*\fsize{\dimexpr\f@size pt\relax}%
  \newcommand*\lineheight[1]{\fontsize{\fsize}{#1\fsize}\selectfont}%
  \ifx\svgwidth\undefined%
    \setlength{\unitlength}{1083.07158703bp}%
    \ifx\svgscale\undefined%
      \relax%
    \else%
      \setlength{\unitlength}{\unitlength * \real{\svgscale}}%
    \fi%
  \else%
    \setlength{\unitlength}{\svgwidth}%
  \fi%
  \global\let\svgwidth\undefined%
  \global\let\svgscale\undefined%
  \makeatother%
  \begin{picture}(1,0.35601262)%
    \lineheight{1}%
    \setlength\tabcolsep{0pt}%
    \put(0,0){\includegraphics[width=\unitlength,page=1]{abc-sim2real-v2.pdf}}%
    \put(0.2348102,0.03653039){\makebox(0,0)[t]{\smash{\begin{tabular}[t]{c}(a) $32\times 32$ SPAD,\\$2\,\text{m}\times 2\,\text{m}$ aperture\end{tabular}}}}%
    \put(0,0){\includegraphics[width=\unitlength,page=2]{abc-sim2real-v2.pdf}}%
    \put(0.77264277,0.03864825){\makebox(0,0)[t]{\smash{\begin{tabular}[t]{c}(b) $16\times 16$ SPAD,\\$50\,\text{cm}\times 32.5\,\text{cm}$ aperture\end{tabular}}}}%
    \put(0,0){\includegraphics[width=\unitlength,page=3]{abc-sim2real-v2.pdf}}%
  \end{picture}%
\endgroup%

%% file: fig/detect-occluded-v11-pad.pdf_tex
\begingroup%
  \makeatletter%
  \providecommand\color[2][]{%
    \errmessage{(Inkscape) Color is used for the text in Inkscape, but the package 'color.sty' is not loaded}%
    \renewcommand\color[2][]{}%
  }%
  \providecommand\transparent[1]{%
    \errmessage{(Inkscape) Transparency is used (non-zero) for the text in Inkscape, but the package 'transparent.sty' is not loaded}%
    \renewcommand\transparent[1]{}%
  }%
  \providecommand\rotatebox[2]{#2}%
  \newcommand*\fsize{\dimexpr\f@size pt\relax}%
  \newcommand*\lineheight[1]{\fontsize{\fsize}{#1\fsize}\selectfont}%
  \ifx\svgwidth\undefined%
    \setlength{\unitlength}{1078.94764818bp}%
    \ifx\svgscale\undefined%
      \relax%
    \else%
      \setlength{\unitlength}{\unitlength * \real{\svgscale}}%
    \fi%
  \else%
    \setlength{\unitlength}{\svgwidth}%
  \fi%
  \global\let\svgwidth\undefined%
  \global\let\svgscale\undefined%
  \makeatother%
  \begin{picture}(1,0.79125453)%
    \lineheight{1}%
    \setlength\tabcolsep{0pt}%
    \put(0,0){\includegraphics[width=\unitlength,page=1]{detect-occluded-v11-pad.pdf}}%
    \put(0.17037917,0.02073322){\makebox(0,0)[t]{\lineheight{1.25}\smash{\begin{tabular}[t]{c}(a) Setup overview\end{tabular}}}}%
    \put(0.17006217,0.36380502){\makebox(0,0)[t]{\lineheight{1.25}\smash{\begin{tabular}[t]{c}Real setup (3D scan)\end{tabular}}}}%
    \put(0.87865818,0.3629451){\makebox(0,0)[t]{\lineheight{1.25}\smash{\begin{tabular}[t]{c}$f_{2,3}(\xv)$ (top view)\end{tabular}}}}%
    \put(0.87899588,0.77556853){\makebox(0,0)[t]{\lineheight{1.25}\smash{\begin{tabular}[t]{c}Projector $\mathcal{L}_2$\\Camera $\mathcal{S}_3$\end{tabular}}}}%
    \put(0.52999077,0.77556853){\makebox(0,0)[t]{\lineheight{1.25}\smash{\begin{tabular}[t]{c}Projector $\mathcal{L}_1$\\Camera $\mathcal{S}_1$\end{tabular}}}}%
    \put(0.87599819,0.02341413){\makebox(0,0)[t]{\lineheight{1.25}\smash{\begin{tabular}[t]{c}(c) Ours\end{tabular}}}}%
    \put(0.26595889,0.51547275){\makebox(0,0)[t]{\lineheight{1.25}\smash{\begin{tabular}[t]{c}$\lRone$\end{tabular}}}}%
    \put(0,0){\includegraphics[width=\unitlength,page=2]{detect-occluded-v11-pad.pdf}}%
    \put(0.27149654,0.71137511){\makebox(0,0)[lt]{\lineheight{1.25}\smash{\begin{tabular}[t]{l}$\lRright$\end{tabular}}}}%
    \put(0.07485451,0.71101898){\makebox(0,0)[rt]{\lineheight{1.25}\smash{\begin{tabular}[t]{r}$\lRleft$\end{tabular}}}}%
    \put(0,0){\includegraphics[width=\unitlength,page=3]{detect-occluded-v11-pad.pdf}}%
    \put(0.79204377,0.69100706){\color[rgb]{1,0.14901961,0.14901961}\makebox(0,0)[t]{\lineheight{1.25}\smash{\begin{tabular}[t]{c}$\mathcal{L}_2$\end{tabular}}}}%
    \put(0.60194941,0.53366761){\color[rgb]{1,0.14901961,0.14901961}\makebox(0,0)[t]{\lineheight{1.25}\smash{\begin{tabular}[t]{c}$\mathcal{L}_1$\end{tabular}}}}%
    \put(0,0){\includegraphics[width=\unitlength,page=4]{detect-occluded-v11-pad.pdf}}%
    \put(0.95532757,0.69089385){\color[rgb]{0,0.4,1}\makebox(0,0)[t]{\lineheight{1.25}\smash{\begin{tabular}[t]{c}$\mathcal{S}_3$\end{tabular}}}}%
    \put(0.60195285,0.49564147){\color[rgb]{0,0.4,1}\makebox(0,0)[t]{\lineheight{1.25}\smash{\begin{tabular}[t]{c}$\mathcal{S}_1$\end{tabular}}}}%
    \put(0,0){\includegraphics[width=\unitlength,page=5]{detect-occluded-v11-pad.pdf}}%
    \put(0.16830684,0.73807459){\color[rgb]{0,0.00392157,0}\makebox(0,0)[t]{\smash{\begin{tabular}[t]{c}Target\\object\end{tabular}}}}%
    \put(0.52965307,0.3629451){\makebox(0,0)[t]{\lineheight{1.25}\smash{\begin{tabular}[t]{c}$f_{1,1}(\xv)$ (top view)\end{tabular}}}}%
    \put(0,0){\includegraphics[width=\unitlength,page=6]{detect-occluded-v11-pad.pdf}}%
    \put(0.48134867,0.2568443){\color[rgb]{1,1,1}\makebox(0,0)[rt]{\lineheight{1.25}\smash{\begin{tabular}[t]{r}$\lV$\end{tabular}}}}%
    \put(0.65520802,0.25654382){\color[rgb]{1,1,1}\makebox(0,0)[rt]{\lineheight{1.25}\smash{\begin{tabular}[t]{r}$\lV^m$\end{tabular}}}}%
    \put(0,0){\includegraphics[width=\unitlength,page=7]{detect-occluded-v11-pad.pdf}}%
    \put(0.45517447,0.11676033){\color[rgb]{1,0.14901961,0.14901961}\makebox(0,0)[t]{\lineheight{1.25}\smash{\begin{tabular}[t]{c}$\lL$\end{tabular}}}}%
    \put(0.52079875,0.11004692){\color[rgb]{0,0.4,1}\makebox(0,0)[t]{\lineheight{1.25}\smash{\begin{tabular}[t]{c}$\lS$\end{tabular}}}}%
    \put(0,0){\includegraphics[width=\unitlength,page=8]{detect-occluded-v11-pad.pdf}}%
    \put(0.89429781,0.25843554){\color[rgb]{1,1,1}\makebox(0,0)[rt]{\lineheight{1.25}\smash{\begin{tabular}[t]{r}$\lV$\end{tabular}}}}%
    \put(0,0){\includegraphics[width=\unitlength,page=9]{detect-occluded-v11-pad.pdf}}%
    \put(0.84104651,0.10893431){\color[rgb]{1,0.14901961,0.14901961}\makebox(0,0)[t]{\lineheight{1.25}\smash{\begin{tabular}[t]{c}$\mathcal{L}_2$\end{tabular}}}}%
    \put(0.91635369,0.10893431){\color[rgb]{0,0.4,1}\makebox(0,0)[t]{\lineheight{1.25}\smash{\begin{tabular}[t]{c}$\mathcal{S}_3$\end{tabular}}}}%
    \put(0,0){\includegraphics[width=\unitlength,page=10]{detect-occluded-v11-pad.pdf}}%
    \put(0.89286709,0.6046652){\color[rgb]{0.12941176,0.61568627,0}\makebox(0,0)[rt]{\lineheight{1.25}\smash{\begin{tabular}[t]{r}$\lV$\end{tabular}}}}%
    \put(0.41414453,0.67773637){\color[rgb]{0.12941176,0.61568627,0}\makebox(0,0)[rt]{\lineheight{1.25}\smash{\begin{tabular}[t]{r}$\lV$\end{tabular}}}}%
    \put(0.67116957,0.60477207){\color[rgb]{0.12941176,0.61568627,0}\makebox(0,0)[rt]{\lineheight{1.25}\smash{\begin{tabular}[t]{r}$\lV^m$\end{tabular}}}}%
    \put(0.5302828,0.02320021){\makebox(0,0)[t]{\lineheight{1.25}\smash{\begin{tabular}[t]{c}(b) Third bounce ($\lV$) and\\\citet{royo2023virtual} ($\lV^m$)\end{tabular}}}}%
    \put(0,0){\includegraphics[width=\unitlength,page=11]{detect-occluded-v11-pad.pdf}}%
    \put(0.08246253,0.27677044){\makebox(0,0)[rt]{\lineheight{1.25}\smash{\begin{tabular}[t]{r}$\lRright$\end{tabular}}}}%
    \put(0.13107756,0.30522617){\makebox(0,0)[rt]{\lineheight{1.25}\smash{\begin{tabular}[t]{r}$\lRone$\end{tabular}}}}%
    \put(0.16562619,0.15865588){\color[rgb]{1,0.99607843,0.99607843}\makebox(0,0)[lt]{\lineheight{1.25}\smash{\begin{tabular}[t]{l}$\lRleft$\end{tabular}}}}%
    \put(0,0){\includegraphics[width=\unitlength,page=12]{detect-occluded-v11-pad.pdf}}%
    \put(0.09044639,0.12292182){\color[rgb]{0,0.00392157,0}\makebox(0,0)[t]{\lineheight{1.25}\smash{\begin{tabular}[t]{c}Target\end{tabular}}}}%
    \put(0.21545112,0.27750457){\color[rgb]{1,0.14901961,0.14901961}\makebox(0,0)[t]{\lineheight{1.25}\smash{\begin{tabular}[t]{c}$\lL$\end{tabular}}}}%
    \put(0.27675705,0.308972){\color[rgb]{0,0.4,1}\makebox(0,0)[t]{\lineheight{1.25}\smash{\begin{tabular}[t]{c}$\lS$\end{tabular}}}}%
  \end{picture}%
\endgroup%

%% file: fig/resim-v6.pdf_tex
\begingroup%
  \makeatletter%
  \providecommand\color[2][]{%
    \errmessage{(Inkscape) Color is used for the text in Inkscape, but the package 'color.sty' is not loaded}%
    \renewcommand\color[2][]{}%
  }%
  \providecommand\transparent[1]{%
    \errmessage{(Inkscape) Transparency is used (non-zero) for the text in Inkscape, but the package 'transparent.sty' is not loaded}%
    \renewcommand\transparent[1]{}%
  }%
  \providecommand\rotatebox[2]{#2}%
  \newcommand*\fsize{\dimexpr\f@size pt\relax}%
  \newcommand*\lineheight[1]{\fontsize{\fsize}{#1\fsize}\selectfont}%
  \ifx\svgwidth\undefined%
    \setlength{\unitlength}{1217.14025783bp}%
    \ifx\svgscale\undefined%
      \relax%
    \else%
      \setlength{\unitlength}{\unitlength * \real{\svgscale}}%
    \fi%
  \else%
    \setlength{\unitlength}{\svgwidth}%
  \fi%
  \global\let\svgwidth\undefined%
  \global\let\svgscale\undefined%
  \makeatother%
  \begin{picture}(1,0.58573116)%
    \lineheight{1}%
    \setlength\tabcolsep{0pt}%
    \put(0,0){\includegraphics[width=\unitlength,page=1]{resim-v6.pdf}}%
    \put(0.01142104,0.44874295){\rotatebox{90}{\makebox(0,0)[t]{\lineheight{1.25}\smash{\begin{tabular}[t]{c}Real capture\\(\fref{fig:detection_planar})\end{tabular}}}}}%
    \put(0.01142104,0.1876592){\rotatebox{90}{\makebox(0,0)[t]{\lineheight{1.25}\smash{\begin{tabular}[t]{c}Recreation\\in simulation\end{tabular}}}}}%
    \put(0.87553587,0.56873212){\makebox(0,0)[t]{\lineheight{1.25}\smash{\begin{tabular}[t]{c}$f_{2,3}(\xv)$ (top view)\end{tabular}}}}%
    \put(0.28257507,0.57172783){\makebox(0,0)[t]{\lineheight{1.25}\smash{\begin{tabular}[t]{c}\final{$\sum_{\,\xl\in\lL}\!H(\xl, \xs, t)$}\end{tabular}}}}%
    \put(0.41677152,0.00439671){\makebox(0,0)[t]{\lineheight{1.25}\smash{\begin{tabular}[t]{c}(a) Captured and virtual impulse responses\end{tabular}}}}%
    \put(0.87548453,0.0076086){\makebox(0,0)[t]{\lineheight{1.25}\smash{\begin{tabular}[t]{c}(b) Final image\end{tabular}}}}%
    \put(0,0){\includegraphics[width=\unitlength,page=2]{resim-v6.pdf}}%
    \put(0.89308598,0.21476115){\color[rgb]{1,1,1}\makebox(0,0)[rt]{\lineheight{1.25}\smash{\begin{tabular}[t]{r}$\lV$\end{tabular}}}}%
    \put(0,0){\includegraphics[width=\unitlength,page=3]{resim-v6.pdf}}%
    \put(0.84307062,0.08222253){\color[rgb]{1,0.14901961,0.14901961}\makebox(0,0)[t]{\lineheight{1.25}\smash{\begin{tabular}[t]{c}$\mathcal{L}_2$\end{tabular}}}}%
    \put(0.90982752,0.08222253){\color[rgb]{0,0.4,1}\makebox(0,0)[t]{\lineheight{1.25}\smash{\begin{tabular}[t]{c}$\mathcal{S}_3$\end{tabular}}}}%
    \put(0,0){\includegraphics[width=\unitlength,page=4]{resim-v6.pdf}}%
    \put(0.89395477,0.47464699){\color[rgb]{1,1,1}\makebox(0,0)[rt]{\lineheight{1.25}\smash{\begin{tabular}[t]{r}$\lV$\end{tabular}}}}%
    \put(0,0){\includegraphics[width=\unitlength,page=5]{resim-v6.pdf}}%
    \put(0.84393941,0.34210836){\color[rgb]{1,0.14901961,0.14901961}\makebox(0,0)[t]{\lineheight{1.25}\smash{\begin{tabular}[t]{c}$\mathcal{L}_2$\end{tabular}}}}%
    \put(0.91069631,0.34210836){\color[rgb]{0,0.4,1}\makebox(0,0)[t]{\lineheight{1.25}\smash{\begin{tabular}[t]{c}$\mathcal{S}_3$\end{tabular}}}}%
    \put(0,0){\includegraphics[width=\unitlength,page=6]{resim-v6.pdf}}%
    \put(0.45464425,0.34322562){\color[rgb]{0,0,0}\makebox(0,0)[lt]{\lineheight{1.25}\smash{\begin{tabular}[t]{l}$t$\end{tabular}}}}%
    \put(0.45445981,0.08213477){\color[rgb]{0,0,0}\makebox(0,0)[lt]{\lineheight{1.25}\smash{\begin{tabular}[t]{l}$t$\end{tabular}}}}%
    \put(0,0){\includegraphics[width=\unitlength,page=7]{resim-v6.pdf}}%
    \put(0.17318676,0.37781751){\color[rgb]{0.14901961,0.14901961,0.14901961}\rotatebox{90}{\makebox(0,0)[lt]{\lineheight{1.25}\smash{\begin{tabular}[t]{l}{\scriptsize Occ.}\end{tabular}}}}}%
    \put(0.3155416,0.53994135){\color[rgb]{0.14901961,0.14901961,0.14901961}\rotatebox{90}{\makebox(0,0)[rt]{\lineheight{1.25}\smash{\begin{tabular}[t]{r}{\scriptsize $\lRtwo$/$\lRthree$}\end{tabular}}}}}%
    \put(0.40104546,0.5375788){\color[rgb]{0.14901961,0.14901961,0.14901961}\rotatebox{90}{\makebox(0,0)[rt]{\lineheight{1.25}\smash{\begin{tabular}[t]{r}{\scriptsize Target}\end{tabular}}}}}%
    \put(0.17318678,0.11672644){\color[rgb]{0.14901961,0.14901961,0.14901961}\rotatebox{90}{\makebox(0,0)[lt]{\lineheight{1.25}\smash{\begin{tabular}[t]{l}{\scriptsize Occ.}\end{tabular}}}}}%
    \put(0.31554163,0.27885027){\color[rgb]{0.14901961,0.14901961,0.14901961}\rotatebox{90}{\makebox(0,0)[rt]{\lineheight{1.25}\smash{\begin{tabular}[t]{r}{\scriptsize $\lRtwo$/$\lRthree$}\end{tabular}}}}}%
    \put(0.40104549,0.2764877){\color[rgb]{0.14901961,0.14901961,0.14901961}\rotatebox{90}{\makebox(0,0)[rt]{\lineheight{1.25}\smash{\begin{tabular}[t]{r}{\scriptsize Target}\end{tabular}}}}}%
    \put(0.86687904,0.42121591){\color[rgb]{0.96862745,0.96862745,0.96862745}\makebox(0,0)[t]{\lineheight{1.25}\smash{\begin{tabular}[t]{c}Occ.\end{tabular}}}}%
    \put(0.86687904,0.16133007){\color[rgb]{0.96862745,0.96862745,0.96862745}\makebox(0,0)[t]{\lineheight{1.25}\smash{\begin{tabular}[t]{c}Occ.\end{tabular}}}}%
    \put(0,0){\includegraphics[width=\unitlength,page=8]{resim-v6.pdf}}%
    \put(0.61069626,0.57113174){\makebox(0,0)[t]{\lineheight{1.25}\smash{\begin{tabular}[t]{c}\final{$\sum_{\,\mathbf{l}_2\in\mathcal{L}_2}\!\Hp_{2,3}\!(\mathbf{l}_2, \mathbf{s}_3, t)$}\end{tabular}}}}%
    \put(0,0){\includegraphics[width=\unitlength,page=9]{resim-v6.pdf}}%
    \put(0.72822655,0.34262953){\color[rgb]{0,0,0}\makebox(0,0)[lt]{\lineheight{1.25}\smash{\begin{tabular}[t]{l}$t$\end{tabular}}}}%
    \put(0,0){\includegraphics[width=\unitlength,page=10]{resim-v6.pdf}}%
    \put(0.54108108,0.3784538){\color[rgb]{0.14901961,0.14901961,0.14901961}\rotatebox{90}{\makebox(0,0)[lt]{\lineheight{1.25}\smash{\begin{tabular}[t]{l}{\scriptsize $\lRtwo$/$\lRthree$}\end{tabular}}}}}%
    \put(0.5834832,0.3784538){\color[rgb]{0.14901961,0.14901961,0.14901961}\rotatebox{90}{\makebox(0,0)[lt]{\lineheight{1.25}\smash{\begin{tabular}[t]{l}{\scriptsize Target}\end{tabular}}}}}%
    \put(0,0){\includegraphics[width=\unitlength,page=11]{resim-v6.pdf}}%
    \put(0.72822655,0.08189198){\color[rgb]{0,0,0}\makebox(0,0)[lt]{\lineheight{1.25}\smash{\begin{tabular}[t]{l}$t$\end{tabular}}}}%
    \put(0,0){\includegraphics[width=\unitlength,page=12]{resim-v6.pdf}}%
    \put(0.54106574,0.11788751){\color[rgb]{0.14901961,0.14901961,0.14901961}\rotatebox{90}{\makebox(0,0)[lt]{\lineheight{1.25}\smash{\begin{tabular}[t]{l}{\scriptsize $\lRtwo$/$\lRthree$}\end{tabular}}}}}%
    \put(0.58360423,0.11772414){\color[rgb]{0.14901961,0.14901961,0.14901961}\rotatebox{90}{\makebox(0,0)[lt]{\lineheight{1.25}\smash{\begin{tabular}[t]{l}{\scriptsize Target}\end{tabular}}}}}%
    \put(0,0){\includegraphics[width=\unitlength,page=13]{resim-v6.pdf}}%
    \put(0.10230011,0.37044038){\rotatebox{90}{\makebox(0,0)[t]{\lineheight{1.25}\smash{\begin{tabular}[t]{c}0\end{tabular}}}}}%
    \put(0.1229922,0.31460099){\makebox(0,0)[t]{\lineheight{1.25}\smash{\begin{tabular}[t]{c}0\end{tabular}}}}%
    \put(0.24475736,0.31460099){\makebox(0,0)[t]{\lineheight{1.25}\smash{\begin{tabular}[t]{c}$10\;\text{ns}$\end{tabular}}}}%
    \put(0.36657754,0.31460099){\makebox(0,0)[t]{\lineheight{1.25}\smash{\begin{tabular}[t]{c}$20\;\text{ns}$\end{tabular}}}}%
    \put(0.58593478,0.31460099){\makebox(0,0)[t]{\lineheight{1.25}\smash{\begin{tabular}[t]{c}$7\;\text{ns}$\end{tabular}}}}%
    \put(0.1229922,0.05350989){\makebox(0,0)[t]{\lineheight{1.25}\smash{\begin{tabular}[t]{c}0\end{tabular}}}}%
    \put(0.24475736,0.05350989){\makebox(0,0)[t]{\lineheight{1.25}\smash{\begin{tabular}[t]{c}$10\;\text{ns}$\end{tabular}}}}%
    \put(0.36657754,0.05350989){\makebox(0,0)[t]{\lineheight{1.25}\smash{\begin{tabular}[t]{c}$20\;\text{ns}$\end{tabular}}}}%
    \put(0.10173546,0.51110992){\rotatebox{90}{\makebox(0,0)[t]{\lineheight{1.25}\smash{\begin{tabular}[t]{c}15,000\end{tabular}}}}}%
    \put(0.10215332,0.10975413){\rotatebox{90}{\makebox(0,0)[t]{\lineheight{1.25}\smash{\begin{tabular}[t]{c}0\end{tabular}}}}}%
    \put(0.10158867,0.25042367){\rotatebox{90}{\makebox(0,0)[t]{\lineheight{1.25}\smash{\begin{tabular}[t]{c}15,000\end{tabular}}}}}%
    \put(0.58593404,0.05350989){\makebox(0,0)[t]{\lineheight{1.25}\smash{\begin{tabular}[t]{c}$7\;\text{ns}$\end{tabular}}}}%
  \end{picture}%
\endgroup%

%% file: fig/rough-v7.pdf_tex
\begingroup%
  \makeatletter%
  \providecommand\color[2][]{%
    \errmessage{(Inkscape) Color is used for the text in Inkscape, but the package 'color.sty' is not loaded}%
    \renewcommand\color[2][]{}%
  }%
  \providecommand\transparent[1]{%
    \errmessage{(Inkscape) Transparency is used (non-zero) for the text in Inkscape, but the package 'transparent.sty' is not loaded}%
    \renewcommand\transparent[1]{}%
  }%
  \providecommand\rotatebox[2]{#2}%
  \newcommand*\fsize{\dimexpr\f@size pt\relax}%
  \newcommand*\lineheight[1]{\fontsize{\fsize}{#1\fsize}\selectfont}%
  \ifx\svgwidth\undefined%
    \setlength{\unitlength}{1454.83396515bp}%
    \ifx\svgscale\undefined%
      \relax%
    \else%
      \setlength{\unitlength}{\unitlength * \real{\svgscale}}%
    \fi%
  \else%
    \setlength{\unitlength}{\svgwidth}%
  \fi%
  \global\let\svgwidth\undefined%
  \global\let\svgscale\undefined%
  \makeatother%
  \begin{picture}(1,0.79554892)%
    \lineheight{1}%
    \setlength\tabcolsep{0pt}%
    \put(0,0){\includegraphics[width=\unitlength,page=1]{rough-v7.pdf}}%
    \put(0.92798428,0.20802612){\color[rgb]{0,0.4,1}\makebox(0,0)[lt]{\lineheight{1.25}\smash{\begin{tabular}[t]{l}$\mathcal{S}_2$\end{tabular}}}}%
    \put(0,0){\includegraphics[width=\unitlength,page=2]{rough-v7.pdf}}%
    \put(0.3617152,0.21725397){\color[rgb]{1,0.14901961,0.14901961}\makebox(0,0)[rt]{\lineheight{1.25}\smash{\begin{tabular}[t]{r}$\lL$\end{tabular}}}}%
    \put(0.36171521,0.16297135){\color[rgb]{0,0.4,1}\makebox(0,0)[rt]{\lineheight{1.25}\smash{\begin{tabular}[t]{r}$\lS$\end{tabular}}}}%
    \put(0.55949364,0.1143417){\color[rgb]{1,1,1}\makebox(0,0)[rt]{\lineheight{1.25}\smash{\begin{tabular}[t]{r}$\lV$\end{tabular}}}}%
    \put(0.61908606,0.16053299){\color[rgb]{1,1,1}\makebox(0,0)[rt]{\lineheight{1.25}\smash{\begin{tabular}[t]{r}$\lV^m$\end{tabular}}}}%
    \put(0.51256695,0.24253316){\color[rgb]{1,1,1}\makebox(0,0)[t]{\lineheight{1.25}\smash{\begin{tabular}[t]{c}$\lRtwo$\end{tabular}}}}%
    \put(0,0){\includegraphics[width=\unitlength,page=3]{rough-v7.pdf}}%
    \put(0.13186936,0.37016085){\makebox(0,0)[t]{\lineheight{1.25}\smash{\begin{tabular}[t]{c}Real setup (3D scan)\end{tabular}}}}%
    \put(0.49174126,0.3762999){\makebox(0,0)[t]{\lineheight{1.25}\smash{\begin{tabular}[t]{c}$f_{1,1}(\xv)$ (top view)\end{tabular}}}}%
    \put(0.1554676,0.42158557){\makebox(0,0)[t]{\lineheight{1.25}\smash{\begin{tabular}[t]{c}Target object\end{tabular}}}}%
    \put(0.49196884,0.77765584){\makebox(0,0)[t]{\lineheight{1.25}\smash{\begin{tabular}[t]{c}Projector $\lL$\\Camera $\lS$\end{tabular}}}}%
    \put(0.8673637,0.77765584){\makebox(0,0)[t]{\lineheight{1.25}\smash{\begin{tabular}[t]{c}Projector $\mathcal{L}_2$\\Camera $\mathcal{S}_2$\end{tabular}}}}%
    \put(0,0){\includegraphics[width=\unitlength,page=4]{rough-v7.pdf}}%
    \put(0.17561417,0.14195243){\makebox(0,0)[lt]{\smash{\begin{tabular}[t]{l}\emph{Rough}\\$\lRtwo$\end{tabular}}}}%
    \put(0.15090121,0.31646125){\makebox(0,0)[lt]{\lineheight{1.25}\smash{\begin{tabular}[t]{l}$\lRone$\end{tabular}}}}%
    \put(0.10318661,0.0979656){\color[rgb]{0,0.00392157,0}\makebox(0,0)[t]{\lineheight{1.25}\smash{\begin{tabular}[t]{c}Target\end{tabular}}}}%
    \put(0.02756098,0.3118573){\color[rgb]{0,0.4,1}\makebox(0,0)[t]{\lineheight{1.25}\smash{\begin{tabular}[t]{c}$\lS$\end{tabular}}}}%
    \put(0,0){\includegraphics[width=\unitlength,page=5]{rough-v7.pdf}}%
    \put(0.06985294,0.26415889){\color[rgb]{1,0.14901961,0.14901961}\makebox(0,0)[t]{\lineheight{1.25}\smash{\begin{tabular}[t]{c}$\lL$\end{tabular}}}}%
    \put(0.13219682,0.02708601){\makebox(0,0)[t]{\lineheight{1.25}\smash{\begin{tabular}[t]{c}(a) Setup overview\end{tabular}}}}%
    \put(0.49224111,0.02505454){\makebox(0,0)[t]{\lineheight{1.25}\smash{\begin{tabular}[t]{c}(b) Third bounce ($\lV$) and\\\citet{royo2023virtual} ($\lV^m$)\end{tabular}}}}%
    \put(0.86747509,0.02610502){\makebox(0,0)[t]{\lineheight{1.25}\smash{\begin{tabular}[t]{c}(c) Ours\end{tabular}}}}%
    \put(0,0){\includegraphics[width=\unitlength,page=6]{rough-v7.pdf}}%
    \put(0.35925671,0.65273438){\color[rgb]{1,0.14901961,0.14901961}\makebox(0,0)[t]{\lineheight{1.25}\smash{\begin{tabular}[t]{c}$\mathcal{L}_1$\end{tabular}}}}%
    \put(0,0){\includegraphics[width=\unitlength,page=7]{rough-v7.pdf}}%
    \put(0.35270898,0.48478428){\color[rgb]{0,0.4,1}\makebox(0,0)[t]{\lineheight{1.25}\smash{\begin{tabular}[t]{c}$\mathcal{S}_1$\end{tabular}}}}%
    \put(0,0){\includegraphics[width=\unitlength,page=8]{rough-v7.pdf}}%
    \put(0.92046925,0.62382395){\color[rgb]{1,0.14901961,0.14901961}\makebox(0,0)[t]{\lineheight{1.25}\smash{\begin{tabular}[t]{c}$\mathcal{L}_2$\end{tabular}}}}%
    \put(0.95321938,0.59328424){\color[rgb]{0,0.4,1}\makebox(0,0)[t]{\lineheight{1.25}\smash{\begin{tabular}[t]{c}$\mathcal{S}_2$\end{tabular}}}}%
    \put(0.86705738,0.3762999){\makebox(0,0)[t]{\lineheight{1.25}\smash{\begin{tabular}[t]{c}$f_{2,2}(\xv)$ (top view)\end{tabular}}}}%
    \put(0.03678999,0.49608015){\makebox(0,0)[t]{\lineheight{1.25}\smash{\begin{tabular}[t]{c}$\lRone$\end{tabular}}}}%
    \put(0.24644667,0.62666681){\makebox(0,0)[rt]{\lineheight{1.25}\smash{\begin{tabular}[t]{r}\emph{Rough}\\$\lRtwo$\end{tabular}}}}%
    \put(0.57228026,0.4474068){\color[rgb]{0.12941176,0.61568627,0}\makebox(0,0)[rt]{\lineheight{1.25}\smash{\begin{tabular}[t]{r}$\lV$\end{tabular}}}}%
    \put(0.97981752,0.45161171){\color[rgb]{0.12941176,0.61568627,0}\makebox(0,0)[rt]{\lineheight{1.25}\smash{\begin{tabular}[t]{r}$\lV$\end{tabular}}}}%
    \put(0.62239096,0.52579054){\color[rgb]{0.12941176,0.61568627,0}\makebox(0,0)[rt]{\lineheight{1.25}\smash{\begin{tabular}[t]{r}$\lV^m$\end{tabular}}}}%
    \put(0,0){\includegraphics[width=\unitlength,page=9]{rough-v7.pdf}}%
    \put(0.83510841,0.11453791){\color[rgb]{1,1,1}\makebox(0,0)[rt]{\lineheight{1.25}\smash{\begin{tabular}[t]{r}$\lV$\end{tabular}}}}%
    \put(0.92798429,0.24767189){\color[rgb]{1,0.14901961,0.14901961}\makebox(0,0)[lt]{\lineheight{1.25}\smash{\begin{tabular}[t]{l}$\mathcal{L}_2$\end{tabular}}}}%
    \put(0,0){\includegraphics[width=\unitlength,page=10]{rough-v7.pdf}}%
  \end{picture}%
\endgroup%

%% file: journal_modified_tex/80_conclusions_discussion.tex
\section{Conclusions and discussion}
\label{sec:discussion:hidden-walls-as-secondary-relay-walls}
\label{sec:discussion:reconstruction-limits}

Our novel  cascaded NLOS imaging methods allow us to create and concatenate secondary virtual imaging systems. This enables new NLOS imaging beyond the three-bounce assumption used by conventional methods, leveraging fourth- and fifth-bounce illumination. We have provided results in simulation and with a hardware prototype, showcasing the potential of our approach in several scenarios.

We have shown results cascading two imaging systems although, in theory, multiple cascading NLOS imaging systems could be concatenated. We could compute another virtual impulse response $\Hpp_{a,b}(\xlpp, \xspp, t)$ on points $\xlpp, \xspp$ on a \emph{tertiary} relay wall, using \emph{fifth-} to \emph{seventh-bounce} illumination, expanding even further the possibilities of future NLOS imaging systems. 
However, each concatenation imposes a resolution loss due to the reduced signal, limited aperture sizes, longer focus distances, and increased multi-path interference. 

\begin{figure}[t]
    \centering
    \captionsetup{skip=-5pt}
    \def\svgwidth{\columnwidth}
    \begin{small}
        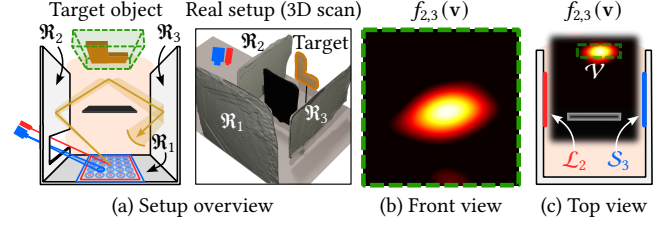
    \end{small}
    \caption{\final{(a) We repeat the setup from \fref{fig:detection_planar} using an L-shaped target. (b) Front view of our imaging result using our real prototype ($16\times16$ SPAD covering a $50\;\text{cm}\times32.5\;\text{cm}$ aperture). (c) Top view of the same result.}}
    \label{fig:image_L}
\end{figure}

In particular, our current prototype uses a $16\times16$ SPAD array \cite{riccardoFastGated16162022} covering a $\qtyproduct{50 x 32.5}{cm}$ area.
\final{This results in a spatial resolution $\Delta x$ for the virtual impulse response of about $20$--$40\,\text{cm}$ depending on the setup (see Appendix B for details on the derivation of both spatial and temporal resolution), which in turn limits the features that can be effectively resolved. This is shown in \fref{fig:image_L}, where we have repeated the same setup from \fref{fig:detection_planar}, this time on an asymmetric, L-shaped target (\fref{fig:image_L}a). We use the imaging procedure explained in \sref{sec:real-experiments:detection}, with filtering parameters $\lambda_c=\sigma=\SI{20}{cm}$ for $H$ and $\Hp_{2,3}$. \fref{fig:image_L}b shows a front view of $f_{2,3}(\xv)$, clearly imaging the target object in the right location, but failing to resolve its asymmetric features. \fref{fig:image_L}c displays a top view of $f_{2,3}(\xv)$, showing how the target is also imaged at the correct depth. Despite the current resolution limitations, note that existing NLOS methods fail to image the target, as we have shown in this paper.}

\final{As Figure \ref{fig:fourthbounce-degradation} illustrates, as larger SPADs become available, the resolution of our cascaded systems will increase.}
From our analysis in Appendix B, a future $128\times128$ SPAD array will theoretically yield a much higher resolution of around \qtyrange[range-phrase=--,range-units=single]{1}{2}{cm}, without changes in our algorithm.
Further research on wave scattering properties may also help; for instance, recent research has found that a photon's Fisher information does not decrease with increasing scattering \cite{hupfl2024continuity}. 
Another interesting path for further analysis is the use of wavefront shaping \cite{cao2022high} to potentially leverage diffuse paths in smoother secondary surfaces without centimeter-scale features.
Last, as more modern laser and sensor technology becomes available, we foresee a progressive increase in imaging capabilities enabled by our cascaded framework.

%% file: fig/image-L-v4.pdf_tex
\begingroup%
  \makeatletter%
  \providecommand\color[2][]{%
    \errmessage{(Inkscape) Color is used for the text in Inkscape, but the package 'color.sty' is not loaded}%
    \renewcommand\color[2][]{}%
  }%
  \providecommand\transparent[1]{%
    \errmessage{(Inkscape) Transparency is used (non-zero) for the text in Inkscape, but the package 'transparent.sty' is not loaded}%
    \renewcommand\transparent[1]{}%
  }%
  \providecommand\rotatebox[2]{#2}%
  \newcommand*\fsize{\dimexpr\f@size pt\relax}%
  \newcommand*\lineheight[1]{\fontsize{\fsize}{#1\fsize}\selectfont}%
  \ifx\svgwidth\undefined%
    \setlength{\unitlength}{1183.79874162bp}%
    \ifx\svgscale\undefined%
      \relax%
    \else%
      \setlength{\unitlength}{\unitlength * \real{\svgscale}}%
    \fi%
  \else%
    \setlength{\unitlength}{\svgwidth}%
  \fi%
  \global\let\svgwidth\undefined%
  \global\let\svgscale\undefined%
  \makeatother%
  \begin{picture}(1,0.32204703)%
    \lineheight{1}%
    \setlength\tabcolsep{0pt}%
    \put(0,0){\includegraphics[width=\unitlength,page=1]{image-L-v4.pdf}}%
    \put(0.49916033,0.14604008){\color[rgb]{1,1,1}\makebox(0,0)[rt]{\lineheight{1.25}\smash{\begin{tabular}[t]{r}$\lRright$\end{tabular}}}}%
    \put(0.36314573,0.25727983){\makebox(0,0)[lt]{\lineheight{1.25}\smash{\begin{tabular}[t]{l}$\lRleft$\end{tabular}}}}%
    \put(0.35010883,0.13282481){\color[rgb]{1,1,1}\makebox(0,0)[t]{\lineheight{1.25}\smash{\begin{tabular}[t]{c}$\lRone$\end{tabular}}}}%
    \put(0,0){\includegraphics[width=\unitlength,page=2]{image-L-v4.pdf}}%
    \put(0.48608392,0.25456413){\color[rgb]{0,0,0}\makebox(0,0)[t]{\lineheight{1.25}\smash{\begin{tabular}[t]{c}Target\end{tabular}}}}%
    \put(0,0){\includegraphics[width=\unitlength,page=3]{image-L-v4.pdf}}%
    \put(0.28773997,0.00332698){\makebox(0,0)[t]{\lineheight{1.25}\smash{\begin{tabular}[t]{c}(a) Setup overview\end{tabular}}}}%
    \put(0.4131754,0.30565162){\makebox(0,0)[t]{\lineheight{1.25}\smash{\begin{tabular}[t]{c}Real setup (3D scan)\end{tabular}}}}%
    \put(0,0){\includegraphics[width=\unitlength,page=4]{image-L-v4.pdf}}%
    \put(0.15339954,0.30565164){\color[rgb]{0,0.00392157,0}\makebox(0,0)[t]{\smash{\begin{tabular}[t]{c}Target object\end{tabular}}}}%
    \put(0,0){\includegraphics[width=\unitlength,page=5]{image-L-v4.pdf}}%
    \put(0.67202003,0.00332698){\makebox(0,0)[t]{\lineheight{1.25}\smash{\begin{tabular}[t]{c}(b) Front view\end{tabular}}}}%
    \put(0.90909462,0.00332698){\makebox(0,0)[t]{\lineheight{1.25}\smash{\begin{tabular}[t]{c}(c) Top view\end{tabular}}}}%
    \put(0,0){\includegraphics[width=\unitlength,page=6]{image-L-v4.pdf}}%
    \put(0.67231677,0.30565162){\makebox(0,0)[t]{\lineheight{1.25}\smash{\begin{tabular}[t]{c}$f_{2,3}(\xv)$\end{tabular}}}}%
    \put(0.90920784,0.30565162){\makebox(0,0)[t]{\lineheight{1.25}\smash{\begin{tabular}[t]{c}$f_{2,3}(\xv)$\end{tabular}}}}%
    \put(0,0){\includegraphics[width=\unitlength,page=7]{image-L-v4.pdf}}%
    \put(0.92677304,0.20970822){\color[rgb]{1,1,1}\makebox(0,0)[rt]{\lineheight{1.25}\smash{\begin{tabular}[t]{r}$\lV$\end{tabular}}}}%
    \put(0,0){\includegraphics[width=\unitlength,page=8]{image-L-v4.pdf}}%
    \put(0.8782383,0.07344856){\color[rgb]{1,0.14901961,0.14901961}\makebox(0,0)[t]{\lineheight{1.25}\smash{\begin{tabular}[t]{c}$\mathcal{L}_2$\end{tabular}}}}%
    \put(0.9468754,0.07344856){\color[rgb]{0,0.4,1}\makebox(0,0)[t]{\lineheight{1.25}\smash{\begin{tabular}[t]{c}$\mathcal{S}_3$\end{tabular}}}}%
    \put(0,0){\includegraphics[width=\unitlength,page=9]{image-L-v4.pdf}}%
    \put(0.24351789,0.10002187){\makebox(0,0)[t]{\lineheight{1.25}\smash{\begin{tabular}[t]{c}$\lRone$\end{tabular}}}}%
    \put(0.23316679,0.27047949){\makebox(0,0)[lt]{\lineheight{1.25}\smash{\begin{tabular}[t]{l}$\lRright$\end{tabular}}}}%
    \put(0.08867005,0.27016731){\makebox(0,0)[rt]{\lineheight{1.25}\smash{\begin{tabular}[t]{r}$\lRleft$\end{tabular}}}}%
    \put(0,0){\includegraphics[width=\unitlength,page=10]{image-L-v4.pdf}}%
  \end{picture}%
\endgroup%

%% file: tex/90_acks.tex
This work has been supported by grants PID2022-141539NB-I00 and PID2025-169453NB-I00, funded by  MICIU/AEI/10.13039/501100011033 and by ERDF, EU; and by grant FA9550-26-1-B169, funded by the Air Force Office of Scientific Research. María Peña was supported by the FPU23/03646 predoctoral grant and Diego Royo was supported by the Government of Aragon CUS/803/2021 predoctoral grant. This material is partially based upon work supported by DE-NA0004196 and through the Enabling Technologies \& Innovation Graduate Fellowship, funded by the Department of Energy/National Nuclear Security Administration. The authors would like to thank Talha Sultan and the members of the Graphics and Imaging Lab for useful discussions. 

%% file: tex/A6_optimizations.tex
\section{Fast computation of our virtual $H'$ using convolutions}
\label{sec:cascaded:focus}

The computation of $\Hp_{a,b}(\xlp, \xsp, t)$ (Equation 11) 
could require heavy computation if implemented naively, due to the large dimensionality of the impulse response $H(\xl, \xs, t)$. Instead, we formulate both $\lL$ and $\lS$ propagations simultaneously with convolutions \cite{Liu2020phasor}. 
This allows us to efficiently compute $\Hp_{a,b}(\xlp, \xsp, t)$ for all pairs of points $\xlp$, $\xsp$ in parallel.

We consider the XYZ coordinates of points on the relay wall $\lRone$, noted as $\xl = [l_x, l_y, 0]$ and $\xs = [s_x, s_y, 0]$, on a plane at depth $z = 0$. The corresponding points on the secondary relay walls, $\xlp = [l_x^\prime, l_y^\prime, l_z^\prime]$ on $\lR_a$ and $\xsp = [s_x^\prime, s_y^\prime, s_z^\prime]$ on $\lR_b$, are both on planes at depths $l_z^\prime$ and $s_z^\prime$. 
With this, we define a 4D convolution kernel $\Fspace{C}$ using RSD operators $\Rf$ from Equation 5:
\begin{equation}
\begin{aligned}
    &\Fspace{C}(d_{lx}, d_{ly}, d_{sx}, d_{sy}; l_z', s_z', \fq) \\
    &\quad \equiv \Rf\left( \sqrt{d_{lx}^2 \!+ d_{ly}^2 \!+\! (l_z')^2}/c, \fq \right)\Rf \left( \sqrt{d_{sx}^2 \!+ d_{sy}^2 \!+\! (s_z')^2}/c, \fq \right) , 
\end{aligned}
\end{equation}
with $d_{lx} = l_x^\prime - l_x$, $d_{ly} = l_y^\prime - l_y$, $d_{sx} = s_x^\prime - s_x$, and $d_{sy} = s_y^\prime - s_y$ representing distances in the $XY$ dimensions,
and parameterized by the plane depths $l_z'$, $s_z'$ and imaging frequency $\fq$.
This allows us to express the frequency-domain counterpart of Equation 11 combining both 2D propagations into a 4D spatial convolution as:
\begin{equation}
\begin{aligned}
    &\Hpf_{a,b} (l_x', l_y', l_z', s_x', s_y', s_z', \fq) \\
    &\quad= \Ifcam \left(s_x^\prime, s_y^\prime, s_z^\prime, \fq; \Ifproj \left(l_x^\prime, l_y^\prime, l_z^\prime, \fq; \Hfilterf \right)\right) \\
    &\quad= \iiiint_{-\infty}^{+\infty} \Fspace{C}\left(l_x' - l_x, l_y' - l_y, s_x' - s_x, s_y' - s_y; l_z', s_z', \fq\right) \\
    &\quad\qquad\qquad\qquad\; \Hfilterf(l_x, l_y, 0, s_x, s_y, 0, \fq)
    \, \mathrm{d}l_x\, \mathrm{d}l_y\, \mathrm{d}s_x\, \mathrm{d}s_y \\
    &\quad= \Fspace{C}(l_x, l_y, s_x, s_y; l_z', s_z', \fq) \convs \Hfilterf(l_x, l_y, 0, s_x, s_y, 0, \fq),
\end{aligned}
\end{equation}
where $\convs$ denotes a 4D spatial convolution over the $XY$ dimensions of both the camera $\lS$ and illumination $\lL$ apertures $l_x, l_y, s_x, s_y$, which becomes a multiplication in the Fourier domain.

Note that this propagation approach also works if the two walls are not coplanar. Assuming for instance that two walls $\lRone$ and $\lRtwo$ are perpendicular (as in Figure 4), we can create a series of virtual walls $\lRtwo{}_i$ parallel to $\lRone$, where each one includes a column of points of $\lRtwo$. We then apply our propagation over multiple $\lRtwo{}_i$, discarding points that do not belong to $\lRtwo$.

%% file: tex/A1_resolution_limits.tex
\section{Resolution limits in cascaded NLOS imaging}
\label{sec:resolution-limits}

We analyze the resolution of our cascaded imaging systems, extending prior work on third-bounce NLOS imaging systems \cite{buttafava2015non, OToole2018confocal, Liu2019phasor}. We split our analysis in two parts. First, in \sref{sec:resolution-limits:virtual-impulse-response}, we reason about the spatial and temporal resolution of our virtual impulse response $H'$. Second, in \sref{sec:resolution-limits:cascaded}, we analyze how the resolution of our virtual impulse response affects the imaging resolution of secondary virtual imaging systems and thus of the entire cascaded imaging system.

\subsection{Resolution of our virtual impulse response}
\label{sec:resolution-limits:virtual-impulse-response}

First, the primary relay wall acts as a virtual imaging system that reconstructs the virtual impulse response $H'$ at a secondary relay wall (Equation 11). We model the resulting imaging resolution of the primary relay wall using the Rayleigh criterion, akin to conventional third-bounce methods.

\paragraph{Spatial and temporal resolution.} The Rayleigh criterion relates the angular resolution $\theta$ of an imaging system to its aperture size $D$ and the imaging wavelength $\lambda$:
\begin{equation}
    \theta = k \frac{\lambda}{D},
    \label{eq:rayleigh-criterion-angle}
\end{equation}
where $k$ is a constant that depends on the shape of the aperture (e.g., for a circular aperture $k \approx 1.22$). We can use a small-angle approximation ($\sin \theta \approx \theta$) to establish a bound for the lateral resolution  $\Delta x$ of our imaging system at a given depth $z$ from the relay wall:%
\begin{equation}
    \Delta x \approx k \frac{\lambda z}{D}.
    \label{eq:rayleigh-criterion}
\end{equation}
For our virtual imaging systems, apertures are typically square-shaped with a side length of $D$, where $k = 1 / \sqrt{2}$ has empirically provided a good estimate of the resolution.

We use $\Delta x$ from \eref{eq:rayleigh-criterion} to model the spatial resolution of our virtual impulse response $H'(\xlp, \xsp, t)$. Thus, when reconstructing $H'$ at several points $\xlp$ or $\xsp$ these can be separated by a distance of $\Delta x$.
Empirically, we found that the temporal resolution of $H'$ can be modeled as $\Delta t = \Delta x / c$, with $c$ the speed of light. Later, we show an experiment which validates this last claim.

\paragraph{Scanning spacing constraints.} The spacing of points $\xs$ at the relay walls sets a theoretical minimum for imaging wavelengths $\lambda$. By the Nyquist-Shannon sampling theorem, to reproduce a phasor field $\Hf(\xl, \xs, \fq)$ with wavelength $\lambda = c / \fq$ over a sparsely sampled aperture (here $\xs \in \lS$ on the relay wall), points $\xs$ must be spaced by a distance $\Delta \xs \leq \lambda/2$. For the phasor-field framework, the scanning resolution effectively provides a lower bound for $\lambda_c$ in the filter $K(t; \lambda_c, \sigma)$ (Equation 3). In practice, $\lambda$ typically exceeds this limit to avoid spatial aliasing of the virtual phasor waves.
For the cases where we also focus the illumination aperture $\lL$ (e.g., Figures~5d,~13~and~15), the same analysis applies to points $\xl \in \lL$ and their spacing $\Delta \xl$.

\paragraph{Resolution constraints using our real prototype.} Our current prototype relies on a $16\times 16$ SPAD array for its illumination aperture $\lL$. Due to physical constraints in our laboratory, our SPAD array only covers a $\SI{50}{cm}\times\SI{32.5}{cm}$ area of the relay wall (i.e., $D = 32.5\,\text{cm}$). Given that imaging with higher-order illumination requires longer imaging wavelengths ($\lambda \geq 10\,\text{cm}$ for our scenes and capture prototype), plugging both $D$ and $\lambda$ into \eref{eq:rayleigh-criterion} gives us a resolution of $\Delta x \approx 22\,\text{cm}$ at $z=1\,\text{m}$. In practice, that resolution allows us to image simple objects in our harder higher-order imaging tasks (see Section 8 of the main paper). Even if our $16\times 16$ SPAD array covered the full $\SI{1.9}{m}\times\SI{1.9}{m}$ relay wall, the loss in detail from longer wavelengths would offset a large part of the resolution gains the larger aperture would have provided.
For comparison, our simulations use a $32\times 32$ SPAD array spanning the entire $\SI{2}{m}\times\SI{2}{m}$ relay wall, resulting in a sharper resolution of $\Delta x \approx 3.5\,\text{cm}$. As we discuss in Section 9 of the main paper, our prototype demonstrates a promising proof-of-concept that will scale significantly with future SPAD technology. %

\begin{figure}
    \centering
    \captionsetup{skip=-3pt}
    \def\svgwidth{\linewidth}
    \begin{small}
        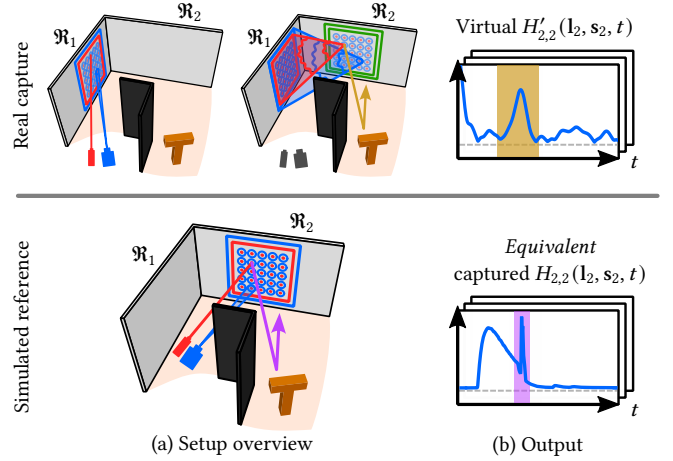
    \end{small}
    \caption{Comparison of our reconstructed virtual impulse response $\Hp_{2,2}$ at the wall $\lRtwo$ with the equivalent captured impulse response $H_{2,2}$ for the setup in Figure 4 of the main paper. (a) Imaging procedures to obtain each impulse response. The top row virtually illuminates and captures points $\mathbf{l}_2$ and $\mathbf{s}_2$ of $\lRtwo$ using our transient projector $\Itproj$ and camera $\Itcam$ operators, respectively. The virtual impulse response uses fifth-bounce illumination from the T letter, highlighted in yellow. In the bottom row, we simulate an equivalent scene where we move the laser and SPAD to point directly towards $\lRtwo$. This results in third-bounce illumination from the T letter, highlighted in purple. (b) Resulting impulse responses for each case.}
    \label{fig:3rd_bounce_compare}
\end{figure}

\paragraph{Validation.}
\label{sec:resolution-limits:experiment}
Here, we validate our virtual impulse response functions computed at secondary relay walls under experimental data captured with our real hardware prototype, and support it with simulations and our theoretical resolution analysis. \fref{fig:3rd_bounce_compare}a (top) shows an overview of the real NLOS scene we use for this validation, with the relay wall $\lRone$, a hidden secondary wall $\lRtwo$, and a hidden T-shaped object.
We capture the impulse response $H$ at $\lRone$. Following Section 4.2 of the main paper, we reconstruct the virtual impulse response $\Hp_{2,2}$ on the secondary wall $\lRtwo$.

To validate our reconstructed $\Hp_{2,2}$, we simulate an equivalent captured impulse response function $H_{2,2}$ as if $\lRtwo$ itself were the \textit{first} relay wall, by pointing the laser and sensor directly to $\lRtwo$, as shown in \fref{fig:3rd_bounce_compare}a (bottom). 
As stated in the main paper and detailed in \sref{sec:noise}, all our simulations include the characteristic temporal uncertainty and noise of real capture systems.

\fref{fig:3rd_bounce_compare}b (top) shows a sample of our virtual impulse response function $\Hp_{2,2}$ reconstructed at $\lRtwo$.  \fref{fig:3rd_bounce_compare}b (bottom) shows the corresponding sample of the equivalent captured impulse response function $H_{2,2}$ at $\lRtwo$. 
We verify that both $\Hp_{2,2}$ and its equivalent $H_{2,2}$ have peaks aligned in time, that correspond to fifth-bounce illumination at $\Hp_{2,2}$ (top, yellow) and third-bounce illumination at $H_{2,2}$ (bottom, purple) from the T-shaped object. 
The peak in $H_{2,2}$ (bottom, purple) has an $\SI{11}{cm}$ Full-Width at Half Maximum (FWHM), which depends on the system temporal uncertainty of our simulated NLOS imaging system pointed at $\lRtwo$. 
The peak in our virtual impulse response function $\Hp_{2,2}$ (top, yellow), computed from the real captured $H$ using $\lambda_c = \SI{15}{cm}$, has a $\SI{40}{cm}$ FWHM. 
This wider peak of $\Hp_{2,2}$ follows \eref{eq:rayleigh-criterion}: plugging $k = 1 / \sqrt{2}$ (square aperture), $\lambda = \lambda_c = \SI{15}{cm}$, $z = \SI{1.2}{m}$ (the distance between $\lRone$ and $\lRtwo$) and $D = \SI{32.5}{cm}$ (shortest side of the area covered by the SPAD array), yields $\Delta x \approx \SI{39.1}{cm}$, which closely matches the measured $\SI{40}{cm}$.

\subsection{Resolution of our cascaded imaging systems}
\label{sec:resolution-limits:cascaded}

Our cascaded imaging systems use our virtual impulse response $H'$ at secondary relay walls to form secondary imaging systems (Equation 14). We measure the resolution of the entire cascaded system using again the Rayleigh criterion (\eref{eq:rayleigh-criterion}) for the secondary imaging system instead, while also taking into account aperture size constraints and resolution constraints from the resolution of our virtual impulse response $H'$.

\paragraph{Scanning spacing constraints.} The Nyquist-Shannon sampling theorem also applies to the scanning spacing at secondary relay walls, when focusing secondary apertures $\mathcal{S}_b$ or $\mathcal{L}_a$. In our cascaded imaging systems, experiments showed that using a scanning resolution $\Delta \mathbf{s}_b = \Delta \xs$ and $\Delta \mathbf{l}_a = \Delta \xl$ for secondary apertures $\mathcal{S}_b$ and $\mathcal{L}_a$ works well in most cases. Finer sampling with shorter $\Delta \mathbf{s}_b$ or $\Delta \mathbf{l}_a$ did not seem to provide any improvements. For experiments with large values of $\lambda_c$, we use coarser resolution with larger $\Delta \mathbf{s}_b$ and $\Delta \mathbf{l}_a$ to reduce execution times.

\paragraph{Effective aperture constraints.} Consider the case where we reconstruct $H'_{a,b}(\xlp, \xsp, t)$ at points $\xsp \in \mathcal{S}_b$ on a secondary relay wall with aperture $\mathcal{S}_b$ with side length $D$. Because surfaces can exhibit \emph{virtual specular} behavior (see \aref{sec:reflectance-general}), some points $\xsp$ might reflect light away from the hidden object. Thus, only part of $\mathcal{S}_b$ has usable signal, which we denote as the effective aperture size $\hat{D}$. To get the exact imaging resolution, one can plug $\hat{D}$ into \eref{eq:rayleigh-criterion} instead of $D$. However, note that exact computation of $\hat{D}$ depends on the scene and the choice of secondary aperture $\mathcal{S}_b$. For most scenes we tested in our work, our rule of thumb is that $\hat{D}$ is roughly half of $D$.

\paragraph{Resolution constraints.} Similar to how the temporal resolution of $H$ imposes a bound on imaging resolution \cite{OToole2018confocal}, the spatial $\Delta x$ and temporal $\Delta t$ resolutions of our virtual impulse response $H'$ affect our cascaded imaging resolution. Exact computation of imaging resolution as in \eref{eq:rayleigh-criterion} depends on the layout of the hidden scene, and thus we do not give a one-size-fits-all formula. To design our experiments, we estimate a lower bound of the minimum resolvable distance $\Delta x'$ of the entire system as $\Delta x' \geq c \Delta t/2 \approx \Delta x/2$.

%% file: fig/3rd_bounce_comparison-v2.pdf_tex
\begingroup%
  \makeatletter%
  \providecommand\color[2][]{%
    \errmessage{(Inkscape) Color is used for the text in Inkscape, but the package 'color.sty' is not loaded}%
    \renewcommand\color[2][]{}%
  }%
  \providecommand\transparent[1]{%
    \errmessage{(Inkscape) Transparency is used (non-zero) for the text in Inkscape, but the package 'transparent.sty' is not loaded}%
    \renewcommand\transparent[1]{}%
  }%
  \providecommand\rotatebox[2]{#2}%
  \newcommand*\fsize{\dimexpr\f@size pt\relax}%
  \newcommand*\lineheight[1]{\fontsize{\fsize}{#1\fsize}\selectfont}%
  \ifx\svgwidth\undefined%
    \setlength{\unitlength}{878.40312808bp}%
    \ifx\svgscale\undefined%
      \relax%
    \else%
      \setlength{\unitlength}{\unitlength * \real{\svgscale}}%
    \fi%
  \else%
    \setlength{\unitlength}{\svgwidth}%
  \fi%
  \global\let\svgwidth\undefined%
  \global\let\svgscale\undefined%
  \makeatother%
  \begin{picture}(1,0.6776837)%
    \lineheight{1}%
    \setlength\tabcolsep{0pt}%
    \put(0.3350229,0.00185906){\color[rgb]{0,0,0}\makebox(0,0)[t]{\lineheight{1.25}\smash{\begin{tabular}[t]{c}(a) Setup overview\end{tabular}}}}%
    \put(0.01962809,0.19293805){\color[rgb]{0,0,0}\rotatebox{90}{\makebox(0,0)[t]{\lineheight{1.25}\smash{\begin{tabular}[t]{c}Simulated reference\end{tabular}}}}}%
    \put(0.01875953,0.55150118){\color[rgb]{0,0,0}\rotatebox{90}{\makebox(0,0)[t]{\lineheight{1.25}\smash{\begin{tabular}[t]{c}Real capture\end{tabular}}}}}%
    \put(0.80822767,0.00177406){\color[rgb]{0,0,0}\makebox(0,0)[t]{\lineheight{1.25}\smash{\begin{tabular}[t]{c}(b) Output\end{tabular}}}}%
    \put(0.8197267,0.64827076){\makebox(0,0)[t]{\smash{\begin{tabular}[t]{c}Virtual $\Hp_{2,2}(\mathbf{l}_2, \mathbf{s}_2, t)$\end{tabular}}}}%
    \put(0,0){\includegraphics[width=\unitlength,page=1]{3rd_bounce_comparison-v2.pdf}}%
    \put(0.95120613,0.43919763){\color[rgb]{0,0,0}\makebox(0,0)[lt]{\lineheight{1.25}\smash{\begin{tabular}[t]{l}$t$\end{tabular}}}}%
    \put(0,0){\includegraphics[width=\unitlength,page=2]{3rd_bounce_comparison-v2.pdf}}%
    \put(0.19910338,0.29457191){\color[rgb]{0,0,0}\makebox(0,0)[t]{\lineheight{0}\smash{\begin{tabular}[t]{c}$\lRone$\end{tabular}}}}%
    \put(0.44373479,0.3456449){\color[rgb]{0,0,0}\makebox(0,0)[t]{\lineheight{0}\smash{\begin{tabular}[t]{c}$\lRtwo$\end{tabular}}}}%
    \put(0.37810244,0.63146961){\color[rgb]{0,0,0}\makebox(0,0)[t]{\lineheight{0}\smash{\begin{tabular}[t]{c}$\lRone$\end{tabular}}}}%
    \put(0.57017591,0.67156979){\color[rgb]{0,0,0}\makebox(0,0)[t]{\lineheight{0}\smash{\begin{tabular}[t]{c}$\lRtwo$\end{tabular}}}}%
    \put(0,0){\includegraphics[width=\unitlength,page=3]{3rd_bounce_comparison-v2.pdf}}%
    \put(0.07763489,0.63146963){\color[rgb]{0,0,0}\makebox(0,0)[t]{\lineheight{0}\smash{\begin{tabular}[t]{c}$\lRone$\end{tabular}}}}%
    \put(0.26970857,0.67156981){\color[rgb]{0,0,0}\makebox(0,0)[t]{\lineheight{0}\smash{\begin{tabular}[t]{c}$\lRtwo$\end{tabular}}}}%
    \put(0,0){\includegraphics[width=\unitlength,page=4]{3rd_bounce_comparison-v2.pdf}}%
    \put(0.8279503,0.30678097){\makebox(0,0)[t]{\lineheight{1.25}\smash{\begin{tabular}[t]{c}\emph{Equivalent}\\captured $H_{2,2}(\mathbf{l}_2, \mathbf{s}_2, t)$\end{tabular}}}}%
    \put(0,0){\includegraphics[width=\unitlength,page=5]{3rd_bounce_comparison-v2.pdf}}%
    \put(0.95151841,0.0550383){\color[rgb]{0,0,0}\makebox(0,0)[lt]{\lineheight{1.25}\smash{\begin{tabular}[t]{l}$t$\end{tabular}}}}%
    \put(0,0){\includegraphics[width=\unitlength,page=6]{3rd_bounce_comparison-v2.pdf}}%
  \end{picture}%
\endgroup%

%% file: tex/A5_scene_knowledge.tex
\section{Impact of estimation error of secondary relay walls}
\label{sec:prior-knowledge}

\begin{figure}
    \centering
    \captionsetup{skip=-6pt}
    \def\svgwidth{\columnwidth} 
    \begin{small}
    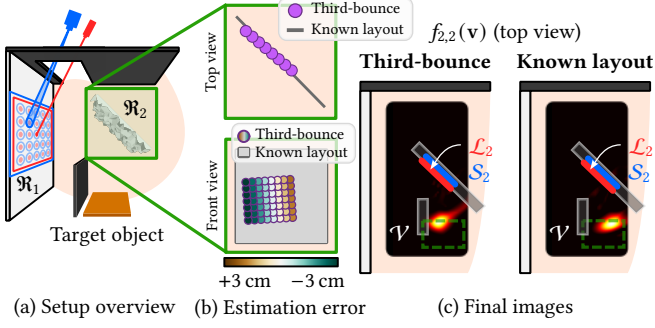
    \end{small}
    \caption{
    (a) We replicate the scene from Figure 15 in the main paper. (b) We compare our estimation of the secondary relay wall $\lRtwo$ using a third-bounce method against a known layout obtained via 3D scans. (c) We analyze the impact of estimation error in the final cascaded imaging results $f_{2,2}(\xv)$, and observe both methods show very similar results.}
    \label{fig:prior-knowledge}
\end{figure}

In Section 4.3 of the main paper, we discuss our strategies to select secondary relay walls for our cascaded imaging systems without needing to know the layout of the hidden scene beforehand. These strategies use third-bounce methods to image potential secondary relay walls from the primary relay wall $\lRone$. Here, we evaluate whether errors in these location estimates actually affect the final image quality, using the NLOS scenario from Figure 15 of the main paper as a practical example.

We show an overview of the setup in \fref{fig:prior-knowledge}a. To obtain the position of the secondary wall $\lRtwo$, we test two settings: (i) we image the hidden scene using third-bounce methods (Equation 6), and then post-process the resulting image to obtain a set of points $\mathbf{l}_2, \mathbf{s}_2$ corresponding to $\lRtwo$ (represented using triangles in \fref{fig:prior-knowledge}b), and (ii) we manually specify the exact positions of $\mathbf{l}_2, \mathbf{s}_2$, based on prior knowledge of $\lRtwo$ (here obtained via a 3D scan) but not of the target object. As shown in \fref{fig:prior-knowledge}b, our third-bounce estimations fall within just $\pm3\,\text{cm}$ of error.

For both settings, we use the same captured impulse response $H$ and follow the same imaging procedure (please see Section 8.2 for the full procedure), only changing the position of $\mathbf{l}_2 \in \mathcal{L}_2$ and $\mathbf{s}_2\in \mathcal{S}_2$. 
The imaging results from both methods are quite similar, as seen in \fref{fig:prior-knowledge}c. In both cases, an object with similar shape appears on the expected location marked with $\lV$ corresponding to the target hidden object.

%% file: fig/compare-prior-v3.pdf_tex
\begingroup%
  \makeatletter%
  \providecommand\color[2][]{%
    \errmessage{(Inkscape) Color is used for the text in Inkscape, but the package 'color.sty' is not loaded}%
    \renewcommand\color[2][]{}%
  }%
  \providecommand\transparent[1]{%
    \errmessage{(Inkscape) Transparency is used (non-zero) for the text in Inkscape, but the package 'transparent.sty' is not loaded}%
    \renewcommand\transparent[1]{}%
  }%
  \providecommand\rotatebox[2]{#2}%
  \newcommand*\fsize{\dimexpr\f@size pt\relax}%
  \newcommand*\lineheight[1]{\fontsize{\fsize}{#1\fsize}\selectfont}%
  \ifx\svgwidth\undefined%
    \setlength{\unitlength}{1426.81017778bp}%
    \ifx\svgscale\undefined%
      \relax%
    \else%
      \setlength{\unitlength}{\unitlength * \real{\svgscale}}%
    \fi%
  \else%
    \setlength{\unitlength}{\svgwidth}%
  \fi%
  \global\let\svgwidth\undefined%
  \global\let\svgscale\undefined%
  \makeatother%
  \begin{picture}(1,0.49416713)%
    \lineheight{1}%
    \setlength\tabcolsep{0pt}%
    \put(0,0){\includegraphics[width=\unitlength,page=1]{compare-prior-v3.pdf}}%
    \put(0.32627877,0.39372803){\rotatebox{90}{\makebox(0,0)[t]{\lineheight{1.25}\smash{\begin{tabular}[t]{c}{\scriptsize Top view}\end{tabular}}}}}%
    \put(0.32834516,0.18637858){\rotatebox{90}{\makebox(0,0)[t]{\lineheight{1.25}\smash{\begin{tabular}[t]{c}{\scriptsize Front view}\end{tabular}}}}}%
    \put(0,0){\includegraphics[width=\unitlength,page=2]{compare-prior-v3.pdf}}%
    \put(0.96319345,0.21274088){\color[rgb]{0,0.4,1}\makebox(0,0)[lt]{\lineheight{1.25}\smash{\begin{tabular}[t]{l}$\mathcal{S}_2$\end{tabular}}}}%
    \put(0,0){\includegraphics[width=\unitlength,page=3]{compare-prior-v3.pdf}}%
    \put(0.87480179,0.11741648){\color[rgb]{1,1,1}\makebox(0,0)[rt]{\lineheight{1.25}\smash{\begin{tabular}[t]{r}$\lV$\end{tabular}}}}%
    \put(0.96319342,0.25316529){\color[rgb]{1,0.14901961,0.14901961}\makebox(0,0)[lt]{\lineheight{1.25}\smash{\begin{tabular}[t]{l}$\mathcal{L}_2$\end{tabular}}}}%
    \put(0,0){\includegraphics[width=\unitlength,page=4]{compare-prior-v3.pdf}}%
    \put(0.71573133,0.21274088){\color[rgb]{0,0.4,1}\makebox(0,0)[lt]{\lineheight{1.25}\smash{\begin{tabular}[t]{l}$\mathcal{S}_2$\end{tabular}}}}%
    \put(0.7746108,0.00411824){\makebox(0,0)[t]{\lineheight{1.25}\smash{\begin{tabular}[t]{c}(c) Final images\end{tabular}}}}%
    \put(0.42616246,0.00341513){\makebox(0,0)[t]{\lineheight{1.25}\smash{\begin{tabular}[t]{c}(b) Estimation error\end{tabular}}}}%
    \put(0.89791599,0.38431978){\makebox(0,0)[t]{\lineheight{1.25}\smash{\begin{tabular}[t]{c}\textbf{Known layout}\end{tabular}}}}%
    \put(0.65045386,0.38431978){\makebox(0,0)[t]{\lineheight{1.25}\smash{\begin{tabular}[t]{c}\textbf{Third-bounce}\end{tabular}}}}%
    \put(0,0){\includegraphics[width=\unitlength,page=5]{compare-prior-v3.pdf}}%
    \put(0.62733967,0.11741648){\color[rgb]{1,1,1}\makebox(0,0)[rt]{\lineheight{1.25}\smash{\begin{tabular}[t]{r}$\lV$\end{tabular}}}}%
    \put(0.7157313,0.25316529){\color[rgb]{1,0.14901961,0.14901961}\makebox(0,0)[lt]{\lineheight{1.25}\smash{\begin{tabular}[t]{l}$\mathcal{L}_2$\end{tabular}}}}%
    \put(0,0){\includegraphics[width=\unitlength,page=6]{compare-prior-v3.pdf}}%
    \put(0.32960005,0.04552489){\makebox(0,0)[lt]{\lineheight{1.25}\smash{\begin{tabular}[t]{l}$+3$\,cm\end{tabular}}}}%
    \put(0.52532908,0.04552489){\makebox(0,0)[rt]{\lineheight{1.25}\smash{\begin{tabular}[t]{r}$-3$\,cm\end{tabular}}}}%
    \put(0,0){\includegraphics[width=\unitlength,page=7]{compare-prior-v3.pdf}}%
    \put(0.1578907,0.11590906){\makebox(0,0)[t]{\lineheight{1.25}\smash{\begin{tabular}[t]{c}Target object\end{tabular}}}}%
    \put(0.03688216,0.19396945){\makebox(0,0)[t]{\lineheight{1.25}\smash{\begin{tabular}[t]{c}$\lRone$\end{tabular}}}}%
    \put(0,0){\includegraphics[width=\unitlength,page=8]{compare-prior-v3.pdf}}%
    \put(0.77418482,0.43058368){\makebox(0,0)[t]{\lineheight{1.25}\smash{\begin{tabular}[t]{c}$f_{2,2}(\xv)$ (top view)\end{tabular}}}}%
    \put(0,0){\includegraphics[width=\unitlength,page=9]{compare-prior-v3.pdf}}%
    \put(0.13891646,0.00441539){\makebox(0,0)[t]{\lineheight{1.25}\smash{\begin{tabular}[t]{c}(a) Setup overview\end{tabular}}}}%
    \put(0,0){\includegraphics[width=\unitlength,page=10]{compare-prior-v3.pdf}}%
    \put(0.47432942,0.4689181){\makebox(0,0)[lt]{\lineheight{0.89999998}\smash{\begin{tabular}[t]{l}{\scriptsize Third-bounce}\end{tabular}}}}%
    \put(0.47436016,0.43602502){\makebox(0,0)[lt]{\lineheight{0.89999998}\smash{\begin{tabular}[t]{l}{\scriptsize Known layout}\end{tabular}}}}%
    \put(0,0){\includegraphics[width=\unitlength,page=11]{compare-prior-v3.pdf}}%
    \put(0.38771777,0.27428124){\makebox(0,0)[lt]{\lineheight{0.89999998}\smash{\begin{tabular}[t]{l}{\scriptsize Third-bounce}\end{tabular}}}}%
    \put(0.38773464,0.24640121){\makebox(0,0)[lt]{\lineheight{0.89999998}\smash{\begin{tabular}[t]{l}{\scriptsize Known layout}\end{tabular}}}}%
    \put(0.2026489,0.31166814){\makebox(0,0)[t]{\lineheight{1.25}\smash{\begin{tabular}[t]{c}$\lRtwo$\end{tabular}}}}%
    \put(0,0){\includegraphics[width=\unitlength,page=12]{compare-prior-v3.pdf}}%
  \end{picture}%
\endgroup%

%% file: tex/A2_virtual_reflectance.tex
\section{Virtual surface reflectance and the missing cone}
\label{sec:reflectance-general}

\final{In this section, we expand on the analysis from Section 6 of the main paper.}
In our NLOS imaging setups, although all surfaces are diffuse under visible light, they may exhibit different reflectance properties when illuminated using phasor-field virtual waves. 
This effect, which we term \textit{virtual surface reflectance}, depends on the subset of imaging frequencies $\fq$ used in the NLOS imaging process.
This virtual surface reflectance in general differs from the \emph{real} reflectance properties of the same surfaces, which respectively depend on their micro- and nano-scale structure.
Under the phasor-field formulation, the impulse response $H(\xl, \xs, t)$ (captured using nanometer-scale laser pulses) is typically convolved with a Morlet wavelet virtual illumination function $K(t;\wlc,\sigma)$ (Equation 3 of the main paper) of central wavelength $\wlc$ and standard deviation $\sigma$. This function acts as a band-pass filter, with a Gaussian-shaped frequency spectrum centered at $\Omega_c = c/\wlc$, and width inversely proportional to $\sigma$. 
Due to limited sampling of the relay wall $\Delta \xs$ (\sref{sec:resolution-limits}) and temporal uncertainty of the imaging system (\sref{sec:noise}), practical $\wlc$ and $\sigma$ values (expressed as spatial instead of temporal measures) are typically in the centimeter range \cite{Liu2019phasor,Liu2020phasor,royo2023virtual}.
As a result of this convolution, the imaged objects exhibit virtual surface reflectance corresponding to a narrow set of centimeter-range wavelengths centered at $\wlc$, which is usually determined by the centimeter-scale structure of such surfaces.

\final{In Section 6 of the main paper, we present two simulated experiments which illustrate the different reflectance behaviors between real and phasor illumination, and their relationship to the missing-cone problem and surface visibility. These insights are especially useful to understand our simulated and real NLOS imaging experiments in Sections 7 and 8 of our paper, respectively.}

\subsection{Missing cone under higher-order illumination}
\label{sec:higher-missing-cone}

\begin{figure}
    \centering
    \captionsetup{skip=-3pt}
    \def\svgwidth{\columnwidth} 
    \begin{small}
    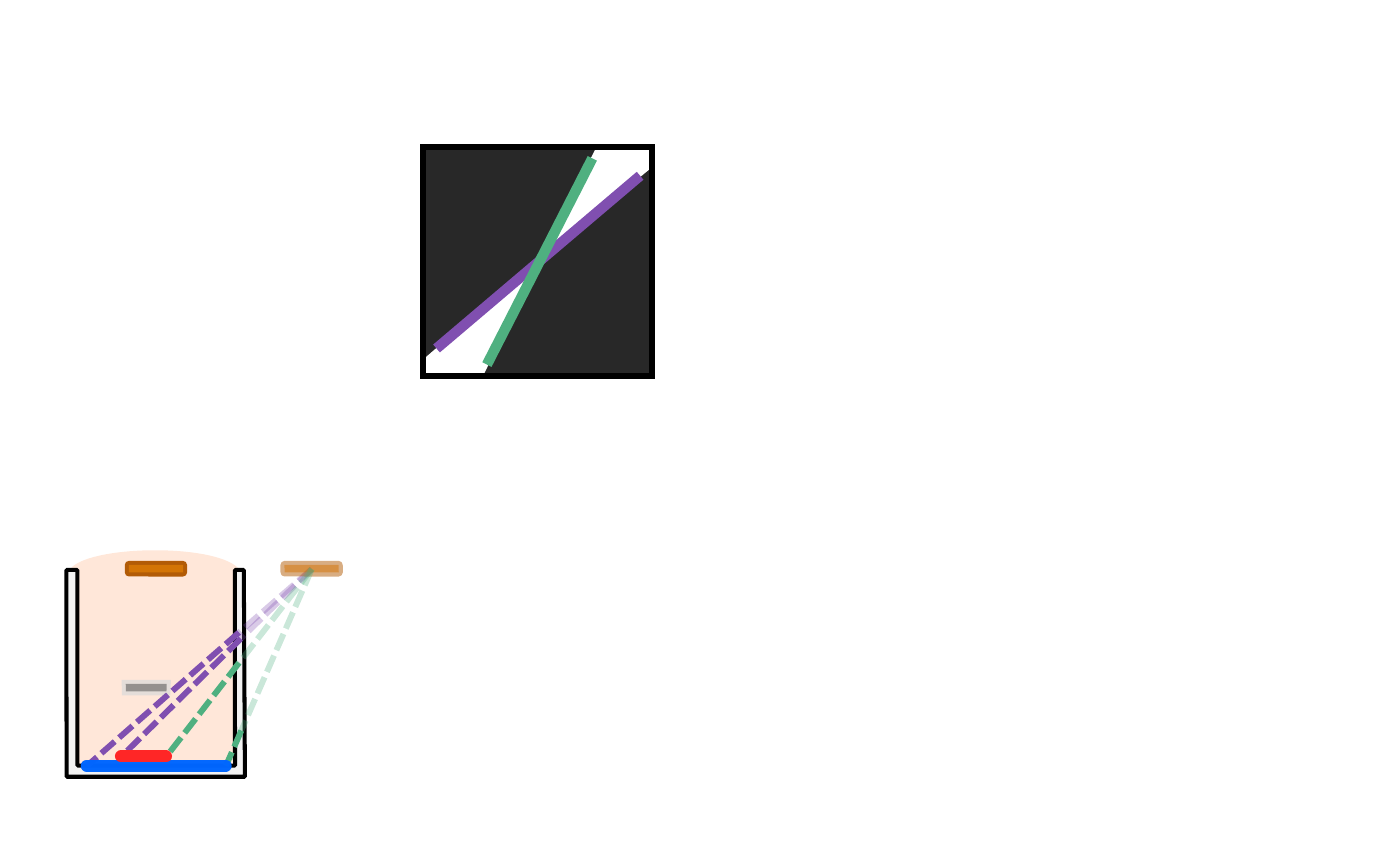
    \end{small}
    \caption{
    We extend the missing cone analysis from \citet{Liu2019analysis} to our scene in Figure 13 of the main paper, where we rotate the target object. The virtual mirrors method from \citet{royo2023virtual} and our method use different apertures $\mathcal{L}$ and $\mathcal{S}$, and as a result each samples different Fourier cones of the NLOS measurement function. (a) The virtual mirrors method is fixed to $\lL$ and $\lS$. In this case, the sampled Fourier cone only works for setups with specific orientations, and thus fails on the bottom row. (b) Our method can create apertures at different parts of the hidden scene ($\mathcal{L}_2$, $\mathcal{L}_3$ and $\mathcal{S}_3$), which in turn sample different parts of the NLOS measurement function. Note how our method can adapt to each configuration, lifting previous restrictions about specific orientations.}
    \label{fig:fig10-missing-cone}
\end{figure}

In this section, we extend the missing-cone analysis from \citet{Liu2019analysis} to our cascaded imaging systems, and show how to leverage our transient projector $\Itproj$ and camera $\Itcam$ operators.

\fref{fig:fig10-missing-cone} illustrates the scene originally presented in Figure 13 of the main paper. There, we show an overview comparing the NLOS imaging approach from \citet{royo2023virtual} with our own. We test for two orientations of the target object, only one of which is favorable to Royo et al.'s method. Specifically, we highlight the different illumination $\mathcal{L}$ and camera $\mathcal{S}$ apertures used for imaging. While the method proposed by Royo et al. can only use the apertures $\lL$ and $\lS$ in the primary relay wall $\lR$, we use our transient projector $\Itproj$ and camera $\Itcam$ operators to form new apertures $\mathcal{L}_2$, $\mathcal{L}_3$, and $\mathcal{S}_3$ on walls $\lRtwo$ and $\lRthree$. We display the corresponding Fourier domain representation of the NLOS measurement function for the hidden target \cite{Liu2019analysis}. While the exact measured time response is a projection along various ellipsoids with foci at points $\textbf{l}\in\mathcal{L}$ and $\textbf{s}\in\mathcal{S}$ (corresponding to purple/green dashed lines), we follow Liu et al. and approximate these projections using planes (corresponding to purple/green arrows). Due to finite aperture sizes, large portions of the Fourier spectrum (represented in black) are not contained in any measurements.

The method proposed by Royo et al. and our own utilize different aperture configurations, resulting in the sampling of distinct regions within the NLOS measurement function. Notably, only our cascaded imaging system samples the specific Fourier components relevant to the hidden target for all cases. Consequently, while our method does not inherently solve the missing-cone problem, we can circumvent it thanks to our $\Itproj$ and $\Itcam$ operators, whereas the approach by Royo et al. remains susceptible to it.

%% file: fig/mcanalysis-v7.pdf_tex
\begingroup%
  \makeatletter%
  \providecommand\color[2][]{%
    \errmessage{(Inkscape) Color is used for the text in Inkscape, but the package 'color.sty' is not loaded}%
    \renewcommand\color[2][]{}%
  }%
  \providecommand\transparent[1]{%
    \errmessage{(Inkscape) Transparency is used (non-zero) for the text in Inkscape, but the package 'transparent.sty' is not loaded}%
    \renewcommand\transparent[1]{}%
  }%
  \providecommand\rotatebox[2]{#2}%
  \newcommand*\fsize{\dimexpr\f@size pt\relax}%
  \newcommand*\lineheight[1]{\fontsize{\fsize}{#1\fsize}\selectfont}%
  \ifx\svgwidth\undefined%
    \setlength{\unitlength}{658.89231284bp}%
    \ifx\svgscale\undefined%
      \relax%
    \else%
      \setlength{\unitlength}{\unitlength * \real{\svgscale}}%
    \fi%
  \else%
    \setlength{\unitlength}{\svgwidth}%
  \fi%
  \global\let\svgwidth\undefined%
  \global\let\svgscale\undefined%
  \makeatother%
  \begin{picture}(1,0.61712527)%
    \lineheight{1}%
    \setlength\tabcolsep{0pt}%
    \put(0.25957736,0.00425875){\makebox(0,0)[t]{\lineheight{1.10000002}\smash{\begin{tabular}[t]{c}(a) Virtual mirrors \cite{royo2023virtual}\end{tabular}}}}%
    \put(0.7508409,0.00425868){\makebox(0,0)[t]{\lineheight{1.10000002}\smash{\begin{tabular}[t]{c}(b) Our cascaded imaging systems\end{tabular}}}}%
    \put(0,0){\includegraphics[width=\unitlength,page=1]{mcanalysis-v7.pdf}}%
    \put(0.39161758,0.53372481){\makebox(0,0)[t]{\lineheight{1.25}\smash{\begin{tabular}[t]{c}$f_y$\end{tabular}}}}%
    \put(0,0){\includegraphics[width=\unitlength,page=2]{mcanalysis-v7.pdf}}%
    \put(0.50542966,0.42237134){\makebox(0,0)[t]{\lineheight{1.25}\smash{\begin{tabular}[t]{c}$f_x$\end{tabular}}}}%
    \put(0,0){\includegraphics[width=\unitlength,page=3]{mcanalysis-v7.pdf}}%
    \put(0.16125146,0.52445658){\makebox(0,0)[t]{\lineheight{0.89999998}\smash{\begin{tabular}[t]{c}Target object (mirror)\end{tabular}}}}%
    \put(0.16171412,0.23229428){\makebox(0,0)[t]{\lineheight{0.89999998}\smash{\begin{tabular}[t]{c}Target object (mirror)\end{tabular}}}}%
    \put(0,0){\includegraphics[width=\unitlength,page=4]{mcanalysis-v7.pdf}}%
    \put(0.57111407,0.12402996){\color[rgb]{1,0.14901961,0.14901961}\makebox(0,0)[rt]{\lineheight{0.89999998}\smash{\begin{tabular}[t]{r}$\mathcal{L}_2$\end{tabular}}}}%
    \put(0.70921733,0.11947689){\color[rgb]{0,0.4,1}\makebox(0,0)[lt]{\lineheight{0.89999998}\smash{\begin{tabular}[t]{l}$\mathcal{S}_3$\end{tabular}}}}%
    \put(0,0){\includegraphics[width=\unitlength,page=5]{mcanalysis-v7.pdf}}%
    \put(0.64115314,0.52218003){\makebox(0,0)[t]{\lineheight{0.89999998}\smash{\begin{tabular}[t]{c}Target object\end{tabular}}}}%
    \put(0,0){\includegraphics[width=\unitlength,page=6]{mcanalysis-v7.pdf}}%
    \put(0.68615661,0.39797942){\color[rgb]{1,0.14901961,0.14901961}\makebox(0,0)[rt]{\lineheight{0.89999998}\smash{\begin{tabular}[t]{r}$\mathcal{L}_3$\end{tabular}}}}%
    \put(0.71149384,0.39797941){\color[rgb]{0,0.4,1}\makebox(0,0)[lt]{\lineheight{0.89999998}\smash{\begin{tabular}[t]{l}$\mathcal{S}_3$\end{tabular}}}}%
    \put(0,0){\includegraphics[width=\unitlength,page=7]{mcanalysis-v7.pdf}}%
    \put(0.80071003,0.28805464){\makebox(0,0)[lt]{\lineheight{0.89999998}\smash{\begin{tabular}[t]{l}Missing cone\end{tabular}}}}%
    \put(0.55172121,0.28721964){\makebox(0,0)[lt]{\lineheight{0.89999998}\smash{\begin{tabular}[t]{l}Sampled cone\end{tabular}}}}%
    \put(0,0){\includegraphics[width=\unitlength,page=8]{mcanalysis-v7.pdf}}%
    \put(0.09842986,0.28650104){\makebox(0,0)[lt]{\lineheight{0.89999998}\smash{\begin{tabular}[t]{l}Target object\end{tabular}}}}%
    \put(0.33971251,0.28694598){\makebox(0,0)[lt]{\lineheight{0.89999998}\smash{\begin{tabular}[t]{l}Boundaries\end{tabular}}}}%
    \put(0.20254892,0.34468173){\color[rgb]{0,0.4,1}\makebox(0,0)[t]{\lineheight{0.89999998}\smash{\begin{tabular}[t]{c}$\lS$\end{tabular}}}}%
    \put(0,0){\includegraphics[width=\unitlength,page=9]{mcanalysis-v7.pdf}}%
    \put(0.86833788,0.53380821){\makebox(0,0)[t]{\lineheight{1.25}\smash{\begin{tabular}[t]{c}$f_y$\end{tabular}}}}%
    \put(0,0){\includegraphics[width=\unitlength,page=10]{mcanalysis-v7.pdf}}%
    \put(0.98214996,0.42245474){\makebox(0,0)[t]{\lineheight{1.25}\smash{\begin{tabular}[t]{c}$f_x$\end{tabular}}}}%
    \put(0.86855313,0.24174663){\makebox(0,0)[t]{\lineheight{1.25}\smash{\begin{tabular}[t]{c}$f_y$\end{tabular}}}}%
    \put(0.98236521,0.13039313){\makebox(0,0)[t]{\lineheight{1.25}\smash{\begin{tabular}[t]{c}$f_x$\end{tabular}}}}%
    \put(0,0){\includegraphics[width=\unitlength,page=11]{mcanalysis-v7.pdf}}%
    \put(0.39161758,0.24165571){\makebox(0,0)[t]{\lineheight{1.25}\smash{\begin{tabular}[t]{c}$f_y$\end{tabular}}}}%
    \put(0,0){\includegraphics[width=\unitlength,page=12]{mcanalysis-v7.pdf}}%
    \put(0.50542966,0.13030228){\makebox(0,0)[t]{\lineheight{1.25}\smash{\begin{tabular}[t]{c}$f_x$\end{tabular}}}}%
    \put(0,0){\includegraphics[width=\unitlength,page=13]{mcanalysis-v7.pdf}}%
    \put(0.11324939,0.60191599){\makebox(0,0)[t]{\lineheight{0.89999998}\smash{\begin{tabular}[t]{c}\textbf{Setup overview}\\\textbf{and apertures}\end{tabular}}}}%
    \put(0.3913485,0.60191599){\makebox(0,0)[t]{\lineheight{0.89999998}\smash{\begin{tabular}[t]{c}\textbf{Measurement}\\\textbf{(Fourier)}\end{tabular}}}}%
    \put(0.63982376,0.60226925){\makebox(0,0)[t]{\lineheight{0.89999998}\smash{\begin{tabular}[t]{c}\textbf{Setup overview}\\\textbf{and apertures}\end{tabular}}}}%
    \put(0.86785873,0.60191599){\makebox(0,0)[t]{\lineheight{0.89999998}\smash{\begin{tabular}[t]{c}\textbf{Measurement}\\\textbf{(Fourier)}\end{tabular}}}}%
    \put(0.07151304,0.38694046){\color[rgb]{1,0.14901961,0.14901961}\makebox(0,0)[t]{\lineheight{0.89999998}\smash{\begin{tabular}[t]{c}$\lL$\end{tabular}}}}%
    \put(0.20264006,0.05200476){\color[rgb]{0,0.4,1}\makebox(0,0)[t]{\lineheight{0.89999998}\smash{\begin{tabular}[t]{c}$\lS$\end{tabular}}}}%
    \put(0.07160419,0.0942635){\color[rgb]{1,0.14901961,0.14901961}\makebox(0,0)[t]{\lineheight{0.89999998}\smash{\begin{tabular}[t]{c}$\lL$\end{tabular}}}}%
    \put(0.64028524,0.22546463){\makebox(0,0)[t]{\lineheight{0.89999998}\smash{\begin{tabular}[t]{c}Target object\end{tabular}}}}%
    \put(0,0){\includegraphics[width=\unitlength,page=14]{mcanalysis-v7.pdf}}%
  \end{picture}%
\endgroup%

%% file: tex/A2_noise.tex
\section{Realistic capture noise models for simulation}
\label{sec:noise}

\begin{figure}[t]
    \centering
    \captionsetup{skip=0pt}
    \def\svgwidth{\columnwidth} 
    \begin{small}
    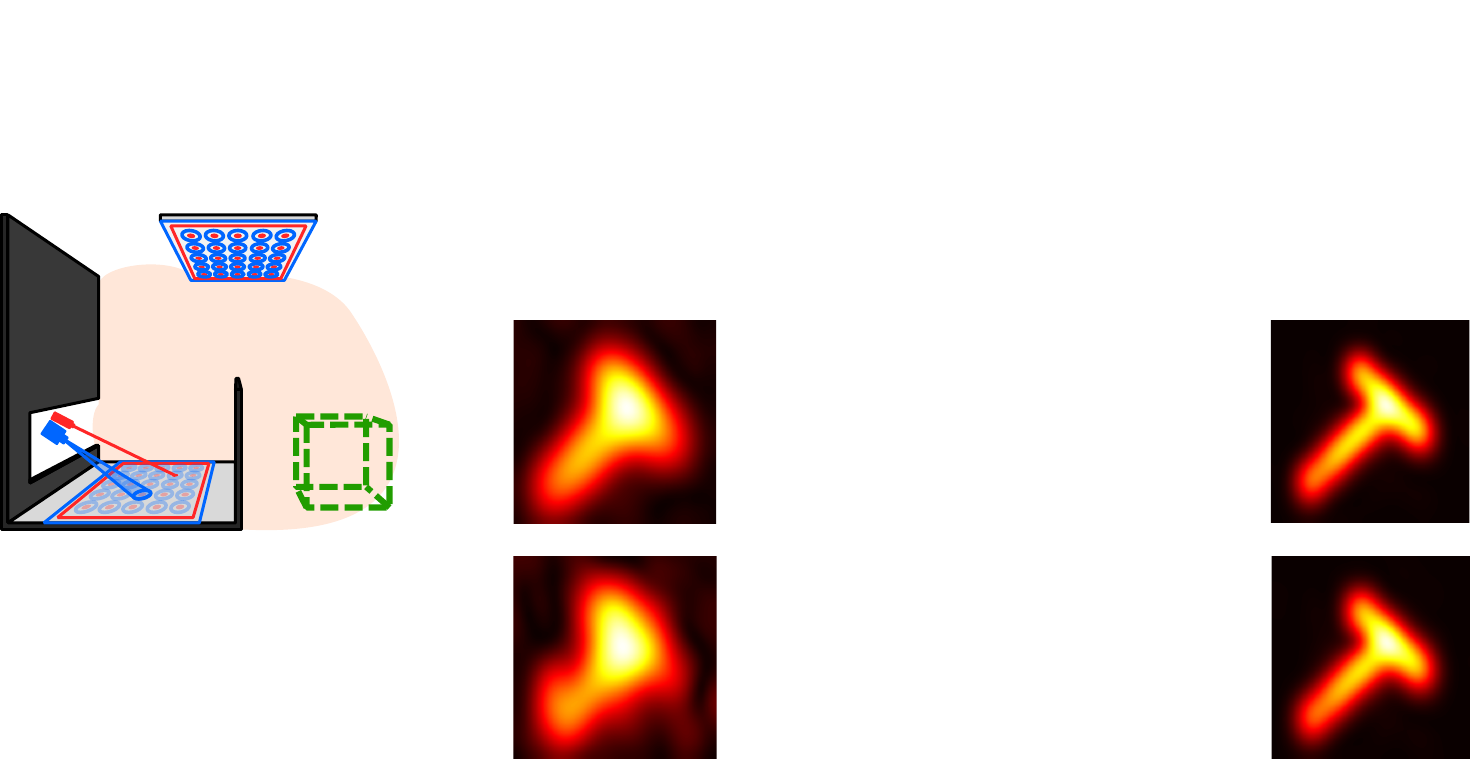
    \end{small}
    \caption{(a) We compare the resolution of our fifth-bounce imaging using $\lRtwo$ and (b) the corresponding third-bounce imaging where the object at $\lV$ is at a similar distance from the relay wall $\lRone$. We simulate both captures with and without noise. The imaging resolution loss from using multiple apertures in $\lRone$ and $\lRtwo$ in the two-corner scene (compared to the single-corner scene) is much more notable than the effect of noise.}
    \label{fig:noise-bounce-comparison}
\end{figure}

This section details how we model realistic noise in our simulations to mimic real capture conditions. In particular, we mimic photon counts, hardware response, and calibration errors from previous works \cite{Liu2019phasor,royo2023virtual, hernandez2017computationalmodelsinglephotonavalanche} and from our own hardware described in Section 8 of the main paper.

We observe that our cascaded NLOS imaging methods are robust to capture noise. %
In \fref{fig:noise-bounce-comparison}a, we simulate an object hidden around two corners and image it using a secondary virtual imaging system at $\lRtwo$. In \fref{fig:noise-bounce-comparison}b we simulate the equivalent single-corner setup, placing the hidden object at the same distance so that values in \eref{eq:rayleigh-criterion} match as much as possible between both scenes. We simulate captures with and without noise. We show that both third-bounce imaging methods and our fifth-bounce imaging method remain robust to noise, while the main change in imaging resolution comes from the fact that fifth-bounce imaging is a much harder problem than the conventional third-bounce problem.

\paragraph{Photon counts and signal-to-noise ratio.} Following prior NLOS imaging work \cite{Liu2019phasor, royo2023virtual} and measurements from our own hardware prototype, we assume that for each illuminated position $\xl$ the sensor captures approximately $10^9$ photons; in those works the entire capture process requires about three minutes of exposure time. 
In our simulated experiments, we add Poisson noise to our simulated impulse response function $H(\xl, \xs, t)$ to mimic similar capture conditions. %
We experimentally observed that the resulting SNR under such conditions is high enough for our cascaded NLOS imaging method, as seen in Figures 13 and 15 of the main paper.

\paragraph{Temporal uncertainty.} Photon arrival times estimated by NLOS imaging devices have inherent temporal uncertainty due to the limitations of laser and sensor hardware.
We mimic temporal uncertainty on our simulated data by applying a convolution over the time domain between the impulse response function $H(\xl, \xs, t)$, the real measured response of a $20\mu m$ CMOS single photon avalanche diode (SPAD) with a temporal resolution of \SI{25}{ps} of full width at half maximum (FWHM) \cite{hernandez2017computationalmodelsinglephotonavalanche}, and a Gaussian-shaped laser pulse with \SI{35}{ps} of FWHM \cite{royo2023virtual}. 
The resulting virtual impulse response functions $\Hp_{a,b}(\xlp, \xsp, t)$ calculated by our cascaded imaging method are thus affected by this temporal uncertainty, along with any inaccuracies inherent to NLOS imaging resolution (\aref{sec:resolution-limits}). %

\paragraph{Calibration error.} 
The real laser emission and sensor pixels of capture devices are not targeted at perfectly zero-dimensional points $\xl, \xs$ on the relay wall, but instead have a finite footprint that may introduce spatial inaccuracies during RSD propagation. 
We model these inaccuracies by randomly jittering the laser and sensor locations $\xl, \xs$ on the relay wall using a Gaussian distribution with standard deviation $\sigma_E = \SI{1}{cm}$, following real capture conditions of previous works \cite{royo2023virtual}.

\begin{figure*}
    \centering
    \def\svgwidth{\textwidth}
        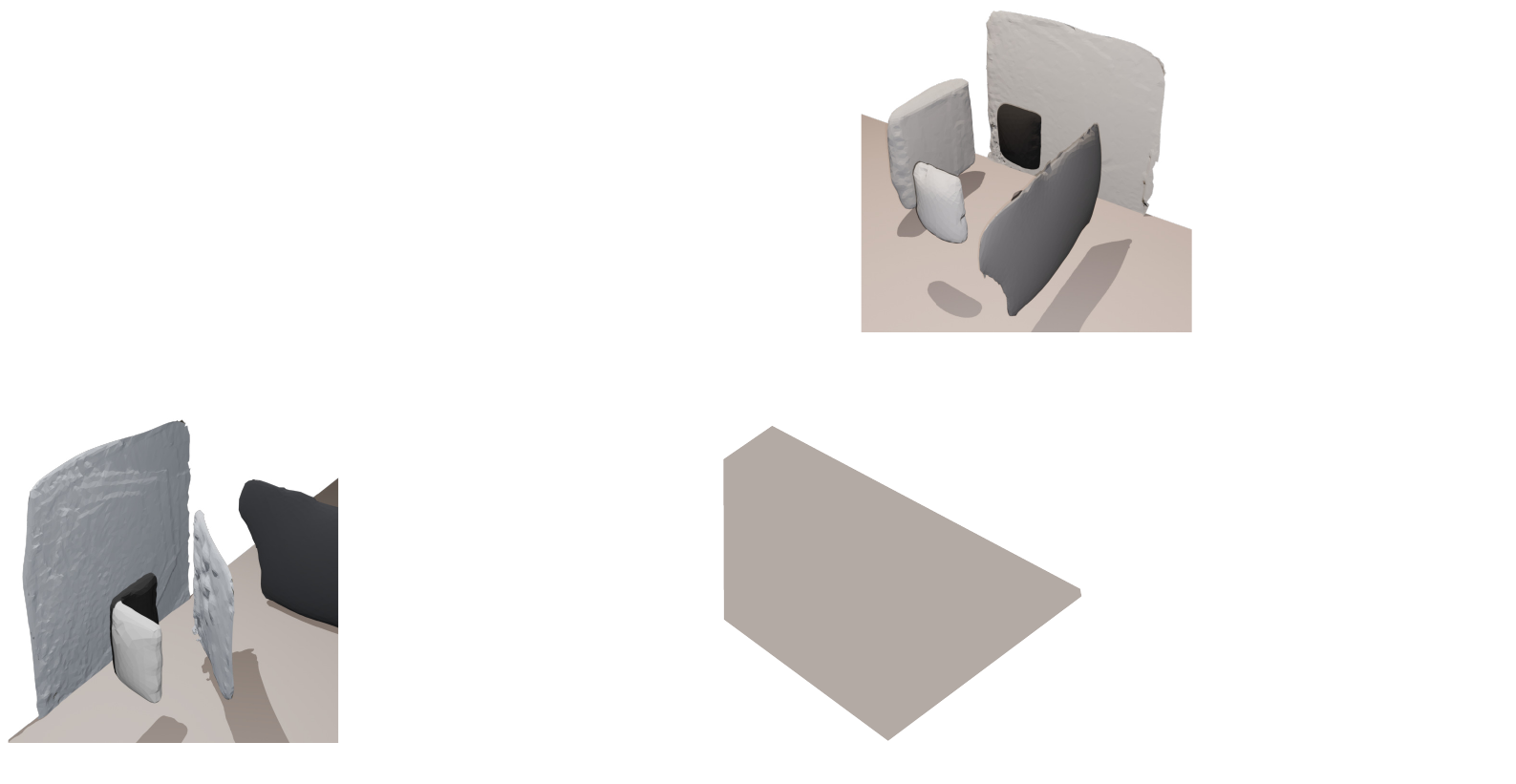
    \caption{\final{3D scans and photographs for the setups using our real prototype in Section 8. Our hardware prototype, visible in (a), is pointed towards the relay wall $\lRone$. In the 3D scans, we highlight the hidden target objects in orange, and the areas where the SPAD array (red, $\lL$) and laser (blue, $\lS$) are focused (b, c).}}
    \label{fig:setup_pictures}
\end{figure*}

\paragraph{Multi-path interference.} NLOS capture devices measure only photon arrival times, leading to multi-path interference (MPI) when two photons follow different paths but arrive simultaneously. Our simulations include MPI from all possible light paths in the hidden scene (see Figures 5 and 8 of the main paper), and our cascaded method remains robust to those, even in real-world cases (Figures 13 and 15 of the main paper).

%% file: fig/noise-v2.pdf_tex
\begingroup%
  \makeatletter%
  \providecommand\color[2][]{%
    \errmessage{(Inkscape) Color is used for the text in Inkscape, but the package 'color.sty' is not loaded}%
    \renewcommand\color[2][]{}%
  }%
  \providecommand\transparent[1]{%
    \errmessage{(Inkscape) Transparency is used (non-zero) for the text in Inkscape, but the package 'transparent.sty' is not loaded}%
    \renewcommand\transparent[1]{}%
  }%
  \providecommand\rotatebox[2]{#2}%
  \newcommand*\fsize{\dimexpr\f@size pt\relax}%
  \newcommand*\lineheight[1]{\fontsize{\fsize}{#1\fsize}\selectfont}%
  \ifx\svgwidth\undefined%
    \setlength{\unitlength}{708.57473778bp}%
    \ifx\svgscale\undefined%
      \relax%
    \else%
      \setlength{\unitlength}{\unitlength * \real{\svgscale}}%
    \fi%
  \else%
    \setlength{\unitlength}{\svgwidth}%
  \fi%
  \global\let\svgwidth\undefined%
  \global\let\svgscale\undefined%
  \makeatother%
  \begin{picture}(1,0.51694441)%
    \lineheight{1}%
    \setlength\tabcolsep{0pt}%
    \put(0,0){\includegraphics[width=\unitlength,page=1]{noise-v2.pdf}}%
    \put(0.16148171,0.38738258){\makebox(0,0)[t]{\lineheight{1.25}\smash{\begin{tabular}[t]{c}$\lRtwo$\end{tabular}}}}%
    \put(0.08231528,0.1239456){\makebox(0,0)[t]{\lineheight{1.25}\smash{\begin{tabular}[t]{c}$\lRone$\end{tabular}}}}%
    \put(0.32290012,0.07135071){\rotatebox{90}{\makebox(0,0)[t]{\lineheight{1.25}\smash{\begin{tabular}[t]{c}Noise\end{tabular}}}}}%
    \put(0.32290012,0.23094527){\rotatebox{90}{\makebox(0,0)[t]{\lineheight{1.25}\smash{\begin{tabular}[t]{c}Noiseless\end{tabular}}}}}%
    \put(0.32290012,0.38985351){\rotatebox{90}{\makebox(0,0)[t]{\lineheight{1.25}\smash{\begin{tabular}[t]{c}Reference\end{tabular}}}}}%
    \put(0.83455735,0.07135071){\rotatebox{90}{\makebox(0,0)[t]{\lineheight{1.25}\smash{\begin{tabular}[t]{c}Noise\end{tabular}}}}}%
    \put(0.83455735,0.23094527){\rotatebox{90}{\makebox(0,0)[t]{\lineheight{1.25}\smash{\begin{tabular}[t]{c}Noiseless\end{tabular}}}}}%
    \put(0.83455735,0.38985351){\rotatebox{90}{\makebox(0,0)[t]{\lineheight{1.25}\smash{\begin{tabular}[t]{c}Reference\end{tabular}}}}}%
    \put(0.93238845,0.48263406){\makebox(0,0)[t]{\lineheight{1.25}\smash{\begin{tabular}[t]{c}$f_{1,1}(\xv)$\end{tabular}}}}%
    \put(0.41647803,0.48262726){\makebox(0,0)[t]{\lineheight{1.25}\smash{\begin{tabular}[t]{c}$f_{2,2}(\xv)$\end{tabular}}}}%
    \put(0.65557272,0.0714279){\makebox(0,0)[t]{\lineheight{1.25}\smash{\begin{tabular}[t]{c}(b) Equivalent\\single-corner scene\\(third-bounce)\end{tabular}}}}%
    \put(0.14037148,0.07177926){\makebox(0,0)[t]{\lineheight{1.25}\smash{\begin{tabular}[t]{c}(a) Two-corner scene\\(fifth-bounce)\end{tabular}}}}%
    \put(0.25401943,0.25119888){\color[rgb]{0.12941176,0.61568627,0}\makebox(0,0)[t]{\lineheight{1.25}\smash{\begin{tabular}[t]{c}$\lV$\end{tabular}}}}%
    \put(0,0){\includegraphics[width=\unitlength,page=2]{noise-v2.pdf}}%
    \put(0.59593218,0.1239456){\makebox(0,0)[t]{\lineheight{1.25}\smash{\begin{tabular}[t]{c}$\lRone$\end{tabular}}}}%
    \put(0.68509881,0.45211018){\color[rgb]{0.12941176,0.61568627,0}\makebox(0,0)[t]{\lineheight{1.25}\smash{\begin{tabular}[t]{c}$\lV$\end{tabular}}}}%
    \put(0,0){\includegraphics[width=\unitlength,page=3]{noise-v2.pdf}}%
  \end{picture}%
\endgroup%

%% file: fig/setup_pics-v3.pdf_tex
\begingroup%
  \makeatletter%
  \providecommand\color[2][]{%
    \errmessage{(Inkscape) Color is used for the text in Inkscape, but the package 'color.sty' is not loaded}%
    \renewcommand\color[2][]{}%
  }%
  \providecommand\transparent[1]{%
    \errmessage{(Inkscape) Transparency is used (non-zero) for the text in Inkscape, but the package 'transparent.sty' is not loaded}%
    \renewcommand\transparent[1]{}%
  }%
  \providecommand\rotatebox[2]{#2}%
  \newcommand*\fsize{\dimexpr\f@size pt\relax}%
  \newcommand*\lineheight[1]{\fontsize{\fsize}{#1\fsize}\selectfont}%
  \ifx\svgwidth\undefined%
    \setlength{\unitlength}{761.43437423bp}%
    \ifx\svgscale\undefined%
      \relax%
    \else%
      \setlength{\unitlength}{\unitlength * \real{\svgscale}}%
    \fi%
  \else%
    \setlength{\unitlength}{\svgwidth}%
  \fi%
  \global\let\svgwidth\undefined%
  \global\let\svgscale\undefined%
  \makeatother%
  \begin{picture}(1,0.51697015)%
    \lineheight{1}%
    \setlength\tabcolsep{0pt}%
    \put(0,0){\includegraphics[width=\unitlength,page=1]{setup_pics-v3.pdf}}%
    \put(0.6039188,0.46313232){\makebox(0,0)[rt]{\lineheight{1.25}\smash{\begin{tabular}[t]{r}$\lRright$\end{tabular}}}}%
    \put(0.64450649,0.48688941){\makebox(0,0)[rt]{\lineheight{1.25}\smash{\begin{tabular}[t]{r}$\lRone$\end{tabular}}}}%
    \put(0.67532044,0.35664097){\color[rgb]{1,0.99607843,0.99607843}\makebox(0,0)[lt]{\lineheight{1.25}\smash{\begin{tabular}[t]{l}$\lRleft$\end{tabular}}}}%
    \put(0,0){\includegraphics[width=\unitlength,page=2]{setup_pics-v3.pdf}}%
    \put(0.61058437,0.33468727){\color[rgb]{0,0.00392157,0}\makebox(0,0)[t]{\lineheight{1.25}\smash{\begin{tabular}[t]{c}Target\end{tabular}}}}%
    \put(0.71494823,0.4637452){\color[rgb]{1,0.14901961,0.14901961}\makebox(0,0)[t]{\lineheight{1.25}\smash{\begin{tabular}[t]{c}$\lL$\end{tabular}}}}%
    \put(0.7661313,0.49001672){\color[rgb]{0,0.4,1}\makebox(0,0)[t]{\lineheight{1.25}\smash{\begin{tabular}[t]{c}$\lS$\end{tabular}}}}%
    \put(0.2642219,0.27392401){\color[rgb]{0,0,0}\makebox(0,0)[t]{\lineheight{0}\smash{\begin{tabular}[t]{c}(a) Setup for Figure 11 (third- and fourth-bounce imaging)\end{tabular}}}}%
    \put(0.2141824,0.00345139){\color[rgb]{0,0,0}\makebox(0,0)[t]{\lineheight{0}\smash{\begin{tabular}[t]{c}(c) Setup for Figure 15 (fifth-bounce imaging)\end{tabular}}}}%
    \put(0.73773748,0.00345139){\color[rgb]{0,0,0}\makebox(0,0)[t]{\lineheight{0}\smash{\begin{tabular}[t]{c}(d) Setup for Figure 16 (fifth-bounce imaging)\end{tabular}}}}%
    \put(0.78287722,0.27392401){\color[rgb]{0,0,0}\makebox(0,0)[t]{\lineheight{0}\smash{\begin{tabular}[t]{c}(b) Setup for Figure 13 (fifth-bounce imaging)\end{tabular}}}}%
    \put(0,0){\includegraphics[width=\unitlength,page=3]{setup_pics-v3.pdf}}%
    \put(0.39673739,0.42513903){\color[rgb]{0,0,0}\makebox(0,0)[t]{\lineheight{0}\smash{\begin{tabular}[t]{c}$\lRleft$\end{tabular}}}}%
    \put(0.27993745,0.38978178){\color[rgb]{0,0,0}\makebox(0,0)[t]{\lineheight{0}\smash{\begin{tabular}[t]{c}$\lRone$\end{tabular}}}}%
    \put(0,0){\includegraphics[width=\unitlength,page=4]{setup_pics-v3.pdf}}%
    \put(0.45397834,0.36787105){\color[rgb]{1,1,1}\makebox(0,0)[t]{\lineheight{0}\smash{\begin{tabular}[t]{c}Targets\end{tabular}}}}%
    \put(0,0){\includegraphics[width=\unitlength,page=5]{setup_pics-v3.pdf}}%
    \put(0.33564664,0.48849455){\color[rgb]{1,1,1}\makebox(0,0)[t]{\smash{\begin{tabular}[t]{c}Laser/\\SPAD\end{tabular}}}}%
    \put(0,0){\includegraphics[width=\unitlength,page=6]{setup_pics-v3.pdf}}%
    \put(0.11278777,0.48455331){\color[rgb]{0,0,0}\makebox(0,0)[t]{\lineheight{1.25}\smash{\begin{tabular}[t]{c}Targets\end{tabular}}}}%
    \put(0.05288856,0.47933942){\makebox(0,0)[lt]{\lineheight{1.25}\smash{\begin{tabular}[t]{l}$\lRleft$\end{tabular}}}}%
    \put(0.17070741,0.37008235){\color[rgb]{1,1,1}\makebox(0,0)[t]{\lineheight{1.25}\smash{\begin{tabular}[t]{c}$\lRone$\end{tabular}}}}%
    \put(0,0){\includegraphics[width=\unitlength,page=7]{setup_pics-v3.pdf}}%
    \put(0.94310164,0.46480158){\color[rgb]{0,0,0}\makebox(0,0)[t]{\lineheight{0}\smash{\begin{tabular}[t]{c}$\lRone$\end{tabular}}}}%
    \put(0,0){\includegraphics[width=\unitlength,page=8]{setup_pics-v3.pdf}}%
    \put(0.83094172,0.32264523){\color[rgb]{1,1,1}\makebox(0,0)[t]{\lineheight{0}\smash{\begin{tabular}[t]{c}Target\end{tabular}}}}%
    \put(0.85157903,0.441162){\color[rgb]{0,0,0}\makebox(0,0)[t]{\lineheight{0}\smash{\begin{tabular}[t]{c}$\lRright$\end{tabular}}}}%
    \put(0.98158745,0.42989627){\color[rgb]{0,0,0}\makebox(0,0)[t]{\lineheight{0}\smash{\begin{tabular}[t]{c}$\lRleft$\end{tabular}}}}%
    \put(0,0){\includegraphics[width=\unitlength,page=9]{setup_pics-v3.pdf}}%
    \put(0.16016543,0.07738853){\makebox(0,0)[lt]{\smash{\begin{tabular}[t]{l}\emph{Rough}\\$\lRtwo$\end{tabular}}}}%
    \put(0.12975353,0.21273611){\makebox(0,0)[lt]{\lineheight{1.25}\smash{\begin{tabular}[t]{l}$\lRone$\end{tabular}}}}%
    \put(0.07035363,0.04866984){\color[rgb]{0,0.00392157,0}\makebox(0,0)[t]{\lineheight{1.25}\smash{\begin{tabular}[t]{c}Target\end{tabular}}}}%
    \put(0.0291001,0.21481543){\color[rgb]{0,0.4,1}\makebox(0,0)[t]{\lineheight{1.25}\smash{\begin{tabular}[t]{c}$\lS$\end{tabular}}}}%
    \put(0,0){\includegraphics[width=\unitlength,page=10]{setup_pics-v3.pdf}}%
    \put(0.06625846,0.17315867){\color[rgb]{1,0.14901961,0.14901961}\makebox(0,0)[t]{\lineheight{1.25}\smash{\begin{tabular}[t]{c}$\lL$\end{tabular}}}}%
    \put(0,0){\includegraphics[width=\unitlength,page=11]{setup_pics-v3.pdf}}%
    \put(0.30281084,0.04680655){\color[rgb]{1,1,1}\makebox(0,0)[t]{\lineheight{0}\smash{\begin{tabular}[t]{c}Target\end{tabular}}}}%
    \put(0.32290707,0.18565996){\color[rgb]{0,0,0}\makebox(0,0)[t]{\lineheight{0}\smash{\begin{tabular}[t]{c}$\lRone$\end{tabular}}}}%
    \put(0,0){\includegraphics[width=\unitlength,page=12]{setup_pics-v3.pdf}}%
    \put(0.39768305,0.19111916){\color[rgb]{1,1,1}\makebox(0,0)[t]{\lineheight{0}\smash{\begin{tabular}[t]{c}$\lRtwo$\end{tabular}}}}%
    \put(0,0){\includegraphics[width=\unitlength,page=13]{setup_pics-v3.pdf}}%
    \put(0.88549455,0.08899391){\color[rgb]{1,1,1}\makebox(0,0)[t]{\lineheight{0}\smash{\begin{tabular}[t]{c}Target\end{tabular}}}}%
    \put(0.80725492,0.1749143){\color[rgb]{0,0,0}\makebox(0,0)[t]{\lineheight{1.25}\smash{\begin{tabular}[t]{c}$\lRone$\end{tabular}}}}%
    \put(0.79096742,0.16322336){\color[rgb]{0,0,0}\makebox(0,0)[rt]{\lineheight{1.25}\smash{\begin{tabular}[t]{r}$\lRright$\end{tabular}}}}%
    \put(0.93225285,0.14773403){\makebox(0,0)[lt]{\lineheight{1.25}\smash{\begin{tabular}[t]{l}$\lRleft$\end{tabular}}}}%
    \put(0,0){\includegraphics[width=\unitlength,page=14]{setup_pics-v3.pdf}}%
    \put(0.67133868,0.12739508){\color[rgb]{1,1,1}\makebox(0,0)[rt]{\lineheight{1.25}\smash{\begin{tabular}[t]{r}$\lRright$\end{tabular}}}}%
    \put(0.56772993,0.22307495){\makebox(0,0)[lt]{\lineheight{1.25}\smash{\begin{tabular}[t]{l}$\lRleft$\end{tabular}}}}%
    \put(0.51847337,0.13880425){\color[rgb]{1,1,1}\makebox(0,0)[t]{\lineheight{1.25}\smash{\begin{tabular}[t]{c}$\lRone$\end{tabular}}}}%
    \put(0,0){\includegraphics[width=\unitlength,page=15]{setup_pics-v3.pdf}}%
    \put(0.66288983,0.21470953){\color[rgb]{0,0,0}\makebox(0,0)[t]{\lineheight{1.25}\smash{\begin{tabular}[t]{c}Target\end{tabular}}}}%
    \put(0,0){\includegraphics[width=\unitlength,page=16]{setup_pics-v3.pdf}}%
    \put(0.04784974,0.3288676){\color[rgb]{1,1,1}\makebox(0,0)[t]{\smash{\begin{tabular}[t]{c}Laser/\end{tabular}}}}%
    \put(0.04807289,0.31298476){\color[rgb]{1,1,1}\makebox(0,0)[t]{\smash{\begin{tabular}[t]{c}SPAD\end{tabular}}}}%
    \put(0,0){\includegraphics[width=\unitlength,page=17]{setup_pics-v3.pdf}}%
  \end{picture}%
\endgroup%

%% file: tex/A3_setup_pictures.tex
\section{\final{Setup photographs}}
\label{sec:setup_pictures}
\final{
Here we include photographs of each scene setup in \fref{fig:setup_pictures}. Our hardware prototype, visible in \fref{fig:setup_pictures}a, uses a $16\times16$ SPAD array sensor focused on a \qtyproduct{50x32.5}{cm} area on $\lRone$ (red rectangle in the scans in Figures~\ref{fig:setup_pictures}b and c), and a laser that covers a \qtyproduct{1.9x1.9}{m} area on $\lRone$ (blue rectangle in the scans in Figures~\ref{fig:setup_pictures}b and c). \final{We use colored tape on the floor to mark the position of hidden objects; this does not have any practical effect on the captured signal.}}